\documentstyle{hydthesis}
\begin{document}
\thesistitle{ASPECTS OF POLYNOMIAL ALGEBRAS AND THEIR PHYSICAL APPLICATIONS}
\author{V. Sunilkumar}
\degreetitle{Doctor of Philosophy}
\department{School of Physics}
\submitdate{February 2002}
\titlepage
\pagenumbering{roman}
\vspace{0.7cm}
\begin{center}
{\bf  CERTIFICATE }\\
\end{center}
\vspace{1.1cm}
This is to certify that the work contained in this thesis entitled,
``Aspects of Polynomial Algebras and their Physical Applications", has
been carried out by Mr. V. Sunilkumar, under my supervision for the full
period prescribed under Ph. D. ordinance of the University and the same
has not been submitted for the research degree of any university.
\vspace{2cm}\\
Place: Hyderabad\\
Date:  \hspace{7cm}  (Thesis supervisor: Bindu A. Bambah) 
\vspace{2.5cm}\\

\noindent Dean:\\
School of Physics\\
University of Hyderabad\\
Hyderabad - 500 046\\
INDIA.
\pagebreak

\tolerance=10000
\hbadness=10000
\vbadness=10000

\begin{center}
{\bf Acknowledgment}
\end{center}
It is my greatest pleasure to thank my thesis supervisor, Dr. Bindu A Bambah
for her enthusiastic approach and  innovative guidance  
towards my thesis problem and the constant demand for the research output
which enabled me to finish my work in a short span of time. I am
also thankful to her for the financial support that was provided
when I was in a difficult situation.\\

I thank Prof. S.N Kaul, Dean, School of physics for the facilities provided to me
for my research work and also for the financial support provided to
me at the time of submission. I would also like to thank Prof. A. K. Bhatnagar, co-ordinater, SAP, for
some financial help.
\\

I would especially like to express my extreme gratitude to Prof. Jagannathan for
his invaluable help
during my entire period of research. His suggestions and discussions helped me
a lot to shape my work in the right direction.\\

I am extremely grateful to Prof. Chandrasekhar Mukku, Prof. V. Srinivasan,
Prof. A. K. Kapoor, Prof. S. Chaturvedi, Dr. P. K Panigrahi, and
Dr. M. Sivakumar for valuable suggestions and  help. The affection and
the moral support that I received from them helped me a lot to
overcome many difficult situation in life.\\

Also I would like to express my extreme gratification towards Dr. M. Sivakumar,
Dr. A. S. Samantha and Dr. Satyan Saha for providing the facility to make the
printed version of my thesis.\\

I am much grateful to my M. Sc teachers for bringing me much closer to
the ocean of physics and teaching me to appreciate the beauty of
its waves.\\

I thank Prof. Balasubramaniam, The Director of Institute of Mathematical Sciences, Chennai
for allowing me to work in the Institute. \\

I am extremely grateful to Prof. B. Sakita for spending some
time with me and also for the valuable suggestions made by him. \\

I thank Prof. K. Srinivasa Rao for providing me a chance to attend the
workshop on Mathematica and Hypergeometric series held at
Institute of Mathematical Science, Chennai.\\

I am also thankful to Dr. S. K. Soni and Dr. Subrata Chakrabarty
for various facilities provided in SERC preparatory and main schools in
High Energy Physics.\\

I thank Dr. Manjit Kaur, Dr. Anjan  Giri and Dr. Rukmani Mohanta for various
help that I received during the DAE symposium on High energy physics
held at Panjab university.\\

I would like to thank Prof. S. Datta Gupta and Dr. Prakash Matthews
for helping me to avail the computer facilities in the School.\\

I recollect here the friendly relationship with Dr. P.K. Suresh, 
Dr. C. Nagaraj Kumar and Dr. K. C. James Raju.\\

I thank Dr. E. Harikumar and Dr. N. Gurappa for various fruitful discussions
which had motivated me a lot. \\

I thank Dr. C. Raghu, K. V. Sajesh and  Subrata Pal for sending many journal papers
whenever I needed them during my Ph.D. work.\\

I would like to express my extreme gratitude towards Ms. Raishma Krishnan and Mr. Dhananjay Bambah Mukku
for the proof reading of my thesis at the final stage.\\

I am much grateful to all the office staffs of School of Physics
especially Mr. Abraham and Mrs. Saramma George for their help.\\

I also thank my friends Abey, Anup, Charan, Sankaran, Solomon, Phani and Ajith,
for providing various help during my thesis submission.\\

My father passed away when I was in the final stages of my work.
It is the moral support of my mother, sister, brother in law and  relatives that
enabled me to recover fast and to complete the work. \\

Finally I would like to recollect that it was the moral support of my parents
that gave me the strength to pursue my Ph.D. work. Also I am extremely grateful
to them for providing the full financial aid to do my Ph.D., without
which  I could not have completed the work. \\

\setcounter{page}{11} \tableofcontents
\newcommand{\al}{\alpha}
\newcommand{\be}{\begin{equation}}
\newcommand{\ee}{\end{equation}}
\newcommand{\bea}{\begin{eqnarray}}
\newcommand{\cc}{{\cal C}}
\newcommand{\ck}{{\cal K}}
\newcommand{\cj}{{\cal J}}
\newcommand{\cl}{{\cal L}}
\newcommand{\da}{\dagger}
\newcommand{\ddz}{\frac{d}{dz}}
\newcommand{\ddzt}{\frac{d^2}{dz^2}}
\newcommand{\ddzth}{\frac{d^3}{dz^3}}  
\newcommand{\ba}{\begin{array}}
\newcommand{\ea}{\end{array}}
\newcommand{\eea}{\end{eqnarray}}
\newcommand{\fr}{\frac}
\newcommand{\cp}{\cal P}
\newcommand{\Gm}{\Gamma}
\newcommand{\la}{\langle}
\newcommand{\lb}{\label}
\newcommand{\ld}{\ldots}
\newcommand{\lcb}{\left\{}
\newcommand{\lla}{\left\langle}
\newcommand{\lmd}{\left|}
\newcommand{\lra}{\longrightarrow} 
\newcommand{\lrb}{\left(}
\newcommand{\lsb}{\left[}
\newcommand{\nn}{\nonumber}
\newcommand{\ra}{\rangle}
\newcommand{\rmd}{\right|}
\newcommand{\rra}{\right\rangle}
\newcommand{\rcb}{\right\}}
\newcommand{\rrb}{\right)}
\newcommand{\rsb}{\right]}
\newcommand{\sq}{\sqrt}
\newcommand{\sm}{\sum_n}
\newcommand{\tl}{\tilde}
\newcommand{\half}{\fr{1}{2}} 
\newcommand{\cP}{\cal P}
\newcommand{\md}{\mid}
\chapter{Introduction}
\markboth{}{Chapter 1.  Introduction}

\section{A brief review of non-linear polynomial algebras}

Symmetry is an important concept in physics due to its connection
 with conservation  laws.
We define  the   symmetry
transformation on a physical system  as any  transformation
that leaves the  system  invariant.
Most symmetry operations in physics are elements of certain groups,
hence, the mathematical tool that describes symmetry is the group theory.
Group theory aids in classifying and analyzing systematically a
multitude of different symmetries which appear in nature.
If  there  are  a  continuous
infinity of transformations that leave the system
invariant, these correspond to infinite groups and the symmetry  is
classified as continuous.
Continuous symmetries are described  globally in the language of
Lie groups and locally as Lie algebras.
If the system is invariant only under  a
finite set of transformations, then the symmetry is classified as a
discrete symmetry and these transformations correspond to a finite
group. Examples  of  continuous  symmetries  are  rotations and
translations.  Examples  of  discrete  symmetries  are  reflection
symmetry(parity) and time reversal invariance.
The root of all symmetry principles lies in the connection between the
conserved quantities, the
implied invariance under a  mathematical  transformation  and  the
physical consequences of conservation laws or selection rules. This
can aid us in studying very complicated systems even when the details about forces and interactions of the system are unknown.
Symmetry methods have played a special role in quantum physics
 because they are useful
computational tools for examining many physical problems
where invariance principles provide
formulations of dynamical laws and classification of quantum states.
 The symmetry structure enables us to make
precise theoretical predictions about the various physical theories that
can be tested experimentally. The theory of relativity is based on the
symmetry that physical laws are the same for all inertial observers.
The general theory of relativity is based on the symmetry that the
physical laws will remain invariant under a general coordinate
transformation. In particle physics, the interactions between elementary
particles are determined through the principle of local gauge invariance.\\

For a long time the theory of
symmetry was restricted only to linear cases i.e. Lie groups and Lie
algebras.
The formalism of Lie groups and Lie algebras especially, generalized coherent
states and related techniques, yield simple and elegant solutions to
spectral and evolution problems \cite{P,Solomon}. In quantum mechanics,
one distinguishes two  types of
physical symmetries depending on the behaviour of the Hamiltonian $H$ under
study with respect to symmetry transformations. These are the symmetries associated
 with invariance groups $G(H)\, ([G,H]=0)$ of the
Hamiltonian. This symmetry, known as invariance symmetry, with its associated symmetry algebra
describes degeneracies of energy spectra
within fixed irreducible representations (IRs) of $G(H)$.
The other symmetry known as dynamical symmetry  connected with  spectrum generating algebras ${\it L(G)}$
 is used to give spectral decompositions
of Hilbert spaces $\cal H $ of quantum systems into invariant subspaces
$\cal H_{\lambda }$ (with $\lambda$ being labels of IRs $D(\lambda)$)
which describe certain (macroscopic) stationary states, i.e., stable
sets of states evolving in time independently under actions of $H$.

 Recently, it was realized that restricting the invariance and dynamical symmetry algebras to linear Lie algebras might be 
a narrow concept. Non-linear symmetries have been observed in string theories
and solvable models. One type of non-linear algebra that has recieved much attention
is the quantum algebra (or often called a quantum group),
which was introduced by Sklyanin \cite{s} and independently by Kulish and Reshetikhin \cite{k}
in their work on the Yang-Baxter equation.
The adjective "quantum" was given to these algebras because they arose from a quantum mechanical
problem in statistical mechanics
and also they involve the construction of an algebra from a parameter
dependent deformation of an ordinary Lie Algebra, much the same
as quantum mechanics can be considered as an $h$ dependent deformation of classical mechanics.\\

Prior to the advent of quantum groups, another class of non-linear algebras were existent in literature known as the
{\it Polynomial Algebras}.
These arose in physical systems in which it was realized
that the physical operators relevant for defining the dynamical algebra 
of a system need not form a linear (Lie) algebra, but might 
obey a nonlinear algebra.  Such nonlinear algebras are, in general, 
characterized by commutation relations of the form 
\be
\lsb N_i , N_j \rsb = C_{ij} \lrb N_k \rrb\,, 
\lb{cr} 
\ee 
where the functions $C_{ij}$ of the generators $\lcb N_k \rcb$ are constrained 
by the Jacobi identity 
\be
\lsb N_i , C_{jk} \rsb + 
\lsb N_j , C_{ki} \rsb + 
\lsb N_k , C_{ij} \rsb = 0\,.  
\ee
The functions $C_{ij}$ can be an infinite power series in $\lcb N_k \rcb$ as is 
in the case of quantum algebras 
and $q$-oscillator algebras.  When $\lcb C_{ij} \rcb$ are polynomials of the 
generators one gets the so called polynomially nonlinear, or simply polynomial 
algebras.  A special case of interest is when the commutation relations 
(\ref{cr}) take the form 
\be
\lsb N_i , N_j \rsb = c_{ij}^k N_k\,, \quad 
\lsb N_i , N_\al \rsb = t_{i\al}^\beta N_\beta\,, \quad 
\lsb N_\al , N_\beta \rsb = f_{\al\beta} \lrb N_k \rrb\,,  
\ee 
containing a linear subalgebra.  Simplest examples of such algebras occur 
when one gets 
\be
\lsb N_0 , N_\pm \rsb = \pm N_\pm\,, \qquad
\lsb N_+ , N_- \rsb = f\lrb N_0 \rrb\,. 
\lb{fn} 
\ee 
In general, the Casimir operator of this algebra (\ref{fn}) is seen to be 
given by 
\be 
\cc = N_+ N_- + g \lrb N_0 - 1 \rrb = N_- N_+ + g \lrb N_0 \rrb\,, 
\lb{genC}
\ee
where $g \lrb N_0 \rrb$ can be determined from the relation 
\be
g \lrb N_0 \rrb - g \lrb N_0 - 1 \rrb = f \lrb N_0 \rrb\,.
\lb{ggf}
\ee  

If $f\lrb N_0 \rrb$ is quadratic in $N_0$ we have a quadratic algebra and  
if $f\lrb N_0 \rrb$ is cubic in $N_0$ we have a cubic algebra.
In general a polynomial algebra is characterized by the degree of the polynomial
$f(N_0 )$.  These nonlinear
algebras, in particular the quadratic and cubic algebras, and their 
representations which we will study in this thesis,   arise in several problems in 
quantum mechanics, statistical physics, field theory, Yang-Mills type 
gauge theories and two-dimensional integrable systems.\\
Historically, the first class of polynomial algebras to be constructed were that of the cubic variety.
These were first observed in physics by
Higgs in 1979, based on the work of Lakshmanan and Eswaran (\cite{el,H}). These arose in the study of the dynamics of a particle
moving non-relativistically on a two dimensional surface embedded in
a three dimensional Euclidean space. 
Higgs demonstrated that classically and quantum mechanically the constants of motion associated with the curved analog of the Kepler problem and the
isotropic harmonic oscillator could be written in a form that explicitly revealed the dynamical symmetry of the problem. In the Kepler problem
($\lambda \neq 0 $), the dynamical algebra was constructed with the two components  of the Runge-Lenz $R_{\pm}$
vector and the diagonal component of the angular momentum $L_z$ and it was found to be a cubic algebra,
which is now known as  the Higgs algebra. The defining equations are:
\begin{eqnarray}
\left[ L_z , R_{\pm} \right] &=& \pm R_{\pm} \nonumber  \\
\left[ R_+, R_- \right] &=& (-4H + \frac{\lambda}{2} )L_z + 4\lambda L_{z}^3,
\end{eqnarray}
where $H$ is the Hamiltonian of the system.
The Runge-Lenz vectors of the system are defined as
\be
R_i = \fr{1}{2} (L_{ij} \pi _{j} -  \pi _{i} L_{ji} ) \hspace{0.7cm}, i,j =1,2 ,\\
\ee
where the canonical momentum $\pi _i =p+ \fr{\lambda }{2}(x(x.p) +(p.x)x )$
and $R_{\pm} = R_1 \pm i R_2 $, and $\lambda $ is the curvature of space.\\

For the isotropic oscillator on the sphere, the corresponding
dynamical algebra was constructed with the components of symmetric second rank Fradkin
tensor $S_{\pm}$ and the angular momentum,
\begin{eqnarray}
\left[ S_z , S_{\pm} \right] &=& \pm R_{\pm} \nonumber  \\
\left[ S_+, S_- \right] &=& 4(\lambda H + \omega ^2 -\frac{\lambda ^2}{4} )L_z - \lambda ^2  L_{z}^3.
\end{eqnarray}
where $ S_{\pm} = \frac{1}{2} (S_{11} - S_{22}) \pm i  S_{12} $ and
$ S_{ij} = p_i p_j + \omega ^2 x_i x_j $.

Both the classical and quantum cases have been studied.
In the classical case the dynamical cubic algebra is a cubic Poisson bracket algebra.
The transition to the quantum mechanical case is connected out by replacing the Poisson
brackets with angular momentum  algebra.
The Hamiltonian is expressed as a function of the Casimir of the angular momentum algebra.\\
Rewriting the Higgs algebra as
\bea
\left[ L_z , R_{\pm} \right] &=& \pm R_{\pm} \nn \\
\left[ R_+ , R_- \right]     &=& 2L_z (a + 2\lambda L_{z}^2 )
\eea
where $a=-2E +\fr{\lambda}{4}$\\
Zhedanov showed that $\lambda $ plays the role of a deformation
parameter for the $SU(2)$ and $SU(1,1)$ algebra which is retrieved
as $\lambda \longrightarrow 0 $. The Higgs algebra was shown to be a second order approximation
to the well known quantum algebra $SU_q (2) $. This can be seen by taking the
following realization of $SU_q(2) $,
\bea
\left[ J_0 , J_{\pm } \right] &=& \pm J_0 \nn \\
\left[ J_+ , J_- \right]      &=& \fr{\sin h (2\nu J_0 ) }{\sin h (\nu)}
\eea
Then a second order expansion in the parameter $\nu  $ gives the Higgs algebra.\\

The Higgs algebra was later studied by Zhedanov as a finite
deformation of $SU_q (2)$\cite{zhed}.
This simple realization enabled him to find finite dimensional representations of a certain class of Higgs algebras based
on the work of Curtwright and Zachos \cite{c} and Polychronakos\cite{p} on $SU_q(2)$.
A general deformation of the $SU(2)$ algebra was later studied by Rocek \cite{r}.
This algebra mimics the undeformed counterparts in many features and is characterized by
a deformation function. The Casimir operator of these algebras was found to be
a deformation of the quadratic Casimir of the undeformed algebra. This
deformation function is the key object for developing the representation
theory . Different types of finite dimensional representations depending on the
peculiar nature of the deformation function were found.\\

In a recent work by Floreanini et.al \cite{flo} a cubic algebra was constructed as
the dynamical symmetry algebra of rational two dimensional Calogero Model.
This algebra was used to explain the degeneracy of energy eigen states
and the explicit introduction of the wave function. Later in this thesis we shall
use the representation theory of cubic algebras that we have developed to
construct the coherent states of the Calogero-Sutherland model.\\

Quadratic algebras were first introduced by Sklyanin in the context of
Yang-Baxter equations \cite{sk1,sk2}.
The algebra that he considered was
\begin{eqnarray}
\left[ S_0 ,S_\alpha \right] &=& iJ_{\beta \gamma}(S_\beta S_\gamma + S_\gamma S_\beta ) \nonumber \\
\left[S_\alpha , S_{\beta }\right] &=& i(S_0 S_\gamma + S_\gamma S_0 ),
\end{eqnarray}
where $\alpha ,\beta $ and $\gamma$ assume the values $1,2,3$.

The structure constants are constrained by the relations,
\begin{equation}
J_{12}-J_{23} + J_{31} -J_{12}J_{23}J_{31} =0
\end{equation}
The finite and infinite dimensional representations for this particular
algebra in terms of elliptic functions were constructed.
Sklyanin first established the connection between integrable
systems and quadratic algebra.\\
The quadratic algebra was obtained as a hidden dynamical symmetry
in Coulomb and isotropic oscillator with a anisotropic term $ \frac{1}{r^2 \sin \theta}$
\cite{z1,z2}. In the two cases, the additional term does not completely destroy
the $SO(4)/SU(3) $ accidental degeneracy. It is known that in the Coulomb and the isotropic
oscillator case, the corresponding accidental symmetry enables one to separate the equation
of motion in three different coordinate system. In the anisotropic case,
it was shown that the Schr$\ddot{o} $dinger equation separates in two different coordinate systems.
Another example of the system in which this occurs is the Hartmann potential\cite{z1}.  
For the Hartmann potential,
\be
u(r)= -\fr{\alpha}{r} + \fr{\beta}{r^2 \sin ^2 \theta },
\ee
  three operators are found to commute with the
Hamiltonian. One is the modified angular momentum $ \tilde{L^2} =L^2 + \frac{2\beta } {\sin^{2} \theta} $
and the second the modified Runge Lenz vector, $ \tilde{A_z} =[\partial _z , \frac{L^2}{2} -\alpha r ]$.
The third one is the $L_z$, the diagonal component of angular momentum, which also commutes with the other two generators.
The hidden symmetry algebra in this case obeyed by the modified operators,
\begin{equation}
T_1 = \tilde{L^2} ,~~~~~T_2= (-2E)^{1/2}\tilde{A_z}, ~~~~~~~T_3=L_z
\end{equation}
is the quadratic Hahn algebra given by:
\begin{eqnarray}
\left[ T_1,T_2 \right] &=&T_3 \nonumber \\
\left[ T_3, T_1 \right] &=& 2(T_1 T_2 + T_2 T_1) \nonumber \\
\left[ T_2, T_3 \right] &=& 2 T^{2}_2 + 4T_1 -(2m^2 + 2\beta -1-\frac{\alpha ^2}{E}),
\end{eqnarray}
where $a=-8,\,\,b_1=12,\,\,c=4(2\epsilon +1),\,\,d_1 =4[3m+\epsilon(1+2m^2)],\,\,
d_2=4(\epsilon +m^2 ),\,\, \epsilon=\fr{E}{\omega }-\fr{3}{2} $.\\
$E$ is the eigen value of the Harmonic oscillator.
In the case of anisotropic oscillator the two integrals that commute with
the Hamiltonian are $ T_1 = L^2 + \frac{2\beta}{\sin^2 \theta }$ and
$ T_2 = \frac{1}{2} (p^2 + \omega ^2 r^2) -\frac{1}{2} $.
These generators along with $T_3=L_z$ also satisfy a quadratic algebra,
\begin {eqnarray}
\left[ T_1,T_2 \right] &=& T_3 \nonumber \\
\left[ T_1, T_3 \right] &=& a(T_1 T_2 + T_2 T_1) + b_1 T_2 + c T_1 + d_1 \nonumber \\
\left[ T_3, T_2 \right] &=& aT^{2}_1 + bT_1 + cT_2 + d_2 ,
\end{eqnarray}
where $a,b_1,b_2 ,c, d_1$ and $d_2 $ are structure constants. The finite dimensional representation
corresponding to the energy degeneracy were constructed. The overlap function between
the two coordinate systems in which the
Schr$\ddot{o} $dinger equation separates is a Hahn polynomial apart from the constant factor
and the vacuum amplitude.\\

Another important non-linear polynomial  algebra that has been studied in physics
is the $W$-algebra \cite{tj}. These are constructed from a Kirillov Poisson
algebra of a Lie algebra by Poisson reduction techniques.
These W-algebras are a special case of the deformed non-linear algebra.
The finite W algebra obtained by three generators has been identified as the
symmetry algebra of an anisotropic oscillator with
the frequency ratio $2:1$. It is a quadratic algebra of the form \cite{tj},
\begin{eqnarray}
\left[ N_0 ,N_{\pm} \right] &=& \pm 2N_{\pm} , \nonumber \\
\left[ N_+ ,N_- \right] &=& N^2_{0} +C,
\end{eqnarray}
 which is a special case of  $W$-algebra called $W^{(2)}_3 $ algebra. The recent developments in $W$-algebra can be
found in \cite{tj}(and references there in ).\\

Many of the nonlinear algebras that arise in physical problems are of the form of
real form of the complex algebra $SL(2)$ with a nonlinear term .
A study in that direction was done by Abdesselam et.al \cite{ab}.
The algebra studied by the authors are given by,
\begin{eqnarray}
\left[ N_0 , N_+ \right] &=& N_{\pm} \nonumber \\
\left[ N_+ , N_- \right] &=& \sum_p \beta _{p} (2N_0 )^{2p +1 }
\end{eqnarray}
where $p=0,1,2...$.
 For $p=1$ one can map this algebra to the Higgs algebra \cite{H}.
The finite dimensional representations were investigated in the angular momentum basis.
This is done by finding a realization of the generators of the nonlinear algebra in
terms of the angular momentum generators. Such a mapping has been studied by many authors in
literature(\cite{p},\cite{r},\cite{c},\cite{vs},\cite{G}) in different contexts. The representations
found were restricted to the finite dimensional representation.
By introducing various deformation parameters different classes of
finite dimensional representations have been found.
Some are peculiar to the nonlinear algebra and do not exist
in the case of linear algebra. Abdessalam et.al further showed that
the non-linear algebras can be equated with a Hopf structure through the
knowledge of the undeformed $SL(2)$ co-product structure. \\

In super symmetric quantum mechanics many conditionally exactly solvable(CES) systems
associated with an exactly solvable potential are found to contain a non-linear algebraic
structure. Junker and Roy on their study on CES potential showed that the non-linear algebra
do exist (\cite{j1},\cite{j2}). The CES potential which are SUSY partners of linear oscillator gives a
quadratic algebra, while the CES potential corresponding to a radial harmonic oscillator gives rise
to a cubic algebra. More work in this direction has been done by Sukhatme, Datt and co workers \cite{sukh}.

It has been observed by Vinet and co-workers that the superintegrable system have a non-linear algebra
known as a dynamical invariance algebra \cite{vin}. It has been shown that many of the
quantum superintegrable system that have been constructed from this classical
counterpart possesses a deformed oscillator algebra. For such systems eigenvalues
of the state with finite dimensional degeneracies have been then calculated algebraically\cite{b3}. The superintegrable system of two dimensional anisotropic
oscillator having quadratic algebra as a dynamical invariance algebra
has a cubic algebra as it's spectrum generating algebra. From these superintegrable
system one can get quasi exactly solvable systems by a dimensional reduction
which has to do with the underlying polynomially deformed symmetry algebra.
The generators of this symmetry algebra of the superintegrable systems have the
property that it can be taken as the product of known Lie algebras such as
$SU(2)$ or $SU(1,1)$ so that the representation space of the non linear algebras
can be taken as the product space of Lie algebras under some constraints
which will reduce the degree of freedom in the product space. This approach
has been examined in detail in the thesis. It was found that the extra conserved
quantities form the constraints required to close the non-linear algebra.
In this way one can get the representation  economically. One can
also map these to various non-linear algebras.\\

This thesis is devoted to a study of polynomial algebras, their representations and applications.
 In the next section we present a review of polynomial algebras and their relation to Lie algebras.
\section{Polynomial algebras and Lie Algebras}

The Polynomial algebras in a loose sense are a generalization of Lie algebras, but also
differ from them in that they do not form a vector space.
The  detailed theory of Lie algebra has been explored in great detail by both mathematicians
and physicists \cite{gl,hm,wy,fuch}(references there in), and is
are out of scope of this thesis. A very short description of main features of the
Lie algebra is given below. This also establishes the terms and notation used in the subsequent chapters.

{\bf Definition:}

                 A Lie algebra $g$ is an algebra with a bilinear mapping, called Lie bracket,
                 satisfying the following properties:\\
For $x,y,z \in g$ and the Lie bracket $\left[ ,\right] $ 
\begin{eqnarray}
&i)& \left[x,y\right] = z \in g ~~~~(\mbox{closure}) \nonumber \\
&ii)& \left[x,y\right] = -\left[y,x \right]~~~~ (\mbox{antisymmetry}) \nonumber \\
&iii)& \left[\alpha x+ \beta y,z \right] = \alpha \left[x,z\right] + \beta \left[y,z\right] \nonumber \\
  &  &\left[x,\alpha y+ \beta z \right] = \alpha \left[x,y \right] + \beta \left[ x,z \right]~~~~ (\mbox{bilinearity} )\nonumber \\
&iv)& \left[x,\left[y,z \right]\right] + \left[z,\left[x,y \right]\right]+
\left[y,\left[z,x \right]\right] =0 ~~~~(\mbox{Jacobi identity})
\end{eqnarray}

Given any associative algebra {\it U} with a product *, one can obtain an associated Lie Algebra , by considering {\it U } as a vector space
and defining a Lie Bracket as the {\it commutator} with respect to the original multiplication i.e
for $x,y,\in g$  $\left[x ,y\right]=x*y-y*x $. One, hence constructs a Lie algebra on the same vector space {\it U}.
The dimension  $d=dim(g)$ of a Lie algebra is the dimension of g considered as a vector space .
For a finite dimensional Lie algebra one uses the notation
${\it B}=\{X_i | i=1,2...d\}$ to denote any basis {\it B} of $g$ and refers to  $X_i$
as the generators of the Lie algebra.
The generators $X_i$ span the Lie Algebra g linearly and one defines the Lie bracket of the generators by the relation,
\begin{equation}
\left[ X_i ,X_j \right] = C^{k}_{ij} X_{k}. ~~~~~k=1,2....d,
\end{equation}
where $C^{k}_{ij}$ are known as structure constants of the Lie algebras.
They are constrained by the Jacobi identity,
\begin{equation}
\sum_{m}^{d} \lrb C^{m}_{jk} C^{l}_{jm} + C^{m}_{ij} C^{l}_{km} +C^{m}_{ki} C^{l}_{jm} \rrb =0.
\end{equation}

Consider a subalgebra {\it h} such that for all $H_i  \in {\it h}
 \left[ H_i,X_j\right] \in {\it h}$  for every $ X_j \in {\it g} $,
then ${\it h}$ is called the ideal of the Lie Algebra ${\it g}$.
An Abelian Lie Algebra is one for which $ \left[X_i,X_j\right]=0 $
for every $X_i \in {\it g}$ and a simple Lie algebra contains only
Abelian ideals.
A direct sum of simple Lie algebras is called {\it semi-simple}.\\

For a semisimple Lie Algebra, the convenient basis chosen for physical applications is
the {\it Cartan-Weyl} basis defined in the following way.
Let ${\it h}$ be the subalgebra of ${\it g}$ containing  ${H_i} $, the maximum
number of mutually commuting elements of ${\it g}$.
${\it h}$ is called the Cartan subalgebra of ${\it g}$.
Then the elements of the Cartan basis are $H_i$ and $E_\alpha$ such that
\bea
\lsb H_i , H_j \rsb &=& 0 \nn \\
\lsb H_i , E_\alpha  \rsb &=& \alpha^i E_{\alpha}
\eea
The $r$ dimensional vectors $\alpha^i$ are called root vectors or root
of $G$.\\
The simplest  example of a three dimensional Lie algebra is the angular momentum algebra
known as $SU(2)$ in literature .The defined commutation relation is,
\begin{equation}
\left[ J_i ,J_j \right] = i\epsilon _{ijk} J_{k} ,~~~~~i,j,k =1,2,3.
\end{equation}
The {\it Cartan Weyl} basis  of $SU(2)$ is given by $J_z, J_+,J_-$, where
$J_z=J_3$,
$J_+=\frac{J_1+iJ_2}{\sqrt{2}}$
and $J_-=\frac{J_1-iJ_2}{\sqrt{2}}$ \\
Lie algebras are linear algebras because the bracket between the two elements of
the Lie algebra give
a linear combination of all the generators.
It should be noted that any higher order term(quadratic,cubic etc.)
is not a member of the Lie algebra but a member of the Universal
Enveloping Algebra defined as follows:\\

{\bf Definition:}\\

A universal enveloping algebra(UEA) is generated by all the possible ordered polynomials of the generators of the Lie Algebra,
subject to the condition that two elements of the UEA are equal if they satisfy the basic commutation relations of the Lie algebra.
For example the universal enveloping algebra  $U(SU(2))$ for the Lie algebra SU(2)  with generators $L_+, L_-$ and $ L_z$ is generated by the basis elements
$(L_{-})^n(L_z)^m(L_+)^l, \forall \,\,\, l,m,n \in Z$.
Thus it is an infinite dimensional algebra. The Lie algebra is a
subalgebra of the universal enveloping algebra.
The Casimir operator C is an element of the UEA such
that $\left[C,L_{\pm.z}\right]=0$.\\

After this introduction to Lie Algebras we proceed to define the
polynomial algebras on similar lines.
A polynomial algebra is a sub-algebra of the Universal Enveloping algebra
with the properties\\

{\bf Definition}:\\

                 A polynomial algebra is a infinite dimensional Lie algebra
                 associated with a bilinear bracket satisfying the following
                 properties. For $x,y,z \in g$ and a polynomial function
                 $h(z)$
\begin{eqnarray}
&i)& \left[x,y\right] = h(z) \nonumber \\
&ii)& \left[x,y\right] = -\left[y,x \right]~~~~ (\mbox{antisymmetry}) \nonumber \\
&iii)& \left[\alpha x+ \beta y,z \right] = \alpha \left[x,z\right] + \beta \left[y,z\right] \nonumber \\
&&    \left[x,\alpha y+ \beta z \right] = \alpha \left[x,y \right] + \beta \left[ x,z \right] ~~~~ (\mbox{bilinearity} )\nonumber \\
&iv)& \left[x,\left[y,z \right]\right] + \left[z,\left[x,y \right]\right]+
\left[y,\left[z,x \right]\right] = 0 ~~~~(\mbox{Jacobi identity})
\end{eqnarray}
Properties (ii) and (iv) together
implies that the nonlinear algebra is a non commutative non associative algebra.
Unlike the Lie algebras non-linear algebras cannot be identified directly with a vector space.
Since the commutation in general is a non linear function of generators
which cannot be an element of the algebra(the product of two generators is
not defined in this algebra). They satisfy the following properties.

In similarity with Lie algebras one can define a basis
${\cp}=\{N_i,\,\,\, i=1,2...d\}$ which  is finite dimensional, but does not span
a linear vector space. One considers the polynomial algebra as the sub algebra
of a Universal enveloping algebra of a corresponding linear Lie algebra.
This enables us to define structure constants of the polynomial algebra as the
\be
[N_i \, , \, N_j ]= f^{k}_{ij} (N_l)(N_k)
\ee
This looks superficially like a Lie algebra except that the structure functions
are a function of $N_l$ rather than constants. If the $f_{ij}^k$ are polynomials
in $N_l$ the non linear algebra is termed as a polynomial algebra. If $f^{k}_{ij}$ are linear in $N_l $
we get a quadratic algebra. If they are quadratic in $N_l$ we get a cubic algebra.

The $f_{ij}^k $ are still constrained by  $ii$ and $iv$.
\begin{eqnarray}
&a)&  f_{ij}^k(N_l) = -f_{ji}^k(N_l) \nonumber \\
&b)& \left[N_i , f_{jk}^i(N_l) \right] + \left[N_j , f_{ki}^j( N_l) \right]+ \left[N_k , f_{ij}^k(N_l)\right] = 0
\end{eqnarray}
In many cases these non-linear algebra admit a coset structure i.e among the generators there is a linear subalgebra and the commutator of the remaining generators give a symmetric function of the generators of the linear algebra.
$P$ of the $d$ generators $N_i$
satisfy
\begin{equation}
\left[ N_i ,N_j \right] = C^{k}_{ij} N_k , ~~~~~~~i,j=1...P.
\end{equation}
The remaining $N-P$ generators satisfy,
\begin{eqnarray}
\left[ N_i ,N_{\alpha} \right]& = &t^{\beta}_{i \alpha}N_{\beta} , \nonumber \\
\left[ N_{\alpha} ,N_{\beta} \right] &=& f_{\alpha \beta} (N_{\gamma}) ,
\end{eqnarray}
A simplest example of such algebra is
the case with $N=1$.
\begin{eqnarray}
\left[ N_0 ,N_{\pm} \right] &=& \pm N_{\pm} \nonumber \\
\left[ N_+ ,N_{-} \right] &=& f(N_0)
\end{eqnarray}
This is also known as polynomial $SU(2)$ or $SU(1,1)$ algebra. Even
though the set ${N_i}$ does not close as a vector space because of nonlinear term coming from
the polynomial algebra, we will call this algebra, in a loose sense, as a
$d$ dimensional polynomial algebra. So the above algebra defines a general three dimensional polynomial algebra.
When $f(N_0)$ is quadratic it is quadratic algebra.
A study of such algebras can be carried out in parallel to Lie algebras to some
extent but contain more features.\\

For a three dimensional polynomial algebra with a deformation function $f$
the Casimir operator is defined as
\begin{equation}
C=N_- N_+ + g(N_0 ) = N_+ N_- =g(N_0 -1)
\lb{casimir}
\end{equation}
where $g(N_0 )$ given by $g(N_0 ) -g(N_0 -1 )=f(N_0 ) $ is called the structure function
of the algebra.\\
 From the definition of $g(N_0)$ it is clear that $g(N_0)$ will be a
 polynomial of one degree greater than $f(N_0)$.
The $SU(2)$ algebra is a special case for which$f_{ij}^k=i\epsilon _{ijk}$.

\section{Other non-linear algebras in literature}

Since much of the work done on the representation theory of polynomial
non-linear algebras relies heavily on the work done for q-deformed
algebras and quantum groups, for completeness
we briefly review the history of these and other non-linear algebras which appear in the literature.

Quantum groups were first studied by Drinfeld \cite{d} and Jimbo \cite{ji}
 in
terms of a deformation of the universal enveloping algebra (UEA) of any simple complex Lie Algebra.
The enveloping algebra is a Hopf algebra i.e one can define an action of the UEA (A) onto  itself known as the adjoint action
commonly denoted by a Hopf product $\Delta:A \longrightarrow A \times A\,\,\,\forall \,\,A \in UEA $, known as the co-product.
For example for U(SL(2)), the Hopf product is the commutator:
\be
(ad L^{\pm,z})A=L^{\pm}A-AL^{\pm,z} ~~~~\forall A \in U(Sl(2))
\ee
 The quantum  algebra is obtained from the
universal enveloping algebra of a Lie algebra by a quantization procedure
\cite{d,ji}. The whole process of getting a quantum algebra from a Lie algebra $g$
can be represented by the diagram,
$$
g  \longrightarrow  U( g) \longrightarrow  U_h (g)
\longrightarrow  g_h
$$
The quantum universal
enveloping algebra is obtained by deforming the UEA by
a deformation parameter $q=e^h $. The UEA associated with the quantum group is  a Hopf algebra \cite{ch} with a deformed Hopf structure.
The standard quantum algebra corresponding to SL(2) is given by the commutation relations:
\begin{eqnarray}
L_+ L_-  - L_- L_+ &=& \frac{q^{L_0} -q^{-L_0 }}{q-q^{-1}} \nonumber\\
L_0 L_\pm  - L_\pm  L_0 &=& \pm L_\pm
\end{eqnarray}
Since the commutator of $ L_+$ and $L_-$ is a power series in $L_0$, 
it is a non-linear algebra. This is still a Hopf algebra\cite{ch} with a deformed Hopf structure given by
\bea
(ad L_{\pm})A &=&L_{\pm}Aq^{\frac{L_z}{2}}-q^{\pm}q^{\frac{L_z}{2}}AL_{\pm} \\ \nn
(ad L_z)A &=&L_{z}A-AL_{z} ~~~~\forall A \in U(SU(2))
\eea
An interesting 
$ U_{q}(SL(2)) $ is generated  by  $L_{h}^{\pm}$ and $L^{h}_0 $  defined by,
\begin{eqnarray}
L_{h}^{\pm } &=& \lrb\frac{2}{(q+q^{-1})}\rrb^{1/2}q^{-L_0 /2} L^{\pm}, \nonumber \\
L^{h}_0 &=& \frac{2}{q+q^{-1}} (qL^+ L^- -q^{-1} L^- L^+ ),
\end{eqnarray}
In this realization the adjoint action replaces the role of the commutator and we may find a three dimensional subspace which is closed under the adjoint action.
The Hopf structure is preserved in this particular realization in which the adjoint action  goes over to the deformed adjoint action and
the commutation relations show the property,
\begin{eqnarray}
\left[ L^{+}_h ,L^{-}_h \right]_h &=& L^{h}_0 , \nonumber \\
\left[ L^{h}_0 ,L^{\pm}_h \right]_h &=& \pm 2q^{\pm 1} L^{\pm}_h ,\nonumber \\
\left[ L^{h}_h ,L^{0}_h \right]_h &=& 2(q-q^{-1} )L^{h}_0 .
\end{eqnarray}
The above algebra is closed under a quantum Lie bracket which plays the role
of the quantum adjoint action and will go to the
classical Lie bracket under the classical limit
$q \rightarrow 1 $, i.e $0^{th}$ order in $h$ (thus justifying the adjective "quantum").\\

The quantum algebra corresponding to the Heisenberg group
was found by Bidenharn and Macfarlane(\cite{bi},\cite{m}), known as the q-oscillator algebra. The q-oscillator
algebra is defined by q-boson creation and annihilation operators which
act on a q-vacuum. The q-oscillator algebra is given by,
\bea
\left[ N_q,a_q \right] &=& -a_q \nn \\
\left[ N_q, a^{\dagger}_q \right] &=& a^{\dagger}_q \nn \\
a^{\dagger}_q a_q -q^{1/2} a_q a^{\dagger}_q &=& q^{-N_q /2}
\eea

Another important deformed algebra which has obtained a lot of attention in physics
is the generalized deformed oscillator and deformed para fermionic algebra.
The deformed oscillator algebra was introduced by Daskaloyannis \cite{das}. It is characterized by a
structure function $F$ satisfying,
\begin{eqnarray}
\left[a ,a^{\dagger} \right] &=& F(N+1) - F(N), \nonumber \\
\left[a ,N  \right] &=&  a  \nonumber \\
\left[a^{\dagger}, N  \right] &=&  -a^{\dagger}.
\end{eqnarray}
$N$ is the number operator satisfying $a a^{\dagger} = F(N) $ and $ a^{\dagger}a = F(N+1) $.
$F(x) $ is a positive analytical function and $F(0) =0 $.
The eigenvalues and eigen states of the deformed oscillator algebra have been calculated.
This deformation is related to a Rocek type of $SU(2)$ deformation \cite{r}
by a Schwinger type realization, where bosonic operators are
replaced by generators of a deformed oscillator algebra. This generalized deformed
oscillator was used to describe the parabose and para fermi quantization scheme.\\

The deformed fermionic algebra was constructed by Bonatsos and Daskaloyannis \cite{b}.
In this case the structure function $F(x)$ is a positive analytical function
defined on the closed interval $[0,2]$ obeying , $ F(0)=F(2)=0 $ and $F(1)=1$.
A polynomial realization of the fermionic algebra
was also constructed. A peculiar feature of the deformed fermionic algebra is that
all the realizations are mutually equivalent and also equivalent to the undeformed
fermionic algebra. So only one type of fermion can exist which is the usual (undeformed)
fermion, while different kinds of bosons can exist with different kinds of deformation (\cite{b1},\cite{b2}).\\

Various deformations of parabosons and parafermions have been studied in literature
(\cite{f},\cite{ce},\cite{od},\cite{kr}).
Some of them are related to the quantum superalgebra $Osp_q (1/2,R)$ (\cite{f},\cite{ce}).
Generalized deformed parafermions were studied by Quesne \cite{q}. There exists a mapping between
deformed parafermionic algebra and the nonlinear deformation of $SO(3)$. Such an
$SO(3)$ deformation is used as a spectrum generating algebra for the
modified P$\ddot o $schl-Teller and Morse potentials \cite{sukh}.\\
\section{Outline of the Thesis}

In this thesis a unified approach for constructing polynomial
algebras and their representations of these algebras is given. Two
different cases, the quadratic and cubic algebras, are studied
extensively. The different realizations of the generators of these
algebras in the space of analytic functions are found. Various
coherent states of  systems that possess the non linear algebra
as a symmetry algebra are constructed, in a manner that is similar
to   the Lie algebraic approach. For this, a mapping of the
polynomial algebra to Lie algebras is defined and used to find
coherent states by  a generalization of the method in ref
\cite{vs}. These coherent states are used to find the eigenspectrum and
dynamic evolution of  quantum optical systems.

 In the first chapter (Introduction), a brief review of  the different
 kinds
of nonlinear algebras appearing in physics is given. For the sake
of clarity, some aspects of  Lie algebras are mentioned . Then,
the definition of  non linear algebras is given and  the main
features are explained.\\

The second chapter begins with a detailed study of  quadratic
algebras which are  special cases  of nonlinear algebras. These
algebras appear as  the hidden symmetry algebras of many
Hamiltonian systems. Bosonic realizations of these algebras are
constructed by taking a product space of three Fock states. The
algebras are studied in the real form. These quadratic algebras
have the structure of $SU(2) $ or $SU(1,1)$ deformations. It is
found that such an algebraic structure can be generated by taking
combinations of the generators of the Heisenberg algebra and
$SU(1,1)$ or $ SU(2)$ generators. This provides a unified way of
producing a large class of quadratic algebras. The structure
constants of the algebra contain the Casimir operator of the
component algebra $SU(2)$ or $SU(1,1) $. The quadratic algebra is
not defined completely by the three generators but is closed only
when augmented by an operator that commutes with all the other
generators. The Casimir of the component algebra  also commutes
with all generators . The representations of the quadratic algebras
are constructed using the representations of the individual
algebras and taking a product basis. The operators that commute
with the generators constrain the space and the representation
space of the quadratic algebra is a projected product Hilbert
space. Both finite dimensional and infinite dimensional unitary
irreducible representations are possible. The unitary nature of
the representation depends on the value of the parameter that
comes from the individual algebras. The condition for different
unitary irreducible representations is found.

Chapter 3 begins with the construction of cubic algebras. This can
be done by a Jordan-Schwinger type of construction. All the
operators that commute with the generators of the cubic algebra
are found. These operators are used to construct the
representations. The representation space is a projection of the
product space of two $SU(2)$ or/and $SU(1,1)$ representations. A
four mode cubic representation of the algebra is also possible.
Another important thing observed is that the cubic algebras are
generated not only by the linear $SU(2)$ and $SU(1,1)$ algebras,
but, also by taking a Heisenberg algebra and quadratic algebra.
The differential realizations of the generators of the various
cubic algebras are also constructed.\\

It is observed that the non-linear algebras have a product
structure containing a lower order algebra. This leads to a
generalization of Jordan-Schwinger mapping which is presented in
fourth chapter. The well known Jordan Schwinger realization of the
$SU(2)$ algebra comes as a special case of this construction.
Starting from the Heisenberg algebra of order zero a chain of
higher order algebras can be constructed.A way to map the
polynomial algebras to the linear Lie algebras such as
$SU(2),SU(1,1)$ and Heisenberg algebras is done by defining a
deformation function. These mappings are realized in the
representation space of the polynomial algebras.
\\

The fifth chapter is dedicated to the application of the nonlinear
algebras . A way of constructing coherent state of these algebras
is obtained with the help of a mapping to the Heisenberg and
$SU(2) $ algebras.The Barut Girardello type coherent states are
constructed. The overcompleteness property and the resolution of
the identity of the coherent states  are shown. The construction
of Perelomov type coherent states is difficult  for the nonlinear
algebras because there is no disentanglement formula to obtain the
disentanglement as in the case of $SU(2)$ coherent states. We
overcome this difficulty by making use of the mapping given in
chapter 4. The algebras are mapped to $SU(2)$ and $SU(1,1)$
algebras and  the Perelomov coherent states are calculated by
making use of the techniques of the $SU(2)$(or $SU(1,1)$)
Perelomov states. These coherent states are appropriate to
describe various multiphoton processes in quantum optics. \\

In chapter 6, some of the physical applications of the quadratic
and cubic algebras are given. Some interesting mathematical
properties of the quadratic algebra are highlighted. The case of
the quantum anharmonic oscillator is considered and the connection
of its degeneracy structure with the theory of partitions is
given. Quantum optical processes such as the Dicke model of N two
level atoms interacting with a radiation field and generic three
photon processes are studied within the framework of the
polynomial algebras. The connection of cubic and quadratic
algebras with superintegrable systems is revealed. A new cubic
algebra symmetry of the two body Calogero model is examined and
coherent states are constructed. The construction of the algebraic
coherent states of the polynomial algebras and their connection
with the construction of quasi exactly solvable quantum systems is
presented.

Chapter 7 is the concluding section of this theses . Further
possible applications of nonlinear algebra are proposed.



\setcounter{equation}{0}
\chapter{Quadratic Algebras: Construction and Representations.}
\markboth{}{Chapter 2.   Quadratic Algebras....}
We have seen in chapter 1 that quadratic algebras  appear as
the dynamical symmetry algebras of many quantum mechanical systems \cite{Ga,G}.
Depending on the particular application of the algebra some finite dimensional
representations have been constructed \cite{L,R}. A natural question that arises is whether any infinite dimensional representations
for such algebras exist. If so, a systematic approach is needed to get all possible
representations. It has observed been that the quadratic algebras have a linear Lie
algebraic structure inherent in it, in the sense that they can be generated from
more than one Lie algebra.
 In section 1 of this chapter a general method of construction
for a class of three dimensional quadratic algebras that admit a coset structure is given. The representation theory is
investigated in section 2 and the differential realization on the space of analytic functions is found in section 3.\\

A  general three dimensional quadratic algebra is defined by,
\bea
\left[ N_0, N_{\pm} \right] &=& \pm N_{\pm} \nonumber \\
\left[ N_+, N_- \right] &=& a N_{0}^{2} + b N_0 + c ,
\lb{qgn}
\eea
where the structure constants $a,b$ and $c$ are constants which will take
constant values in any irreducible representation.
The Casimir operator for the algebra (\ref{qgn}) is given by,
\bea
\cc &=& N_+ N_- + g(N_0) \nonumber \\
\cc &=& Q_+Q_- + \fr{1}{3} a Q_0^3 - \half(a-b)Q_0^2
             + \fr{1}{6}(a-3b+6c)Q_0 - c\,. 
\eea

The above definition gives
a spectrum of
quadratic algebras corresponding to different values for the
parameters $a, b,$ and $c$. These parameters can be related to the physical parameter
of the system if we identify this quadratic algebra as the symmetry algebra of the
physical system. In this section a method to generate four different classes
of three dimensional quadratic algebras are given.
\section{Construction of three dimensional quadratic algebras}

In this section a method to construct four different classes of quadratic algebras are given
The construction is made out of the linear algebras in the following way,

Let us consider the generators of the $SU(2)$ algebra and Heisenberg algebra
satisfying the commutation relation,
\bea
\lsb J_0 , J_\pm \rsb &=& \pm J_\pm\,, \quad \lsb J_+ , J_- \rsb = 2J_0\,.  \\
\lsb a , a^{\da} \rsb &=& 1 \,, \quad  \quad \lsb N , a \rsb =-a \,, \quad \lsb N , a^{\da} \rsb = a^{\da}
\eea
We consider 4 seperate cases.\\
{\bf Case a: $ Q^-$ (2)}\\

Now consider the following operators constructed out of the above six generators.
Let,
\bea
Q_0 &=& \half \lrb J_0 - N \rrb\,, \qquad Q_+ = J_+ a\,, \qquad Q_- = J_- a^\da\,, \nonumber \\
\cl &=& \half\lrb J_0 + N \rrb\,, \qquad \cj = J_+J_- + J_0\lrb J_0 - 1 \rrb = J^2 . 
\eea
Then we have the following quadratic algebra,
\bea 
\lsb Q_0 , Q_\pm \rsb & = & \pm Q_\pm\,, \nn \\
\lsb Q_+ , Q_- \rsb & = & -3Q_0^2 - \lrb 2\cl - 1\rrb Q_0 
                          + \lrb \cj + \cl \lrb\cl+1\rrb \rrb\,, 
\eea  
and
\be
\lsb \cl , \cj \rsb = 0\,, \qquad 
\lsb \cl , Q_{0,\pm} \rsb = 0\,, \qquad 
\lsb \cj , Q_{0,\pm} \rsb = 0\,.
\ee 
Here $\cl$ and $\cj $ are the constant operators(central elements) of the algebra.
The Casimir operator of the algebra is then function of
these operators given by,
\be
\cc = Q_+Q_- - Q_0^3 - (\cl-2)Q_0^2 + \lrb \cj+\cl^2+2\cl-1 \rrb Q_0 
             - \lrb \cj + \cl \lrb \cl+ 1 \rrb \rrb\,. 
\ee
The above algebra can be  identified with the general quadratic algebra (\ref{qgn}) by identifying
\be
a=-3\mu , b=-\lrb 2\cl -1 \rrb\mu {\mbox{and}} c=\mu \lrb\cj +\cl \lrb \cl +1 \rrb \rrb .
\ee
From the representation theory of $SU(2)$ algebra we know that $\cj$ can
take only the values, $j \lrb j +1 \rrb $, where, $j=\fr{1}{2},1.\fr{3}{2} \, \ld $.
From the definition of $ \cl $ it it follows that $\cl $ can take values  $\fr{1}{2}
\lrb \pm \fr{m}{2} + n \rrb$, where $m=1,2 \, $ and $n=0,1 \, \ld $.\\

{\bf Case b: $Q^+$ (2)}\\
Now consider the generators 
\bea
Q_0 &=& \half \lrb J_0 + N \rrb\,, \qquad Q_+ = J_+ a^{\da}\,, \qquad Q_- = J_- a\,, \nonumber \\
\cl &=& \half\lrb J_0 - N \rrb\,, \qquad \cj = J_+J_- + J_0\lrb J_0 - 1 \rrb = J^2 . 
\eea
They obey the quadratic algebra,                
\bea 
\lsb Q_0 , Q_\pm \rsb & = & \pm Q_\pm\,, \nn \\
\lsb Q_+ , Q_- \rsb & = & 3Q_0^2 + \lrb 2\cl + 1\rrb Q_0 
                          - \lrb \cj + \cl \lrb\cl+1\rrb \rrb\,. 
\lb{q+2}
\eea

Here $\cl$ and $\cj$ are central elements.
The Casimir of the algebra is,
\be
\cc = Q_+Q_- + Q_0^3 + (\cl-1)Q_0^2 - \lrb \cj+\cl^2 \rrb Q_0 
             + \lrb \cj + \cl\lrb\cl-1\rrb \rrb\,. 
\lb{cb}
\ee

{\bf Case c: $Q^-$ (1,1)}\\
Now we will consider the Generators of the $SU(1,1)$ algebra instead of
the $SU(2)$ algebra  given by,
\bea
\lsb K_0 , K_\pm \rsb &=& \pm K_\pm \nn \\
\lsb K_+ , K_- \rsb &=& - 2K_0\,.  \\
\eea

Then consider the following operators,
\bea
Q_0 &=& \half \lrb K_0 -N \rrb\,, \qquad Q_+ = K_+ a\,, \qquad Q_- = K_- a^{\da}\,, \nonumber \\
\cl &=& \half\lrb K_0 + N \rrb\,, \qquad \ck = K_+K_- - K_0\lrb K_0 - 1 \rrb = K^2 . 
\eea
By taking the commutators of the above opertors it is
found that they closes as a quadratic algebra. The quadratic algebra satisfied by the above generators is,
\bea 
\lsb Q_0 , Q_\pm \rsb & = & \pm Q_\pm\,, \nn \\ 
\lsb Q_+ , Q_- \rsb & = & 3Q_0^2 + (2\cl-1) Q_0 
                        + \lrb \ck - \cl\lrb\cl+1\rrb \rrb\,. 
\lb{q-11}
\eea
The Casimir operator for the above algebra is given by,
\be
\cc = Q_+Q_- + Q_0^3 + (\cl-2)Q_0^2 + \lrb \ck-\cl^2-2\cl+1 \rrb Q_0 
             - \lrb \ck - \cl\lrb\cl+1\rrb \rrb\,. 
\lb{cc}
\ee

{\bf Case d: $Q^+$ (1,1)}\\

The fourth class of quadratic algebra is obtained by choosing the operators,
\bea
Q_0 &=& \half \lrb K_0 + N \rrb\,, \qquad Q_+ = K_+ a^{\da}\,, \qquad Q_- = K_- a\,, \nonumber \\
\cl &=& \half\lrb K_0 - N \rrb\,, \qquad \ck = K_+K_- + K_0\lrb K_0 - 1 \rrb = K^2 . 
\eea
The quadratic algebra satisfied in this case is given by,
\bea 
\lsb Q_0 , Q_\pm \rsb & = & \pm Q_\pm\,, \nn \\ 
\lsb Q_+ , Q_- \rsb & = & -3Q_0^2 - (2\cl + 1)Q_0 
                          - \lrb\ck - \cl\lrb\cl-1\rrb \rrb\,,
\lb{q+11}
\eea
and the casimir operator by,
\be
\cc = Q_+Q_- - Q_0^3 - (\cl-1)Q_0^2 - \lrb \ck-\cl^2 \rrb Q_0 
             + \lrb \ck - \cl\lrb\cl-1\rrb \rrb\,. 
\lb{cd}
\ee
This method of generating the quadratic algebra resembles the
well-known Jordan-Schwinger realization of su(2) algebra.
This point will be studied in detail in the fourth chapter. For
simplicity let us call these four classes of algebras as
$Q^- (2) , Q^+ (2) , Q^- (1,1)$ and $Q^+ (1,1)$. The number in the
bracket indicate the generator that is combined with the Harmonic operator generators.
The$+$ or $-$ fixes form of $Q_0$ (and therefore $Q_{\pm}$). In the next
sections we will study the representations and the differential realization of
the quadratic algebras generated above

\section{Representations of $Q^- (2)$ }

We have found that the four classes of quadratic algebra
contains  central elements which commute with all other generators  and also
among themselves. From the definition of the generators of quadratic algebra
it is obvious that, if we consider the product space of the unitary irreducible
representation space of $SU(2)$(or $SU(1,1)$) and the Fock space of the
Heisenberg algebra, the representation space of the quadratic algebra will be a projected
space of this product space. Consider the product space of $S(2)$ and Heisenberg algebra.
\be
\lmd j,m,n \rra = \lmd j,m \rra \lmd n \rra\,,
\lb{qprd}
\ee

Let $l$ be the value of $\cl$ in this space. Then one have,
\be
l=\fr{m+n}{2}.
\lb{con-2}
\ee
 Therefore, a subspace
with a fixed value for $l$ will be the irreducible space for the
Quadratic algebra, labeled by the value of $l$. A basis state for
such a space is obtained by imposing the constrained (\ref{con-2}).
The basis states are given by,
\be
\lmd j,l,m \rra = \lmd j,m \rra \lmd 2l-m \rra\,,
\ee
such that 
\be
\cj\lmd j,l,m \rra = j(j+1)\lmd j,l,m \rra\,, \qquad 
\cl\lmd j,l,m \rra = l\lmd j,l,m \rra\,,
\ee

Since $m$ has a upper bound and lower bound in $SU(2)$ representation,
it is obvious that the Irreducible representation of the Quadratic algebra will be a
finite dimensional representation. The dimension depends on the value
of $j$ and $l$. For a general $g$ and $l$ the representation are given by,
\bea
Q_0\lmd j,l,m \rra & = & (m-l)\lmd j,l,m \rra\,, \nn \\
Q_+\lmd j,l,m \rra & = & \sqrt{(j-m)(j+m+1)(2l-m)}\lmd j,l,m+1 \rra\,, \nn \\
Q_-\lmd j,l,m \rra & = & \sqrt{(j+m)(j-m+1)(2l-m+1)}\lmd j,l,m-1 \rra\,, \nn \\ 
  &   & \qquad \qquad m = -j, -j+1,\,\ld\,j-1, j\,.  
\lb{q2-rep1}
\eea 
Two cases are possible with the same form (\ref{q2-rep1}).\\
\noindent 
{\it Case}-I\,: $2l-j \geq 0$

\medskip 

In this case the set of basis states 
\be
\lmd j,l,m \rra = \lmd j,m \rra \lmd 2l-m \rra\,, \quad 
m = -j, -j+1,\,\ld\,j-1, j\,,  
\ee
carry the $(j,l)$-th, $(2j+1)$-dimensional, unitary irreducible representation  

Now as an example consider the two dimensional representation.
This corresponding to $j = 1/2$ and for each value of $l$ $=$ $1/4,3/4,5/4,\,\ld\,,$ there is a 
two dimensional representation given by 
\bea 
Q_0 & = & \lrb \ba{cc}
               -\half-l & 0   \\
                     0   & \half-l \ea \rrb\,, \nn \\  
Q_+ & = & \lrb \ba{cc} 
                   0 & 0 \\
           \sqrt{2l+\half} & 0 \ea \rrb\,, \qquad 
Q_- = \lrb \ba{cc}
          0 & \sqrt{2l+\half}  \\
          0 & 0          \ea \rrb\,, \nn \\ 
\cj & = & 3/4\,, \qquad \cl = l\,, \qquad \cc = (-4l^3+7l+3)/4\,, 
\eea
as can be verified directly. \\
\noindent 
{\it Case}-II\,: $2l-j < 0$

\medskip 

In this case the set of basis states 
\be
\lmd j,l,m \rra = \lmd j,m \rra \lmd 2l-m \rra\,, \quad 
m = -j, -j+1,\,\ld\,2l\,,  
\ee
carry the $(j,l)$-th unitary irreducible representation of dimension $j+2l+1$ 
In this case for any $j > 1/2$ there is a two dimensional representation 
corresponding to $l = (1-j)/2$.  Explicitly, 
\bea 
Q_0 & = & \half \lrb \ba{cc}
                         -j-1 & 0   \\
                           0  & 1-j \ea \rrb\,, \nn \\  
Q_+ & = & \lrb \ba{cc} 
                   0 & 0 \\
           \sqrt{2j} & 0 \ea \rrb\,, \qquad 
Q_- = \lrb \ba{cc}
          0 & \sqrt{2j}  \\
          0 & 0          \ea \rrb\,, \nn \\ 
\cj & = & j(j+1)\,, \quad \cl = (1-j)/2\,, \quad 
\cc = (-3j^3+5j^2+11j+3)/8\,, \nn \\ 
    &   &  
\eea
as can be verified directly. \\

Note that unlike the $SU(2)$ case the quadratic algebras have many two dimensional
basic representations.

By using the two-mode bosonic realization of $su(2)$ we can write down the 
three-mode bosonic realization of $Q^-(2)$ as 
\be
Q_0 = \fr{1}{4}\lrb a_1^\da a_1-a_2^\da a_2-2a_3^\da a_3 \rrb\,, \quad 
Q_+ = a_1^\da a_2a_3\,, \quad Q_- = a_1a_2^\da a_3^\da\,.
\lb{q2-bose}
\ee
Correspondingly we can take the basis states of the irreducible 
representations as the three-mode Fock states 
\be
\lmd j,l,m \rra = \lmd j+m \rra\lmd j-m \rra \lmd 2l-m \rra\,. 
\ee 
Then the action of $\lrb Q_0,Q_+,Q_- \rrb$ defined by (\ref{q2-bose}) on these basis states 
leads to the irreducible representations (\ref{q2-rep1}) in the two cases $2l-j$ $\geq$ $0$ and $2l-j$ $<$ $0$.  

Let us now make the association
\be
\lmd j,m,l \rra \lra \fr{z_2^{2j}z_3^{2l+j}\lrb z_1/z_2z_3\rrb^{j+m}}
                     {\sqrt{(j+m)!(j-m)!(2l-m)!}}\,.
\ee
Since $j$ and $l$ are constants, this shows that the set of functions
\be
\psi^-_{j,l,n}(z) = \fr{z^n}{\sqrt{n!(2j-n)!(2l+j-n)!}}\,, \quad 
 n = 0,1,2,\,\ld\,,2j,\ {\mbox{or}}\ 2l+j\,, 
\ee
forms the basis for the representation (\ref{q2-rep1})  
corresponding to the single variable realization 
\bea
Q_0 & = & z\fr{d}{dz}-j-l\,, \nn \\ 
Q_+ & = & z^3\fr{d^2}{dz^2}-(2l+3j+1)z^2\fr{d}{dz}+2j(2l_j)z\,, \quad
Q_- = \fr{d}{dz}\,.                        
\eea
\section{Representations of $Q^+(2)$} 
For the algebra $Q^+ (2)$ defined by (\ref{q+2}) the constraint in the
product space(\ref{qprd})
is given by,
\be
l=\fr{m-n}{2}.
\lb{q-2 con}
\ee
The basis states for the
irreducible representation are given by.
\be
\lmd j,l,m \rra = \lmd j,m \rra \lmd m-2l \rra\,, 
\lb{q2+base}
\ee
such that 
\be
\cj\lmd j,l,m \rra = j(j+1)\lmd j,l,m \rra\,, \qquad 
\cl\lmd j,l,m \rra = l\lmd j,l,m \rra\,
\ee
The corresponding finite dimensional representation is given by,
\bea
Q_0\lmd j,l,m \rra & = & (m-l)\lmd j,l,m \rra\,, \nn \\
Q_+\lmd j,l,m \rra & = & \sqrt{(j-m)(j+m+1)(m-2l+1)}\lmd j,l,m+1 \rra\,, \nn \\
Q_-\lmd j,l,m \rra & = & \sqrt{(j+m)(j-m+1)(m-2l)}\lmd j,l,m-1 \rra\,, \nn \\
  &   & \qquad \qquad m = -j, -j+1,\,\ld\,j-1, j\,.  
\lb{q2+rep}
\eea 
It can seen from (\ref{q-2 con}) that $m-2l$ will be always positive.
So the unitary irreducible representations are of dimension $2j +1$.
The Casimir operator (\ref{cb}) takes the value 
$(1-l)\lsb j(j+1)-l(l+1) \rsb$ in this representation.\\
The two dimensional representations correspond to $j = 1/2$ and $l$ $=$ 
$-1/4,-3/4,-5/4,\,\ld\,,$ and are given by 
\bea 
Q_0 & = & \lrb \ba{cc}
                -\half-l & 0   \\
                          0  & \half-l \ea \rrb\,, \nn \\  
Q_+ & = & \lrb \ba{cc} 
                   0 & 0 \\
           \sqrt{\half-2l} & 0 \ea \rrb\,, \qquad 
Q_- = \lrb \ba{cc}
          0 & \sqrt{\half-2l}  \\
          0 & 0          \ea \rrb\,, \nn \\ 
\cj & = & 3/4\,, \qquad \cl = l\,, \qquad 
\cc = \fr{1}{4}\lrb 4l^3-7l+3 \rrb\,.  
\eea

By using the two-mode bosonic realization of $su(2)$ we can write down the 
three-mode bosonic realization of $Q^+(2)$ as 
\be
Q_0 = \fr{1}{4}\lrb a_1^\da a_1-a_2^\da a_2+2a_3^\da a_3 \rrb\,, \quad 
Q_+ = a_1^\da a_2a_3^\da\,, \quad Q_- = a_1a_2^\da a_3\,.
\lb{q2+bose}
\ee
Correspondingly we can take the basis states (\ref{q2+base}) of the irreducible 
representations as the three-mode Fock states 
\be
\lmd j,l,m \rra = \lmd j+m \rra\lmd j-m \rra \lmd m-2l \rra\,. 
\ee 
Then the action of $\lrb Q_0,Q_+,Q_- \rrb$ defined by (\ref{q2+bose}) on these basis 
states leads to the irreducible representation (\ref{q2+rep}).  

Now, we can make the association
\be
\lmd j,m,l \rra \lra \fr{z_2^{2j}z_3^{-(2l+j)}\lrb z_1z_3/z_2\rrb^{j+m}}
                     {\sqrt{(j+m)!(j-m)!(m-2l)!}}\,.
\ee
Since $j$ and $l$ are constants, this shows that the set of functions
\be
\psi^+_{j,l,n}(z) = \fr{z^n}{\sqrt{n!(2j-n)!(n-2l-j)!}}\,, \quad 
 n = 0,1,2,\,\ld\,,2j\,, 
\ee
forms the basis for the representation (\ref{q2+rep}) corresponding to the 
single variable realization 
\be
Q_0 = z\fr{d}{dz}-j-l\,, \quad
Q_+ = -z^2\fr{d}{dz} + 2jz\,, \quad
Q_- = z\fr{d^2}{dz^2} - (2l-j-1)\fr{d}{dz}\,.
\ee 
.
\section{Representations of $Q^-(1,1)$}

For the algebra $Q^-(1,1)$ defined in (\ref{q-11}), instead of $SU(2)$
representation we will consider $SU(1,1)$ representations.
The corresponding product state is given by,
\be
\lmd k , l , n ,m \rra   \lmd k , l , n \rra \lmd m \rra
\ee
In this case
$l$ obey the constraint,
\be
l=\fr{k+m+n}{2}
\ee
The basis are the set of states,
\be
\lmd k , l , n \rra = \lmd k,n \rra \lmd 2l-k-n \rra\,, \quad 
\lb{q11-base}
\ee
\be
2l-k = 0,1,2,\,\ld\,, \qquad k = 1/2, 1, 3/2,\,\ld\,, 
\ee  
and 
\be 
\ck \lmd k , l , n \rra = k(1-k) \lmd k , l , n \rra\,, \quad 
\cl \lmd k , l , n \rra = l \lmd k , l , n \rra\,. 
\ee 
The unitary irreducible representations are given by,
\bea 
Q_0 \lmd k , l , n \rra & = & (k-l+n) \lmd k , l , n \rra\,, \nn \\ 
Q_+ \lmd k , l , n \rra & = & 
  \sqrt{(n+2k)(n+1)(2l-k-n)}\,\lmd k , l , n+1 \rra\,, \nn \\
Q_- \lmd k , l , n \rra & = & 
  \sqrt{(n+2k-1)n(2l-k-n+1)}\,\lmd k , l , n-1 \rra\,. \nn \\ 
 &  & \qquad \qquad n = 0,1,2,\,\ld\,,(2l-k)\,. 
\lb{q11-rep} 
\eea 
Since $2l-k$ is always positive the unitary irreducible representations are
finite dimensional of dimension $2l-k+1$.\\

The Casimir operator (\ref{cc}) has the value
$(l+1)\lsb k(1-k) + l(l-1)\rsb$ in this representation. \\

For this algebra there is a two dimensional representation for each value of 
$k = 1/2,1,3/2,\ld\,,$ as given by 
\bea 
Q_0 & = & \half \lrb \ba{cc}
                     k-1 & 0   \\
                     0   & k+1 \ea \rrb\,, \quad 
Q_+ = \lrb \ba{cc} 
                   0 & 0 \\
           \sqrt{2k} & 0 \ea \rrb\,, \quad 
Q_- = \lrb \ba{cc}
          0 & \sqrt{2k}  \\
          0 & 0          \ea \rrb\,, \nn \\ 
\ck & = & k(1-k)\,, \quad 
\cl = l = \frac{1}{2}(k+1)\,, \quad   
\cc = \frac{1}{8}\lrb -3k^3 - 5k^2 + 11k - 3 \rrb\,, \nn \\
  &   &  
\eea
as can be verified directly. 

By using the two-mode bosonic realization of $su(1,1)$ we can write down the 
three-mode bosonic realization of $Q^-(1,1)$ as 
\be
Q_0 = \fr{1}{4} \lrb a_1^\da a_1 + a_2^\da a_2 
           - 2a_3^\da a_3 + 1 \rrb\,, \quad  
Q_+ = a_1^\da a_2^\da a_3\,, \quad 
Q_- = Q_+^\da = a_1 a_2 a_3^\da\,.  
\lb{q11-bose}
\ee
Translating the basis states (\ref{q11-base}) into the three-mode Fock states it 
is found that the action of $\lrb Q_0, Q_+, Q_- \rrb$ defined by (\ref{q11-bose}) 
on the basis states 
$\lcb\lmd k,l,n \rra\right.$ $=$ 
$\left.\lmd n+2k-1 \rra \lmd n \rra \lmd 2l-k-n \rra\rcb$ 
leads to the representation (\ref{q11-rep}).  

As before, let us make the association 
\be
\lmd k , l , n \rra \lra \fr{z_1^{2k-1} z_3^{2l-k} \lrb z_1z_2/z_3 \rrb^n}
                            {\sqrt{(n+2k-1)!n!(2l-k-n)!}}\,. 
\ee
Since $k$ and $l$ are constants for a given representation we can take 
\be
\phi^-_{k,l,n}(z) = \fr{z^n}{\sqrt{(n+2k-1)!n!(2l-k-n)!}}\,, \quad 
  n = 0,1,2,\,\ld\,,(2l-k)\,, 
\ee
as the set of basis functions for the single variable realization 
\be 
Q_0 = z \ddz + k - l\,, \quad 
Q_+ = -z^2 \ddz + (2l-k)z\,, \quad 
Q_- = z \ddzt + 2k \ddz\,,
\label{q11-an}
\ee
leading to the representation (\ref{q11-rep}).  

\section{Representations of $Q^+(1,1)$}
In this case the constraint is given by,
\be
l=\fr{k+n-m}{2}
\lb{q+11 con}
\ee
The basis states for the irreducible representations are given by,
\be
\lmd k , l , n \rra = \lmd k,n \rra \lmd n+k-2l \rra , 
\lb{q11+base}
\ee
where,
\be
k-2l = 0,1,2,\ld\,, \quad k = 1/2, 1, 3/2, \ld\,, 
\ee
and 
\be
\ck \lmd k,l,n \rra = k(1-k) \lmd k,l,n \rra\,, \qquad
\cl \lmd k,l,n \rra = l \lmd k,l,n \rra\,. 
\ee            0
In this case  from (\ref{q+11 con} it is clear that only infinite dimensional
unitary irreducible representations are possible, given by,
\bea 
Q_0 \lmd k,l,n \rra & = & (k-l+n) \lmd k,l,n \rra\,, \nn \\ 
Q_+ \lmd k,l,n \rra & = & \sqrt{(n+2k)(n+1)(n+k-2l+1)}\,\lmd k,l,n+1 \rra\, \nn \\  
Q_- \lmd k,l,n \rra & = & \sqrt{(n+2k-1)n(n+k-2l)}\,\lmd k,l,n-1 \rra\,, \nn \\ 
   &   & \qquad \qquad n = 0,1,2,\ld\,. 
\lb{q11+rep}
\eea 
The Casimir operator (\ref{cd}) has the value $l\lrb l-k^2 \rrb$ in this 
representation.\\ 

In terms of three bosonic modes the realization of $\lrb Q_0, Q_+, Q_- \rrb$
is given by 
\be
Q_0 = \fr{1}{4} \lrb a_1^\da a_1 + a_2^\da a_2 
           + 2a_3^\da a_3 + 1 \rrb\,, \quad  
Q_+ = a_1^\da a_2^\da a_3^\da\,, \quad 
Q_- = Q_+^\da = a_1 a_2 a_3\,. 
\lb{q11+bose} 
\ee 
Application of this realization on the set of three-mode Fock states 
\be
\lmd k , l , n \rra = \lmd n+2k-1\rra \lmd n \rra \lmd n+k-2l \rra\,, \quad 
 n = 0, 1, 2,\,\ld\,, 
\ee
leads to the $(k,l)$-th irreducible representation (\ref{q11+rep}).  

From the association 
\be 
\lmd k , l , n \rra \lra \fr{z_1^{2k-1} z_3^{k-2l}\lrb z_1 z_2 z_3 \rrb^n} 
                         {\sqrt{(n+2k-1)!n!(n+k-2l)!}} 
\ee 
it is clear that we can take 
\be 
\phi^+_{k,l,n}(z) = \fr{z^n}{\sqrt{(n+2k-1)!n!(n+k-2l)!}} 
\ee
as the set of basis functions for the single variable realization associated 
with the representation (\ref{q11+rep}).  The corresponding realization is 
\bea
Q_0 & = & z \ddz + k - l\,, \qquad Q_+ = z\,, \nn \\  
Q_- & = & z^2 \ddzth + (3k-2l+2) z \ddzt + \lrb 2k^2-4kl+2k \rrb \ddz\,.
\eea
To conclude this chapter , starting with $su(2)$, $su(1,1)$, and an oscillator algebra, we 
have constructed four classes of three dimensional quadratic algebras of 
the type 
\be
\lsb Q_0 , Q_\pm \rsb = \pm Q_\pm\,, \quad 
\lsb Q_+ , Q_- \rsb = a Q_0^2 + b Q_0 + c\,.  
\ee  
In each class the structure constants $(a,b,c)$ take particular series of 
values.  We have also found for these algebras the three-mode bosonic 
realizations, corresponding matrix representations, and single variable 
differential operator realizations.
Some interesting physical applications of these algebras will be given in chapter 6.


\setcounter{equation}{0}
\chapter{Cubic Algebras: Generation and Representations.}
\markboth{}{Chapter 3.   Cubic Algebras ....}

In the previous chapter we have seen a method to generate
and classify  three dimensional quadratic algebras with a coset structure.
The present chapter is devoted to the construction and representation of
cubic algebras, and also their differential realizations.
Examples of cubic algebras are
 the well-known Higgs algebra,
which arises in the study of the dynamical symmetries of the
 the Coulomb problem in a space of constant
curvature 
and symmetry algebras of many exactly solvable quantum mechanical problems
of the Calogero-Sutherland type\cite{h,flo}.

A  general three dimensional cubic algebra  with a coset structure is given by,
\bea
\left[ C_0, C_{\pm} \right] &=& \pm C_{\pm} \nonumber \\
\left[ C_+, C_- \right] &=& a C_{0}^{3} + b C_{0}^{2} + cC_{0} +d ,
\eea
The structure constants $a,b,c$ and $d$ are constants. The Casimir operator
for this algebra is given by\ref{casimir}
\be
\cc = C_- C_+ + \fr{a}{4} C_{0}^{4} + (\fr{b}{3} - \fr{a}{2} )C_{0}^{3} +
(\fr{a}{4} + \fr{b}{2} + \fr{c}{2} )C_{0}^{2} + (\fr{b}{6} -\fr{c}{2} +d)C_0
\ee
The definition of Casimir operator follows from the definition of polynomial
algebras.
In the following sections on the cubic algebra, we generate the cubic algebra
following an algorithm analogous
to that given for quadratic algebra.
\section{Construction of three dimensional cubic algebras}
In this section we present a general method to construct a different
classes of cubic algebra.
The algebraic and transparent way  how this has done will greatly facilitate the physical applications of
these algebras.
We will first construct the cubic algebras by taking two commuting $SU(1,1)$ algebras generated by $L$ 
and $M$ and  two commuting $SU(2) $ algebras generated by $J$ and $P$ and
then construct the Jordan-Schwinger
type realizations with them.
We label the various cases
$C_{\alpha}(a,b)$
 where $ a,\,\,b=11,\,\, 2,\,\,q_{\pm} 1,\,\,
q_{\pm}2, h$ depending on the
subalgebras used to construct the cubic algebra.
For example $11$ corresponds to a $SU(1,1)$ algebra, $2$ corresponds
to a $ SU(2)$ algebra and $q_{\pm} 1$ and $q_{\pm} 2$
correspond to the quadratic algebras dicussed in chapter 2. 
The $\alpha $ can take either $+$ or $-$ sign. If $\alpha=+$ then $C_0 $ is the sum of the
diagonal operators and if $\alpha =-$ then $C_0$ is the difference of the diagonal operators.
The notations will be clear in the following discussions of the different cases.
Consider two commuting SU(1,1) generators $(L_0,L_{\pm})$ and $(M_0,M_\pm)$.
One can construct two distinct cubic algebras out of the above  $SU(1,1)$ algebras.
The construction follows the same philosophy as the quaudratic algebra case. The two
different cases are given below.\\
{\bf Case 1: $ C_- (11,11)$}\\
Consider the operators
\begin{eqnarray}
K &=& \left(L_0+M_0\right)/2\,,  \nonumber \\
C_0 &=& \left(L_0-M_0\right)/2\,, \nonumber \\
C_+ &=& \mu L_+M_- \,,\nonumber \\
C_- &=& \mu L_-M_+ \,, \nonumber \\
C_1&=&L_+L_-+L_0(L_0-1)=L^2,\;\;\; C_2=M_+M_-+M_0(M_0-1)=M^2 .
\end{eqnarray}
The cubic algebra obtained by the operators $\lrb C_0 , C_+ , C_- \rrb $
are given by 
\bea 
\lsb C_0 , C_\pm \rsb & = & \pm C_\pm  \nonumber  \\
\lsb C_+ , C_- \rsb  &=&( -4\mu^2 C_0^3 +C_0 (4 K^2\mu^2 - \sigma )+\lambda K ) \nonumber \\
\lsb C_{1,2} , K \rsb & = & 0  \;\;\;
\lsb C_{1,2} , C_{0,\pm} \rsb =0 \;\;\;  
\lsb K , C_{0,\pm} \rsb = 0, 
\eea
where,
$ \sigma  = 2\mu^2 (C_1 +C_2 )$
 and $\lambda =  2\mu^2 (C_1 - C_2 ) $.

Here $\mu$ is a constant introduced to identify the given cubic algebra
with the symmetry algebra of some physical problem. This will be clear in
chapter 6 where we will consider some physica systems. The $\mu $ will appear
in the algebra as an overall multiplication factor. The Casimir operator
of the algebra can be calculated by finding the structure function $g(C_0 )$
given by,
$g(C_0)= -\mu^2 C_0^2(C_0+1)^2 + +C_0^2(2K^2\mu^2-\frac{1}{2}\sigma)+C_0(2\mu^2K^2-\frac{1}{2}\sigma+\lambda K)$.

Then Casimir operator of the algebra is given by
\bea
\cc= C_- C_+ 
 -\mu^2 C_0^2(C_0+1)^2 + C_0^2(2K^2\mu^2-\frac{1}{2}\sigma) 
 +C_0(2\mu^2K^2-\frac{1}{2}\sigma+\lambda K) .
\eea
By taking a suitable choice one can reduce the above algebra to the
well known Higgs algebra.
\bea 
\lsb C_0 , C_\pm \rsb & = & \pm C_\pm\,,  \nonumber \\
\lsb C_+ , C_- \rsb  &= &
 h C_0^3 +2aC_0 \\
\eea
This is done by choosing,
$C_1=C_2\;\; , -4\mu^2=h \;\; a=\mu^2(2K^2-C_1)$.\\
{\bf Case 2: $C_+ (11,11)$} .\\

Now consider the same $SU(1,1)$ algebras in a different combination given by
\bea
C_0 &=& \half \lrb L_0 + M_0 \rrb\,, \qquad C_+ = L_- M_-\,, \qquad C_- = L_+ M_+\,, \nonumber \\
\ck &=& \half\lrb L_0 - M_0 \rrb \nonumber \\
C_1 &=& L_+L_- + L_0\lrb L_0 - 1 \rrb = L^2 \, , \qquad C_2 = M_+M_- + M_0\lrb M_0 - 1 \rrb = M^2 .
\eea
The above operators  also satisfy a cubic algebra given by
\bea 
\lsb C_0 , C_\pm \rsb & = & \pm C_\pm\,, \nn \\
\lsb C_+ , C_- \rsb & = & \lrb -4 C_{0}^{2} + C_{0} \lrb 4\ck^2  -\sigma \rrb - \lambda \ck \rrb ,
\eea  
where, $\sigma = 2 \lrb C_1 + C_2 \rrb $ and $ \lambda =  \lrb C_1 - C_2 \rrb $.

The Casimir operator of the algebra is given by

\be
\cc = C_- C_+ - C_{0}^{2} \lrb C_0 +1\rrb^2 + C_{0}^{2} \lrb 2 \ck ^2 -\fr{1}{2} \sigma \rrb + C_0 \lrb 2\ck^2 -\fr{1}{2} \sigma - \lambda \ck \rrb .
\ee
A different class of cubic algebra is obtained by replacing the $SU(1,1)$
generators by $SU(2)$ generators.\\ 
{\bf Case 3: $C_- (2,2)$}\\
Let $(J_0 ,J_{\pm})$ and $(P_0 ,P_{\pm})$ be
two sets of $SU(2)$ generators. Consider the following operators
constructed with the $SU(2)$ generators,
\bea
C_0 &=& \half \lrb J_0 - P_0 \rrb\,, \qquad C_+ = J_+ P_-\,, \qquad C_- = J_- P_+\,, \nonumber \\
\ck &=& \half\lrb J_0 + P_0 \rrb \nonumber \\
\cj_1 &=& J_+J_- + J_0\lrb J_0 - 1 \rrb = J^2 \, , \qquad \cj_2 = P_+P_- + P_0\lrb P_0 - 1 \rrb = P^2 .
\eea
The cubic algebra satisfied by these operators is given by
\bea 
\lsb C_0 , C_\pm \rsb & = & \pm C_\pm\,, \nn \\
\lsb C_+ , C_- \rsb & = &  4 C_{0}^{3} - C_{0} \lrb 4\ck^2  +\sigma \rrb - \lambda \ck  ,
\eea  
where, $\sigma = 2 \lrb \cj_1 + \cj_2 \rrb $ and $ \lambda =  \lrb \cj_1 - \cj_2 \rrb $.\\

The Casimir operator of the algebra is given by
\be
\cc = C_- C_+ - C_{0}^{2} \lrb C_0 +1\rrb^2 + C_{0}^{2} \lrb 2 \ck ^2 -\fr{1}{2} \sigma \rrb + C_0 \lrb 2\ck \lrb \ck  -\fr{1}{2}  \lambda \rrb + \fr{1}{2} \sigma \rrb .
\ee

{\bf Case 4: $C_+(2,2)$}\\
The  set of operators are given by,
\bea
C_0 &=& \half \lrb J_0 + P_0 \rrb\,, \qquad C_+ = J_+ P_+\,, \qquad C_- = J_- P_-\,, \nonumber \\
\ck &=& \half\lrb J_0 - P_0 \rrb \nonumber \\
\cj_1 &=& J_+J_- + J_0\lrb J_0 - 1 \rrb = J^2 \, , \qquad \cj_2 = P_+P_- + P_0\lrb P_0 - 1 \rrb = P^2
\eea
 satisfy the following cubic algebra:
\bea 
\lsb C_0 , C_\pm \rsb & = & \pm C_\pm\,, \nn \\
\lsb C_+ , C_- \rsb & = &  4 C_{0}^{3} - C_{0} \lrb 4\ck^2  +\sigma \rrb + \lambda \ck ,
\eea  

where $\lambda$ and $\sigma$ are given by
$\sigma = 2 \lrb \cj_1 + \cj_2 \rrb $ and $ \lambda =  \lrb \cj_1 - \cj_2 \rrb $.\\
The corresponding Casimir operator is given by
\be
\cc = C_- C_+ - C_{0}^{2} \lrb C_0 +1\rrb^2 + C_{0}^{2} \lrb 2 \ck ^2 -\fr{1}{2} \sigma \rrb + C_0 \lrb 2\ck \lrb \ck  -\fr{1}{2}  \sigma \rrb + \fr{1}{2} \sigma \rrb
\ee

One can also take one set of $SU(2)$ algebras ,$ (J_0, J_{\pm} )$, and one
set of $SU(1,1)$ algebras, $(L_0 ,L_{\pm} )$ to construct a cubic algebra

{\bf Case 5: $C_- (2,11)$}\\
Consider the set of operators are given by
\bea
C_0 &=& \half \lrb J_0 - L_0 \rrb\,, \qquad C_+ = J_+ L_-\,, \qquad C_- = J_- L_+\,, \nonumber \\
\ck &=& \half\lrb J_0 + L_0 \rrb ,\nonumber \\
\cj &=& J_+J_- + J_0\lrb J_0 - 1 \rrb = J^2 \, , \qquad C_1 = L_+L_- - L_0\lrb L_0 - 1 \rrb = L^2,
\lb{3case5}
\eea
They obey the cubic algebra,
\bea 
\lsb C_0 , C_\pm \rsb & = & \pm C_\pm\,, \nn \\
\lsb C_+ , C_- \rsb & = &  4 C_{0}^{3} -4\ck^2  C_{0}  + 2\lrb \cj + C_1 \rrb,
\eea  
where
$\sigma = 2 \lrb \cj_1 + \cc_2 \rrb $ and $ \lambda =  \lrb \cj_1 - \cc_2 \rrb $.\\

{\bf Case 6: $C_+ (2,11)$}\\
The  other combination,
\bea
C_0 &=& \half \lrb J_0 + L_0 \rrb\,, \qquad C_+ = J_+ L_+\,, \qquad C_- = J_- L_-\,, \nonumber \\
\ck &=& \half\lrb J_0 - L_0 \rrb \nonumber \\
\cj &=& J_+J_- + J_0\lrb J_0 - 1 \rrb = J^2 \, , \qquad C_1 = L_+L_- - L_0\lrb L_0 - 1 \rrb = L^2 ,
\lb{3case6}
\eea
satisfies the following cubic algebra:
\bea 
\lsb C_0 , C_\pm \rsb & = & \pm C_\pm\,, \nn \\
\lsb C_+ , C_- \rsb & = &  4 C_{0}^{3} -4\ck^2  C_{0}  + 2\lrb \cj - C_1 \rrb.
\eea  
The $\lambda$ and $\sigma$ are given as in case 5.\\
The above two algebras differ only by their central elements. But these elements only take some
specific values obeyed by the Casimir operators of the $SU(2)$ and $SU(1,1)$
algebras. The advantage of this identification  is  that once a Cubic algebra
is given then one can map
the algebra  to any of the classes. We also observed that the cubic algebras are
generated not only from linear algebras but also from non-linear algebras(quadratic algebras).
This is an additional information available in the case of cubic algebra. In the case
of quadratic algebra we have used only linear algebras. In the next few cases
a cubic algebra is generated out of a quadratic algebra and a Heisenberg
algebra. The quadratic algebra used in these cases are those obtained in the
chapter 2. Such a construction seems to be restrictable because the most general
quadratic algebra may not be of the four type constructed in chapter 2.
Even though our construction does not give all the cubic algebra possible
but it allows us to construct  different classes of cubic algebras
that arise in physical problems. 
Four different classes of cubic algebras are constructed out of the quadratic and
Heisenberg algebra.

{\bf Case 7: $ C_+ (q_- (1),h)$}\\
Consider the operators,
\bea
C_0 &=& \half \lrb Q_0 + N \rrb\,, \qquad C_+ = Q_+ a^{\da}\,, \qquad C_- = Q_- a\,, \nonumber \\
\ck &=& \half\lrb Q_0 - N \rrb\,. \qquad  
\eea
Here the operators $(Q_{\pm},Q_0 )$ are the generators of the $Q^- (1,1)$ algebra
given in chapter2 and $N=a^{\dagger}a $. They satisfy the following cubic algebras:
\bea 
\lsb C_0 , C_\pm \rsb & = & \pm C_\pm\,, \nn \\
\lsb C_+ , C_- \rsb & = & - 4 C_{0}^{3} -\lrb 6\ck + 3(\cl +1) \rrb C_{0}^2
                        +\lrb \cl (2\cl -3)- 2C_1 -1-2(\cl +2)\ck \rrb C_0
                        +2\ck^3 \nn \\
                    &   & + (\cl -1) \ck^2 -(\cl +1)\ck  -C_q \, .
\eea  
Here $C_q $ is the Casimir operator of the quadratic algebra given in chapter2.\\
{\bf Case 8: $C_- (q_- 1,h)$}\\
Now consider another form of generators,
\bea
C_0 &=& \half \lrb Q_0 - N \rrb\,, \qquad C_+ = Q_+ a\,, \qquad C_- = Q_- a^{\da}\,, \nonumber \\
\ck &=& \half\lrb Q_0 + N \rrb\,. \qquad  
\eea

They satisfy the following cubic algebras
\bea 
\lsb C_0 , C_\pm \rsb & = & \pm C_\pm\,, \nn \\
\lsb C_+ , C_- \rsb & = &  4 C_{0}^{3} +\lrb 6\ck + \cl -2) \rrb C_{0}^2
                       -\lrb \cl (2\cl +2)+2\ck (\cl -1)\rrb C_0 
                        -2\ck^3 -3\ck^2 \nn \\
                    &   &  -(2\cl +1)\ck +C_1 -\cl(\cl -1)(\ck +1)
                        -(\cl -1)\ck^2 + (C_1 -\cl^2 )K +C_q \, . \nn \\
                    &   &
\eea  

Now replacing $Q^-(1,1) $ with $Q^+ (1,1)$ we can construct another two classes of cubic algebras.

{\bf Case 9: $C_+ (q_+ 1,h)$}\\
For the operators,
\bea
C_0 &=& \half \lrb Q_0 + N \rrb\,, \qquad C_+ = Q_+ a^{\da}\,, \qquad C_- = Q_- a\,, \nonumber \\
\ck &=& \half\lrb Q_0 - N \rrb\,, \qquad  
\eea
the cubic algebra is given by
\bea 
\lsb C_0 , C_\pm \rsb & = & \pm C_\pm\,, \nn \\
\lsb C_+ , C_- \rsb & = &  4 C_{0}^{3} +\lrb 6\ck +3 \cl) \rrb C_{0}^2
                       +\lrb 2C_1 -\cl (2\cl +1)+2(\cl +1)\ck \rrb C_0 \nn \\
                    &   & -2\ck^3  -(\cl -2)\ck^2 +\cl \ck -C_q.
\eea
{\bf Case 10: $C_- (q_+ 1,h)$}\\
The generators,
\bea
C_0 &=& \half \lrb Q_0 - N \rrb\,, \qquad C_+ = Q_+ a\,, \qquad C_- = Q_- a^{\da}\,, \nonumber \\
\ck &=& \half\lrb Q_0 + N \rrb\,, \qquad  
\eea

satisfy the cubic algebra given by
\bea 
\lsb C_0 , C_\pm \rsb & = & \pm C_\pm\,, \nn \\
\lsb C_+ , C_- \rsb & = & - 4 C_{0}^{3} -3\lrb 2\ck +\cl -1 \rrb C_{0}^2
                        +\lrb \cl(2\cl + 3) -2C_1 -1 + 2(\cl +2)\ck \rrb C_0 \nn\\
                    &   & +2\ck^3 + (\cl +1) \ck^2 + (\cl -1)\ck +C_1 + C_q
                        -\cl(\cl +1) \, .
\eea
Thus we have constructed 10 classes of cubic algebras and the invariants
of the algebra are constructed.

\section{Representations of cubic algebras}
In this section we will consider the representation of the cubic algebras.
The first step in the construction will be to form the
product state  out of the basis state of the component algebras.
Since all the cases can be mapped onto each other by suitable
transformations we shall consider  explicitly the method of
construction of cases 1 and 2, the representation of
the finite and infinite dimensional discrete series representations.
The other cases are catalogued and for brevity only the explicit results are shown.
The method is self explanatory.
\subsection{Finite dimensional discrete series representations}
First we construct the generalization of the angular momentum type representations of the cubic algebra and
then followed
by the differential "Fock-Bargmann" type of representations.

Let $|l,k_1>$ and $|n,k_2>$ be
the positive discrete series representations of the two $SU(1,1)$ algebras,
$L_i$ and $M_i$ respectively($i=\pm,0 $). 
Here we will consider the case 1 of the previous section. The representation
space of the algebra is obtained by imposing the constraints
obeyed by the Casimir operator and the constant $K$ on the product space of
the two respective discrete series representations given by $|k_1,k_2 ,l,n >
=|k_1 l > |k_2 , n > $.
The condition that $K$ take constant values in an irreducible representation
fixes the basis state 
 by the relation:
\begin{eqnarray}
K |k_1 , k_2 ,l,n> & = & (L_0 +M_0)/2 \,\,\, |k_1 , k_2 ,l,n>  \nn \\
& = & (k_1+k_2+l+n)/2 \,\,\, |k_1 ,k_2 ,l,n>  \nn \\
& = & k \,\,\, |k_1 ,k_2 ,l,n>
\lb{k1}
\end{eqnarray}
So the basis state of the representation space is given by
\be
\lmd k_1 ,k_2 ,k,n \rra = \lmd k_1 , k_2 ,2k-k_1 -k_2 -l ,n \rra \,.
\ee
Explicitly the representation  on the basis states labeled by $|k_1,k_2,k,n>$ is given by
\bea 
C_0 \lmd k_1 ,k_2 ,k , n \rra & = & (k_1-k+n) \lmd k_1 ,k_2 , n \rra\,, \nn \\ 
C_+\lmd k_1 ,k_2 ,k , n \rra & = & \sqrt{(n+2k_1)(n+1)(2k-k_1 -k_2-n)(2k+k_2 -k_1-1-n)}\times\nn \\
                             &   &  \lmd k_1 ,k_2 ,k , n+1 \rra  , \nn \\
C_- \lmd k_1 ,k_2 ,k , n \rra & = & \sqrt{n(n+2k_1-1)(2k-k_1 -k_2 +1-n)(2k +k_2 -k_1 -n)} \times \nn \\
                              &   & \lmd k_1 ,k_2 ,k , n-1 \rra  . \nn \\
 &  & \qquad \qquad n = 0,1,2,\,\ld\,\,. 
\lb{c11-rep} 
\eea 

 The operators
  $\lrb C_-,C_+,C_0 \rrb$  are given by the operators which can be viewed as a deformation of the SU(1,1) algebra formed from the operators
\begin{eqnarray}
C_0&=& K_0-k .  \nn \\
C_- & = &K_- f(K_0,k,k_1,k_2) ,  \nn  \\
C_+ & = & f(K_0,k_1.k_2,k)K_+
\end{eqnarray}
where $f(K_0,k_1,k_2,k)=\sqrt{(k+k_2- 1-K_0)(k-k_2-K_0)}=\sqrt{(k_2- 1-C_0)(-k_2-C_0)}$.

The dimensionality of the representation is fixed by $C_+C_-\ge 0$ and $C_-C_+ \ge 0$.
The deformed function$f$ will obey the condition
\bea
f^2(K_0,k_1,k_2,k) &\ge & 0 \, ,\nn \\
f^2(K_0-1,k_1,k_2,k) &\ge & 0
\eea
or
\bea
 (2k-k_1 -k_2 -l)(2k+k_2 -k_1 -1-l)\ge 0  \nonumber \\
(2k-k_1 -k_2 +1-l)(2k+k_2 -k_1 -l) \ge 0
\eea
For $ 2k-k_1-k_2\le(2k+k_2-k_1-1)$ (which is satisfied for $k_2>1$)
we have a finite dimensional representation of dimension $2k-k_1-k_2+1$.\\
When $0>k_2 \ge 1 $ the dimension of the representation is $2k+k_2 -k_1 $.
The Casimir operator of this representation is given by
\bea
\cc &=&-k^4 +k^2 +kk_2 k_1^2 -kk_2 k_1 -kk_2 +2k^2 k_2^2 -2k^2 k_2 -kk_2^3 +2kk_2^2  \nn \\
    & &+k_2^2 k_1 -k_2^2 k_1^2 +kk_1  -kk_1^2-k_2k_1+k_2k_1^2
\eea

Recall that the Holstein-Primakoff realization of SU(1,1) is given in the single-mode realization
 by
\bea
{K}_{+}(k) &=& \sqrt{{a}^{\dagger}{a} + 2k - 1}\,
{a}^{\dagger} ,  \nn \\
{K}_{-}(k) &=& {a} \sqrt{{a}^{\dagger}{a} + 2k - 1} ,
 \nn \\
{K}_{0}(k) &=& {a}^{\dagger}{a} + k .
\eea
Here $k$ is the Bargmann index labeling unitary irreducible
representations of the SU(1,1) Lie group.

Similarly the equivalent of the Holstein Primakoff Realization in the
two mode realization of the cubic algebra in this case is given by
\bea
C_{-}(k) &=& {ab}\sqrt{(2k+k_2- 1-K_0)(2k-k_2-K_0)} \nn \\
C_{+}(k) &=& \sqrt{(2k+k_2- 1-K_0)(2k-k_2-K_0)}(ab)^{\dag} \nn \\
C_{0}(k) &=& \frac{1}{2}({{a}^{\dagger}{a}}+{b^{\dagger}b+1}) - k .
\eea
Here $k_1$ is the Bargmann index labeling unitary irreducible
representation of the SU(1,1) Lie group given by the generators ${ab},\,\,a^{\dagger} b^{\dagger},\,\,
\frac{1}{2}({a^{\dagger}a}+{b^{\dagger}b}+1)$ and  $k$ is the additional quantum number labeling the representation.
This gives the manifestly symmetric form for $C_+,\,\,C_-$ and $C_0$.\\
However we may get asymmetric representations by shifting the square root
part in either $C_{+}$ or $C_{-}$ to get two different Fock-Bargmann
realizations in holomorphic co-ordinates given by:
\bea
C_{-}(k) &=& K_-(2k+k_2- 1-K_0)(2k-k_2-K_0) \nn  \\
C_{+}(k) &=& K_+ \nn  \\
C_{0}(k) &=& K_0 - k ,
\eea
or
\bea
C_{+}(k) &=& (2k+k_2- 1-K_0)(2k-k_2-K_0) K_+ \nn  \\
C_{-}(k) &=& K_- \nn   \\
C_{0}(k) &=& K_0 - k .
\eea
Thus we have the  following Fock-Bargmann realizations of the cubic algebra:\\
\begin{eqnarray}
C_0&=& z \frac{d}{dz}+k_1-k ,  \nn   \\
C_- & = & (z\ddz2 +2k_1\ddz) ,  \nn  \\
C_- & = & (z\ddz -2k-k_2-1)(z\ddz-2k+k_2)z ,  \nn  \\
\end{eqnarray}
for the basis function given by the monomials
\bea
\psi_{k_1 k_2 k} (z) &=& \frac{z^m}{\sqrt{m!(2k-k_1 -k_2 -m)!(m+2k_1 -1)!(2k+k_2-k_1-m-1)!}},\nn \\
                     & & 
\eea
and
\begin{eqnarray}
C_0&=& z \frac{d}{dz}+k_1-k .  \nn  \\
C_- & = & (z\ddz2 +2k_1\ddz)(z\ddz-2k-k_2-1)(z\ddz-2k+k_2) ,  \nn \\
C_+ & = & z ,  \nn \\
\end{eqnarray}
for the basis function given by the monomials
\begin{equation}
\psi_{k_1 k_2 k} (z) = \frac{z^m}{\sqrt{m!(-2k+k_1 +k_2 +m)!(m+2k_1 -1)!(-2k-k_2+k_1+m+1)!}}.
\end{equation}
It is instructive to get a more symmetric form related to the $2j+1$ angular momentum basis by defining $j=k-\frac{k_1+k_2}{2}$ and the basis function as
$C_mz^{m}=C_pz^{p+j}$, so that
\begin{eqnarray}
C_0 \psi_{k_1,k_2,j,p} &=& (p+\frac{k_1-k_2}{2})  \psi_{k_1,k_2,j,p}  \, ,\nn \\
C_+ \psi_{k_1,k_2,j,p} &=& \sqrt{ (p+j+1)((p+j+2k_1))(j -p)(j+2k_2 -1-p) } \psi_{k_1,k_2,k,p+1} \, , \nn   \\
C_- \psi_{k_1,k_2,j,p} &=& \sqrt{ (p+j)( p+j+2k_1-1))(j +1-p))(j+2k_2 -p) } \psi_{k_1,k_2,k,p-1}, \nn  \\
\end{eqnarray}
which can be viewed as a deformation of the SU(2) algebra:
\begin{eqnarray}
C_0&=& J_0+\frac{k_1-k_2}{2} .  \nn  \\
C_- & = &J_- f(J_0,j,k_1,k_2) ,  \nn  \\
C_+ & = & f(J_0,j,k_1.k_2)J_+
\end{eqnarray}
The differential representation takes the form equivalent to the holomorphic realization of SU(1,1)
\begin{eqnarray}
C_0&=& z \frac{d}{dz}-j+\frac{k_1-k_2}{2} ,  \nn  \\
C_- & = & (\ddz) ,  \nn   \\
C_- & = & (z\ddz +2k_1-1)(-\ddz+k_2+2j)(-z^2\ddz+2jz) ,  \nn \\
\end{eqnarray}
It should be noted that there are infinitely many unitary irreducible representations
of the same dimension. For example one can have a $2$ dimensional matrix representation
for all values of $ k_{1,2} = 1/2, 1, 3/2, \ld \,$ , given by
\bea 
C_0 & = & \fr{1}{2} \lrb \ba{cc}
                     k_1 -k_2 -1 & 0   \\
                     0   & k_1 -k_2 +1 \ea \rrb\,, \quad 
C_+ = \lrb \ba{cc} 
                   0 & 0 \\
          2\sqrt{k_1 k_2} & 0 \ea \rrb\,, \quad   \nn \\
C_- = \lrb \ba{cc}
          0 & 2\sqrt{k_1 k_2}  \\
          0 & 0          \ea \rrb\,,  \nn  \\
C_{1,2} & = & k_{1,2}(1-k_{1,2})\,, \quad   \nn \\
K & = & \frac{1}{2}(k_1 +k_2 +1) \, .
\eea
The basis states satisfy the following recurrance relation
\bea
C_0 |k_1 ,k_2 ,k,l> & = & l+(k_1 -k_2)/2-1/2 |k_1 ,k_2 ,k,l>   \nonumber  \\
C_+ |k_1 ,k_2 ,k,l> & = & \sqrt{ (l+1)(2k_1 +l)(2 -l)(1+2k_2 -l) } |k_1 ,k_2 ,k,l+1>  \nn  \\
C_- |k_1 ,k_2 ,k,l> & = &\sqrt{ l(2k_1 -1+l)(2-l)(2+2k_2 -l) } |k_1 .k_2 .k.l-1>. \nonumber \\
\eea
Another finite dimensional representation is possible for the case 3, $C_-(2,2)$.
Here we consider the algebra given by
two commuting SU(2) generators, $\lrb J_{0}, J_{\pm} \rrb $ and $\lrb
P_{0},P_{\pm} \rrb$, with the
Casimir operator$ \cj_{1,2} $.

 In this case one can take the basis
states as $|j_1, j_2 , m.n> = |j_1 ,m>*|j_2 ,n  >$.
where $|j_1,m>$ and $|j_2,n>$ are the canonical basis states of $J_i$ and $P_i$ respectively.
\bea
J_0 \lmd j_1 , m \rra & = & m -j_1 \lmd j_1 , m \rra\,,  \nn   \\
J_{+} \lmd j_1 , m \rra & = & 
      \sqrt{(m+1)(2j_1-m)}\,\lmd j_1 , m + 1 \rra\,,  \nn  \\
J_{-} \lmd j_1 , m \rra & = & 
      \sqrt{m(2j_1+1-m)}\,\lmd j_1 , m - 1 \rra\,,  \nn \\
   &  & \qquad \qquad m = 0,1,...,2j_1,
\eea
\bea
P_0 \lmd j_2 , n \rra & = & n -j_2 \lmd j_2 , n \rra\,,  \nn \\
P_{+} \lmd j_2 , n \rra & = & 
      \sqrt{(n+1)(2j_2-n)}\,\lmd j_2 , n + 1 \rra\,,  \nn  \\
P_{-} \lmd j_2 , n \rra & = & 
      \sqrt{(n)(2j_2+1-2k+m)}\,\lmd j_2 , n - 1 \rra\,,  \nn  \\
   &  & \qquad \qquad n = 0,1,...,2j_2,
\eea
\bea
J_0+P_0 \lmd j_1 ,j_2, m,n \rra & = & m -j_1+n-j_2 \lmd j_1 ,j_2, m,n \rra\,,  \nn  \\
& = & 2k \lmd j_1 , m \rra\,.  \nn  \\
\eea
In this case the constant operator $k$ acting on the product space and give the
condition,
\be
n=2k+j_1+j_2-m
\ee
The matrix representation
in this case is given by
\begin{eqnarray}
C_0 |j_1 ,j_2 , k, m>  & = & m-k-j_1  |j_1 ,j_2 ,k ,m>  \nn  \\
C_- |j_1 , j_2 ,k, m>& = & \sqrt{(m+1)(2j_1-m)(2k +j_1+j_2-m)(1-2k-j_1+j_2+m)}\times \nn \\
                     &   & |j_1 , j_2 ,k, m-1>    \nn \\
C_+ |j_1 ,j_2 ,k ,m>& = & \sqrt{m(2j_1+1 -m)(2k+j_1+j_2 -m+1)( -2k-j_1+j_2 +m)} \times \nn \\
                    &   & |j_1 , j_2 ,k, m-2>
\end{eqnarray}
The unitary irreducible representations are possible only if the following matrix
elements are positive.
\bea
<j_1,j_2,k,m|C_- C_- |j_1 j_2 k m> &= & (m+1)(2j_1-m)(2k +j_1+j_2-m)\times \nn \\
                                   &  & (1-2k-j_1+j_2+m)\ge 0 \, . \nn \\
<j_1,j_2,k,m|C_- C_+ |j_1 j_2 k m>& = & m(2j_1+1 -m)(2k+j_1+j_2 -m+1)\times \\
                                  &   & ( -2k-j_1+j_2 +m)\ge 0 \, .
\eea
The above conditions demands that the representations should be
of dimension $2j_1 +1 $ since $2k+j_1 +j_2 \ge  2j_1 . $
The differential realization in monomial basis for the above algebra
will be
\begin{eqnarray}
C_0&=& z \frac{d}{dz}-k-j_1 \, ,  \nn  \\
C_- & = & \ddz ,  \nn \\
C_+ & = & (z\ddz -2j_1-1)(-\ddz+2k+j_1-j_2)(-z^2\ddz+(2k+j_1+j_2)z) ,
\end{eqnarray}

acting on the monomial function,
\begin{equation}
\phi(z) =\fr{z^m}{\sqrt{m!(2j_1 -m)!(2k+j_1+j_2-m)!(1 -2k-j_1+j_2+m)!}} \, .
\end{equation}
\subsection{Infinite dimensional discrete series representations}
The cubic algebras also have an infinite dimensional representation.
In the following we will consider the infinite dimensional 
representation of the cubic algebra that we had constructed.
We will consider the algebra $C_+ (11,11)$. The representation space of the
algebra will be the product state
$|k_1 , m>*|k_2 ,n >$ with the constraint that K is a constant over the
representation which gives
\be
2k= k_1 + k_2 +n +m .
\ee
Thus the basis states can be taken as
\be
\lmd k_1 ,k_2 ,k,m \rra = \lmd k_1 m \rra \lmd k_2 , 2k-k_1 -k_2 -m\rra \, .
\ee
The representation looks explicitly as given below.
\begin{eqnarray}
C_0 |k_1 ,k_2 ,k,m> &=& m+k_1 +k |k_1 ,k_2 ,k,m>  \nn \\
C_+ |k_1 ,k_2 ,k,m> &=& \sqrt{(m+1)(2k_1 +m)(+k_1 -k_2 +2k +m+1)(k_2 +k_1 +2k+m)}\times \nn \\
                    & & |k_1 ,k_2 ,k,m+1>  \nn  \\
C_- |k_1 ,k_2 ,k,m> &=& \sqrt{ m(2k_1 -1+m)(k_1 -k_2 +2k +m)(k_2 +k_1 +2k-1+m) }\times \nn \\
                    & &  |k_1 .k_2 .k.m-1>.
\lb{11inf}
\end{eqnarray}
Since $2k+k_1+k_2>0$ and $2k+k_2-k_1+1>0$
in this case we have infinite dimensional representations.
 The operators
 $\lrb C_-,C_+,C_0 \rrb$  are given by the operators which can be viewed as a deformation of the SU(1,1) algebra formed from the operators
\begin{eqnarray}
C_0 &=& K_0+k \, ,  \nn  \\
C_- & = &K_- f(K_0,k,k_1,k_2) ,  \nn  \\
C_+ & = & f(K_0,k_1.k_2,k)K_+  \, ,
\end{eqnarray}
where $f(K_0,k_1,k_2,k)=\sqrt{(k+k_2- 1+K_0)(k-k_2+K_0)}\, .$

The differential Barut-Girardello realizations are
\begin{eqnarray}
C_0&=& z \frac{d}{dz}+k_1+k .  \nn    \\
C_- & = & (z\ddz2 +2k_1\ddz) ,  \nn  \\
C_- & = & (z\ddz +k+k_2-1)(z\ddz+k+k_2)z ,  \nn   \\
\end{eqnarray}
for the basis function given by the monomials
\bea
\psi_{k_1 k_2 k} (z) &=&  
\frac{z^m}{\sqrt{m!(2k+k_1+ k_2 +m+1)!(m+2k_1 -1)!(2k-k_2+k_1+m+1)!}}\, , \nn \\
                     &&          
\eea
and the differential realization
\begin{eqnarray}
C_0&=& z \frac{d}{dz}+k_1-k .  \nn  \\
C_- & = & (z\ddz2 +2k_1\ddz)(z\ddz-2k-k_2-1)(z\ddz-2k+k_2) ,  \nn  \\
C_+ & = & z ,  \nn  \\
\end{eqnarray}
for the basis function given by the monomials
\begin{equation}
\psi_{k_1 k_2 k} (z) = \frac{z^m}{\sqrt{m!(+2k+k_1 +k_2 +m)!(m+2k_1 -1)!(+2k-k_1+k_2+m+1)!}}\, .
\end{equation}
The case 4 and 6 are very similar to the previous cases. Here we will take the
product state of the basis of the unitary representations of the $SU(2)$ and
$SU(1,1)$ algebra. Here, imposing the constraint obeyed by the operators
$\ck$ in \ref{3case5} and \ref{3case6}, one will get the representation states for the
algebra. For example for the case 5, $C_- (2,11)$, the product states are  $\lmd
j,n \rra \lmd k_1 ,m \rra $. But the operator $\ck$ which commute with
all the other generators are realized in the above product states as
\be
\ck = \frac{1}{2} (j+k_1 +m+n )
\ee
and is a constant. So the new reduced space is taken as (by choosing the $SU(2)$ basis)
\be
\lmd j,k_1 ,k,n \rra = \lmd j,n \rra \lmd k_1 ,2k -j-k_1 -n \rra
\ee
The corresponding unitary irreducible representation is given by 
\bea
C_0 \lmd j ,k_1 ,k , n \rra &=& n-k \lmd j ,k_1 ,k , n \rra \nn \\
C_+\lmd j ,k_1 ,k , n \rra &=&\sqrt{(j-n)(j+n+1)(2k-k_1 -n)(2k+k_1 -1-n)}\,\,\lmd j ,k_1 ,k , n+1 \rra  , \nn \\
C_- \lmd j ,k_1 ,k , n \rra &=&\sqrt{(j+n)(j-n+1)(2k-k_1 +1-n)(2k+k_1-n)}\,\,\lmd j ,k_1 ,k , n-1 \rra  . \nn \\ 
                            &  & \qquad \qquad n = 0,1,2,\,\ld 
\lb{c211-rep}
\eea

The above representations are finite dimensional with the dimension
\bea
2k+j-k_1+1 & \mbox{if}&  j> 2k-k_1 \, ,\nn \\
2j+1       & \mbox{if}&  j< 2k-k_1 \, .
\eea
On a monomial basis state given by
\begin{equation}
\psi_{k_1 j k} (z) = \frac{z^{n+j}}{\sqrt{(j-n)!(j+n)!(2k-k_1-n)!(2k-k_1 -1-n)!}},
\ee
the following differential realizations are possible.
\bea
C_0 &=& z\ddz -k-j \, ,\nn \\
C_+ &=& z^4 \ddzth 4(k+j-1) z^3 \ddzt \, ,\nn \\
    & & + \lrb (2k +j)(2k +j+3) +k_1 (k_1 -1)-(4k +2j -2 )2j \rrb z^2 \ddz  \, ,\nn \\
    & & +2(2k +j -k_1 )(2k +j +k_1 -1 )j z \, ,\nn \\
C_- &=& \ddz  \, .
\eea
In the same way one for the case 6, we will take the basis states in the $SU(1,1)$
basis, given by,
\be
\lmd j, 2k+k-1 -n \rra \lmd k, n \rra
\ee
In this case we have also we have a finite dimensional unitary representation
with the condition
\bea
(j+2k+n)(j-2k-k_1 +1-n ) &\ge& 0 \, ,\nn \\
(j-2k-k_1 -n )(j+2k+k_1 +1 +n) &\ge& 0 \, .
\eea
The dimensionality of the representation is $j-2k-k_1+1$.
The unitary irreducible representations are explicitly given by 
\bea
C_0 \lmd j ,k_1 ,k , n \rra & = & n+k_1 +k \lmd j ,k_1 , n \rra\,, \nn \\ 
C_+\lmd j ,k_1 ,k , n \rra & = & 
  \sqrt{(j-2k-k_1 -n)(j+2k +k_1 +1+n)(n+1)(2k_1 +n)}\times \nn \\&&\lmd  j ,k_1 ,k , n+1 \rra  , \nn \\
C_- \lmd j ,k_1 ,k , n \rra & = & 
  \sqrt{(j-2k-k_1 +1-n)(j+2k +k_1 +n)(n)(2k_1 -1+n)}\times \nn \\&&\lmd j ,k_1 ,k , n-1 \rra  . \nn \\ 
 &  & \qquad \qquad n = 0,1,2,\,\ld 
\lb{c211+rep}
\eea
The differential realization in the monomial basis
\be
\psi_{k_1 j k} (z) = \frac{z^{n}}{\sqrt{n! (n+2k_1 -1 )! (j-2k -k_1 -n )!(j+2k +k_1 +n)!}},
\ee
are given by
\bea
C_0 &=& z\ddz +k_1 +k \, ,\nn \\
C_+ &=& -z^2 \ddz + (j-2k -k-1 )z \, ,\nn \\
C_- &=& z^2 \ddzth + (j+2k +3k_1 +2 )z\ddzt \, ,\nn \\
    & & + (2k_1 ) (j+2k +k_1 +1 ) \ddz \, .
\eea

Now will consider the representation of the cubic algebras constructed
out of quadratic algebra and Heisenberg algebra.
We have seen in chapter 2 the quadratic algebras having finite and infinite dimensional
representation. These algebras are used in cases 7 to 10 to form  cubic algebras.
The construction of the representation is the same as the previous case. Here here we will
take a product space of Fock space and the finite and infinite dimensional
spaces constructed in chapter 2(case 1 and 2). Since the quadratic algebras are
characterized by the quantum numbers $l$ and $k_1 $, for all the four case will consider
the product state as $\lmd m \rra \lmd l,k_1 , n \rra  $. In the all the case we will be forming 
the basis in the quadratic algebra basis by imposing the respective
constraints on each case.

In case 7 the basis states after imposing the constarints are given by,
\be
\lmd l,k_1 , k, n \rra = \lmd k_1 -2l-2k+n\rra \lmd l,k,n \rra   \, .
\ee
The corresponding unitary irreducible representation are given by
\bea 
C_0 \lmd l ,k_1 ,k , n \rra & = & (k_1-l-k+n) \lmd l ,k_1 ,k, n \rra\,, \nn \\ 
C_+\lmd l ,k_1 ,k , n \rra & = & 
  \sqrt{(n+2k_1)(n+1)(k_1 -2l+1+n)(k_1 -l -2k+1+n)}\,\,\lmd l ,k_1 ,k , n+1 \rra  , \nn \\
C_- \lmd l ,k_1 ,k , n \rra & = & 
  \sqrt{n(n+2k_1-1)(k_1 -2l+n)(k_1 -l -2k +n)}\,\,\lmd l ,k_1 ,k , n-1 \rra  . \nn \\ 
 &  & \qquad \qquad n = 0,1,2,\,\ld 
\lb{cq11-rep} 
\eea 
with $k_1 -l-2k =0 ... $ we have infinite dimensional representations. In this case also
monomial differential realizations are also possible which are of forth order.
The coefficient of the differential operator becomes complicated compared to the
previous case. For the monomial basis ,
\be
\psi (z)=\frac{z^{n}}{\sqrt{n!(k-l-2k+n)!(n+2k_1 -1 )!(k-2l+n)!}},
\ee
The differential realizations are given by
\bea
C_0 &=& z\ddz +k_1 -l -k \, ,\nn \\
C_+ &=& z \, ,\nn \\
C_- &=& z^3 \fr{d^4}{dz^4} + (4k_1 -3l -2k-5)z^2 \ddzth \nn \\
    & & (k_1 -l-2k)(3k_1 -2l-1) +(2k_1 -1 )(k_1 -2l ) +9l+6k-12k_1 +26)z\ddzt \nn \\
    & & ((k_1 -l-2k )(2k_1 -l)(k_1 -2l )-2(4k_1 -3l -2k +2) \nn \\
    & & (k_1 -l-2k)(3k_1 -2l +1) +(2k_1 -1)(k_1 -2l) \nn \\
    & & 9l +6k -12k_1 +26)\ddzt  \, .
\eea
In the case 8 the basis states are taken as
\be
\lmd l,k_1 , k, n \rra = \lmd 2k-k_1 +l-n\rra \lmd l,k,n \rra  \, ,
\ee
with the matrix representation
\bea 
C_0 \lmd l ,k_1 ,k , n \rra & = & (k_1-l-k+n) \lmd l ,k_1 , n \rra\,, \nn \\ 
C_+\lmd l ,k_1 ,k , n \rra & = & 
  \sqrt{(n+2k_1)(n+1)(2k-k_1 +l-n)(k_1 -2l +1+n)}\,\,\lmd l ,k_1 ,k , n+1 \rra  , \nn \\
C_- \lmd l ,k_1 ,k , n \rra & = & 
  \sqrt{n(n+2k_1-1)(2k-k_1 +l+1-n)(k_1 -2l +n)}\,\,\lmd l ,k_1 ,k , n-1 \rra  . \nn \\ 
 &  & \qquad \qquad n = 0,1,2,\,\ld 
\lb{cq-11+rep} 
\eea 
which form a $2k-k_1 +l +1 $ dimensional unitary irreducible representation.The
corresponding differential realization
are given by
\bea
C_0 &=& z\ddz +k_1 -l -k \, ,\nn \\
C_+ &=& -z^2 \ddz + (2k-k_1 +l)z \, ,nn  \\
C_- &=& z^2 \ddzth + z\ddzt + 2k-1 (k_1 -2l +1 )\ddz  \, ,
\eea
realized in the monomial basis
\be
\psi(z)\frac{z^{n}}{\sqrt{n!(2k-k_1 +l-n)(n+2k-1)!(n+k_1 -2l)!}}\, .
\ee
In case 9 and 10 we will be using the finite dimensional basis of the quadratic algebra
to form the product state. For case 9 the basis states are
\be
\lmd l,k_1 , k, n \rra = \lmd k_1 -2k-l+n\rra \lmd l,k,n \rra   \, .
\ee
The unitary irreducible representations are given by, 
\bea 
C_0 \lmd l ,k_1 ,k , n \rra & = & (k_1-l-k+n) \lmd l ,k_1 , n \rra\,, \nn \\ 
C_+\lmd l ,k_1 ,k , n \rra & = & 
  \sqrt{(n+2k_1)(n+1)(k_1 -2k_1-l+n)(2l-k -n)}\,\,\lmd l ,k_1 ,k , n+1 \rra  , \nn \\
C_- \lmd l ,k_1 ,k , n \rra & = & 
  \sqrt{n(n+2k_1-1)(2k-k_1 +l+1-n)(k_1 -2l +n)}\,\,\lmd l ,k_1 ,k , n-1 \rra  . \nn \\ 
 &  & \qquad \qquad n = 0,1,2,\,\ld 
\lb{cq+11+rep} 
\eea
They form a finite dimensional representation. The dimension is
\bea
k_1 -2k -l +2 &\mbox{if}& 2k-l+1 < 0 \, ,\nn \\
2l-k_1 +1     &\mbox{if}& 2k-l+1 > 0  \, .
\eea
The differential realization in this case is given by
\bea
C_0 &=& z\ddz +k_1 -l -k \, .\nn \\
C_+ &=& -z^2\ddz + (2l-k)z \, .\nn  \\
C_- &=& z^3\ddzth +(3k_1 -2k -l +2 )z\ddzt +2k_1 (k_1 -2k -l +1 )\ddz  \, ,
\eea
realized in the monomial basis,
\be
\frac{z^{n}}{\sqrt{(k_1 -2k-l +n)!n!(n+2k_1 -1 )!(2l-k +1-n)!}}\, .
\ee
In case 10 the basis states are given by 
\be
\lmd l,k_1 , k, n \rra = \lmd 2k-k_1 +l-n\rra \lmd l,k,n \rra \, .
\ee
The unitary irreducible representations are given by
\bea 
C_0 \lmd l ,k_1 ,k , n \rra & = & (k_1-l-k+n) \lmd l ,k_1 , n \rra\,, \nn \\ 
C_+\lmd l ,k_1 ,k , n \rra & = & 
  \sqrt{(n+2k_1)(n+1)(2k-k_1 +l-n)(2l-k -n)}\,\,\lmd l ,k_1 ,k , n+1 \rra  , \nn \\
C_- \lmd l ,k_1 ,k , n \rra & = & 
  \sqrt{n(n+2k_1-1)(2k-k_1 +l+1-n)(2l-k+1-n)}\,\,\lmd l ,k_1 ,k , n-1 \rra  . \nn \\ 
 &  & \qquad \qquad n = 0,1,2,\,\ld 
\lb{cq+11-rep} 
\eea
which are finite dimensional.
The differential realization in this case are given by
\bea
C_0 &=& z\ddz +k_1 -l -k \, .\nn \\
C_+ &=& z^3 \ddzt + (k_1 -k-3l +1)z^2 \ddz +(2k-k_1 +l)(2l-k)z \, .\nn  \\
C_- &=& z\ddzt + 2k_1 \ddz \, , 
\eea
where the above operators act on the monomial basis,
\be
\frac{z^{n}}{\sqrt{n!(2k-k_1 +l -n)!(n+2k_1 -1 )!(2l-k +1-n)!}}\, .
\ee
Thus we have seen that different classes of cubic algebras can be
generated in a way similar to the bosonic realization of $SU(2)$ and
$SU(1,1)$ algebras.
Besides the 10 cubic algebras explicitly worked out in this chapter, we would like to point out that
4 more classes of cubic algebras can be generated .
These occur when the underlying $SU(1,1)$ and $SU(2)$ Lie algebras in cases 1 to 4
are the same
 (i.e $\{L_{\pm.0}\}=\{M_{\pm,0}\} $ and $\{J_{\pm.0}\}=\{P_{\pm,0}\} $), thereby , generating two additional algebras.
 Another two are generated when the Bosonic realizations of the Heisenberg and the quadratic algebra
 contain a common boson operator.
 These are equivalent to the $a^2, a^{\dag 2}$ and $\frac{1}{4}(a^{\dag}a+aa^{\dag}$ representations of SU(1,1).
 These cases have not been worked out explicitly as they can be easily derived from the  cases already discussed using Fock representations.

The representations constructed
in this chapter are used in chapters 5 and 6.


\setcounter{equation}{0}
\chapter{A Generalisation of the Jordan-Schwinger method for Polynomial Algebras}
\markboth{}{Chapter 4.  A Generalisation of....}

In this chapter we generalise the the method which we have  developed for
quadratic and cubic algebras to a three dimensional polynomial algebra
of arbitrary order.
In chapter 2 we have seen that a class of quadratic algebras can be
constructed by taking products
 of a Heisenberg algebra and an $SU(2)$ or $SU(1,1)$ algebra.
In chapter 3, a similar construction for the cubic algebras  out
of two $SU(2)$ algebras, two $SU(1,1)$ algebras, one $SU(1,1)$ and
one $SU(2)$ algebra and a quadratic and a Heisenberg algebra was
given . The quadratic and cubic algebras can be considered as
higher order generalization of the linear Lie algebras. The method
of construction we have used is reminiscent of the well known
Jordan-Schwinger realization, where the   $SU(2)$ algebra is
constructed out of a Heisenberg algebra generators.
 In this chapter we will show that
 a general higher order algebra can be constructed from lower order algebras
by a method which can be considered as a generalization of the
well known Jordan-Schwinger method.
If we examine the Heisenberg algebra, the $SU(2)$ and $SU(1,1)$ algebras
and the quadratic and cubic algebras (with a coset structure) one can see that
out of the three commutation relations two, which are linear, are common to
all the algebras. Only the third commutation relation is non-linear.
If one considers a  Heisenberg algebra as a polynomial algebra of order zero and
the $SU(2)$ and
$SU(1,1)$ algebra as a polynomial algebra of order one ,then, from the Jordan-
Shcwinger construction and also in the light of the quadratic
and cubic algebras, it is possible
to propose that given two polynomial algebras one can construct
a polynomial algebra of higher order out of them.
For the purpose of clarity we will briefly otline the Jordan-Schwinger
realizations of $SU(2)$ and $SU(1,1)$ algebras. Then we will generalize it to
the polynomial algebras.\\
\section{Jordan Schwinger realization of $SU(2)$ and $SU(1,1)$ algebras: A brief review}
 Let us briefly recall the study of $SU(2)$ and $SU(1,1)$ in terms of two-mode
bosonic realizations, to fix the framework and notations for our work.  Let
$\lrb a_1 , a_1^\da \rrb$ and $\lrb a_2 , a_2^\da \rrb$ be two mutually
commuting boson annihilation-creation operator pairs.  Let
$H_1 = N_1+\half = a_1^\da a_1+\half$ and
$H_2 = N_2+\half = a_2^\da a_2+\half$.  As is well known,
$\lrb J_0, J_+, J_- \rrb$ defined by
\be
J_0 = \half \lrb H_1 - H_2 \rrb\,, \quad
J_+ = a_1^\da a_2\,, \quad
J_- = J_+^\da = a_1 a_2^\da\,,
\lb{js}
\ee
satisfy the $SU(2)$ algebra,
\be
\lsb J_0 , J_\pm \rsb = \pm J_\pm\,, \quad
\lsb J_+ , J_- \rsb = 2J_0\,.
\ee
In this Jordan-Schwinger realization of $su(2)$, $H_1 + H_2$ is seen to be a
central element. Let us define
\be
\cl = \half \lrb H_1 + H_2 \rrb\,,
\ee
then,
\be
\lsb \cl , J_{0,\pm} \rsb = 0\,.
\ee
The usual Casimir operator is
\be
\cc = J^2 = J_+J_- + J_0 \lrb J_0 - 1 \rrb = \cl^2 - \fr{1}{4}\,.
\ee
Consequently, the application of the realization (\ref{js}) on a set of
$2j+1$ two-mode Fock states $\lmd n_1 \rra \lmd n_2 \rra$, with constant
$n_1+n_2$ $=$ $2j$, leads to the $(2j+1)$-dimensional unitary irreducible
representation for each $j = 0,1/2,1,\ld\,$.
Now by redefining $m =j-n_2  $ one can take the basis states as
$ |j,m > =  |j-m >|j+m >$, where $-j \le m \le j $, one gets the $j$-th unitary
irreducible representations
\bea
J_0 \lmd j , m \rra & = & m \lmd j , m \rra\,, \nn \\
J_\pm \lmd j , m \rra & = &
      \sqrt{(j \mp m)(j \pm m+1)}\,\lmd j , m \pm 1 \rra\,, \nn \\
\cl\lmd j , m \rra & = & ( j+\half ) \lmd j , m \rra\,, \quad
J^2\lmd j , m \rra = j(j+1)\lmd j , m \rra\,, \nn \\
   &  & \qquad \qquad \qquad \qquad m = j,j-1,\ld\,,-j\,.
\lb{su2rep}
\eea
In an analogous way, $\lrb K_0, K_+, K_- \rrb$ defined by
\be
K_0 = \half \lrb H_1 + H_2 \rrb\,, \quad
K_+ = a_1^\da a_2^\da\,, \quad
K_- = K_+^\da = a_1 a_2\,,
\lb{su11}
\ee
satisfy the $SU(1,1)$ algebra
\be
\lsb K_0 , K_\pm \rsb = \pm K_\pm\,, \qquad
\lsb K_+ , K_- \rsb = -2K_0\,.
\ee
Now,
\be
\cl = \half \lrb H_1 - H_2 \rrb
\ee
is a central element of the algebra\,:
\be
\lsb \cl , K_{0,\pm} \rsb = 0\,.
\ee
The usual Casimir operator is
\be
\cc = K^2 = K_+ K_- - K_0 \lrb K_0-1 \rrb = \fr{1}{4} - \cl^2\,.
\ee
Consequently, the application of the realization (\ref{su11}) on any infinite
set of two-mode Fock states
$\lcb \lmd k , n \rra = \lmd n+2k-1 \rra ,\,\,\, \lmd n \rra \right.$
$\lmd n = 0,1,2,\,\ld \rcb$, with constant $n_1 - n_2 = 2k-1$, leads to the
infinite dimensional unitary irreducible representation, the so-called positive
discrete representation ${\cal D}^+(k)$\,, corresponding to any
$k = 1/2, 1, 3/2,\,\ld\,.$
\bea
K_0 \lmd k , n \rra & = & \lrb k+n \rrb \lmd k , n \rra\,, \nn \\
K_+ \lmd k , n \rra & = & \sqrt{(2k+n)(n+1)}\,\lmd k , n+1 \rra\,, \nn \\
K_- \lmd k , n \rra & = & \sqrt{(2k+n-1)n}\,\lmd k , n-1 \rra\,, \nn \\
\cl\lmd k , n \rra & = & ( k - \half ) \lmd k , n \rra\,, \quad
K^2\lmd k , n \rra = k(1-k)\lmd k , n \rra\,, \nn \\
  &  & \qquad \qquad \qquad \qquad n = 0,1,2,\,\ld\,.
\lb{su11rep}
\eea
Note that the choice of basis states as
$\lcb \lmd k , n \rra = \lmd n \rra\lmd n+2k-1 \rra \right.$
$\lmd n = 0,1,2,\,\ld \rcb$, with $n_1-n_2 = 1-2k$, is also posssible leading to
the same representation (\ref{su11rep}), with $K^2 = k(1-k)$, but corresponding
to $\cl = \half - k$.\\
\section{Generalization of the Jordan Schwinger method to polynomial algebras}
 In this section a three dimensional polynomial algebra of arbitrary order is defined.
Using this definition a method to construct a higher order polynomial algebra from
a lower order polynomial algebra is given.\\
{\bf Definition:} \\
An  $n^{th}$ order three dimensional polynomial algebra, generated by
${\cp}^{(n)}_+$, ${\cp}^{(n)}_- $ and ${\cp}^{(n)}_0$ is defined as
\bea
\lsb {\cp}^{(n)}_0 , {\cp}^{(n)}_{\pm} \rsb &=&  {\cp}^{(n)}_{\pm} \nn \\
\lsb {\cp}^{(n)}_+ , {\cp}^{(n)}_- \rsb &=& f_{n} ({\cp}^{(n)}_0 )
\lb{4.0.1}
\eea
Here $f_{n} ({\cp}_0 )$ is an $n^{th}$ order polynomial in ${\cp} ^{(n)}_0$.

When $n=0$ and $ f_{0}=1$, the algebra corresponds to the Heisenberg algebra.
When $n=1 $ the algebra corresponds to an $SU(2)$ or $SU(1,1)$
with $f_{1} = +1$ or $-1$ respectively.
The Casimir operator for such algebras are given by
\be
C^{(m)} = {\cp} ^{(m)}_- {\cp}^{(m)}_+  + g_{m+1}({\cp} ^{(m)}_0),
\ee
where $g_{m+1} ({\cp} ^{(m)})$ is a polynomial of order  $m+1$ in ${\cp}^{(m)}_0$

From our construction of quadratic and cubic algebras in chapter 2
and chapter 3 the following conclusions are possible. When a
Heisenberg algebra of order zero combines with a $SU(2)$ or a
$SU(1,1)$ algebra of order 1 we will always get a quadratic
algebra of order 2. In the cubic algebra two algebras of order 1
are used to construct a cubic algebra of order 3. Also a
Heisenberg algebra of order 0 combines with a quadratic algebra to
get an order 3 cubic algebra.
 In all the above
examples when two algebras are combined as in the special way
given in all cases, the resultant algebras are of order one degree
greater than the sum of the order of the two algebras. Thus we
have the following general result:\\

{\it If                  
$({\cp}^{(n)}_{\pm}, {\cp}^{(n)}_0 )$ and $({\cp}^{(m)}_{\pm} ,
{\cp}^{(m)}_0)$ are the two sets of two dimensional polynomial
algebras of order $n$ and $m$ respectively, then 
operators defined by

\bea
\Pi_+ &=& {\cp}^{(m)}_+ {\cp}^{(n)}_+ \,\,\,,\,\,\, \Pi_- = {\cp}^{(m)}_- {\cp}^{(n)}_- ,\nn \\
\Pi_0 &=&  \fr{{\cp}^{(m)}_0 +{\cp}^{(n)}_0 }{2}, \nn \\
\Pi   &=&  \fr{{\cp}^{(m)}_0 -{\cp}^{(n)}_0 }{2},
\eea
will always satisfy a polynomial algebra of order $m+n+1$.}\\
 The form of the algebra will depend on the
form of ${\cp}^{(m)}_{\pm,0}$ and ${\cp}^{(n)}_{\pm,0}$.\\

The proof of this general result is given by the following arguments:\\
Since,
\bea
\lsb \Pi_0 , \Pi_{\pm} \rsb &=& \fr{1}{2} \lsb {\cp}_0^{(m)} + {\cp}_0^{(n)} ,{\cp}^{(m)}_{\pm} {\cp}^{(n)}_{\pm} \rsb \nn \\
                                      &=& \pm {\cp}^{(m)}_{\pm} {\cp}^{(n)}_{\pm} \nn \\
                                      &=& \pm \Pi_{\pm}
\eea
are the commutation relations  in (\ref{4.0.1}), we have
\bea
\lsb \Pi_+ , \Pi_{-} \rsb   &=& \lsb {\cp}^{(m)}_{+} {\cp}^{(n)}_{+}  , {\cp}^{(m)}_{-} {\cp}^{(n)}_{-} \rsb \nn \\
                                      &=& {\cp}^{(m)}_{+} {\cp}^{(m)}_{-} f_{n} ({\cp}^{(n)}_0 ) + {\cp}^{(n)}_{-} {\cp}^{(n)}_{+} f_{m} ({\cp}^{(m)}_0 ) \nn \\
                                      &=& \lrb C^{(m)} -g_{m+1} ({\cp}^{(m)}_0 -1) \rrb f_n ({\cp}^{(n)}_0 )
                                       + \lrb C^{(n)} -g_{n+1} ({\cp}^{(n)}_0 -1) \rrb f_m ({\cp}^{(m)}_0 ) \nn \\
                                      & &
\eea  where, $C^{(m)}$ and $C^{(n)}$ are the Casimir operators. 
Expanding all the polynomials in powers of ${\cp}_0 $ and
substituting for ${\cp}_0$ in terms of $\Pi_0 $ and $\Pi$ we have
\bea RHS &=& C^{(m)}\sum _{l=0}^{n} C_l (\Pi_0 -\Pi )^l -
\sum_{l=0}^{m+1} \sum_{s=0}^{n}
         a_l c_s (\Pi_0 +\Pi)^l (\Pi_0 -\Pi)^s \nn \\
    & & + C^{(k)}\sum_{l=0}^{m} d_l (\Pi_0 + \Pi )^l - \sum_{l=0}^{n+1} \sum_{s=0}^{m}
        b_l d_s (\Pi_0 -\Pi)^l (\Pi_0 +\Pi)^s \nn \\
    &=& f_{m+n+1} (\Pi_0^{m+n+1} ),\eea where,
    $f_{m+n+1} (\Pi_0^{m+n+1}) $ is
  a polynomial in $\Pi_0^{m+n+1}$ of order $m+n+1$ and
  the operator $\Pi$ commutes with all the other generators.
This defines  a polynomial algebra of order $m+n+1$. Hence the result.

Since we have constructed a higher order algebra from a  lower algebra
the representation of the higher order algebra can also be constructed from the
lower order algebra. For example, let $ \mid m,\lambda^{(m)}> $ and  $ \mid n,\lambda^{(n)}> $
be the basis states for the irreducible unitary representations of the polynomial
algebras $({\cp}^{(m)}_0 , {\cp}_{\pm}^{(m)})$ and $({\cp}^{(n)}_0 , {\cp}^{(n)}_{\pm})$ respectively,
where $\lambda^{(m)}$ and $\lambda^{(n)}$ $\in \bf{R}$, labeling the unitary irreducible  representations,
\bea
{\cp}_0^{(m)} \lmd q, \lambda^{(m)} \rra &=& q+ \lambda^{(m)} \lmd q, \lambda^{(m)} \rra \nn \\
{\cp}_{+}^{(m)} \lmd q, \lambda^{(m)} \rra &=& \sqrt{t^{\lambda}_{q+1}} \lmd q+1, \lambda^{(m)} \rra \nn \\
{\cp}_{-}^{(m)} \lmd q, \lambda^{(m)} \rra &=& \sqrt{t^{\lambda}_{q}} \lmd q-1, \lambda^{(m)} \rra
\eea

\bea
{\cp}_0^{(n)} \lmd s, \lambda^{(n)} \rra &=& s+ \lambda^{(n)} \lmd s, \lambda^{(n)} \rra \nn \\
{\cp}_{+}^{(n)} \lmd s, \lambda^{(n)} \rra &=& \sqrt{d^{\lambda}_{s+1}} \lmd s+1, \lambda^{(n)} \rra \nn \\
{\cp}_{-}^{(n)} \lmd s, \lambda^{(n)} \rra &=& \sqrt{d^{\lambda}_{s}} \lmd s-1, \lambda^{(n)} \rra
\eea
where $q,s = 0,\pm 1 , ...$

The fact that $\Pi $ is a constant over the product state $\lmd q, \lambda^{(m)} \rra *\lmd s,
\lambda^{(n)} \rra $ gives the condition,
\be
2\Pi = \lambda^{(m)} + \lambda^{(n)} + q+s
\ee
If we impose the above constraint in the product states the following
representations are possible:
\bea
\Pi_0^{(m)} \lmd q, \lambda^{(m)},\lambda^{(n)},\Pi \rra &=& q+ \lambda^{(m)} -\Pi \lmd q, \lambda^{(m)},\lambda^{(n)},\Pi \rra  \nn \\
\Pi_{+}^{(m)}\lmd q, \lambda^{(m)},\lambda^{(n)},\Pi \rra  &=&
\sqrt{t^{\lambda^{(m)}}_{q+1}d^{\lambda^{(n)}}_{2\Pi -\lambda^{(m)} -\lambda^{(n)} +1-q }}\lmd q+1, \lambda^{(m)},\lambda^{(n)},\Pi \rra  \nn \\
\Pi_{-}^{(m)}\lmd q, \lambda^{(m)},\lambda^{(n)},\Pi \rra  &=&
\sqrt{t^{\lambda^{(m)}}_{q}d^{\lambda^{(n)}}_{2\Pi -\lambda^{(m)} -\lambda^{(n)} -q }}\lmd q-1, \lambda^{(m)},\lambda^{(n)},\Pi \rra
\eea
The dimension of the representations are decided by the condition
\bea
t_{q}^{\lambda^{(m)}} d^{\lambda^{(n)}}_{2\Pi -\lambda^{(m)} -\lambda^{(n)} -q} &\ge& 0 \nn \\
2\Pi -\lambda^{(m)} -\lambda^{(n)} -q \ge 0
\eea
\section{Mapping of polynomial algebras to linear Lie algebras}

In this section we will give a general formalism of mapping a general
polynomial algebras to a linear Lie algebra. Consider a polynomial algebra
defined by (1.24)
\bea
\lsb {\cp}_0 , {\cp}_{\pm} \rsb &=& \pm {\cp}_{\pm}, \nn \\
\lsb {\cp}_+ , {\cp}_- \rsb     &=& f({\cp}_0 ), \nn \\
                        &=& g({\cp}_0 )-g({\cp}_0 -1),
\eea
where the functions $f({\cp}_0)$ and $g({\cp}_0)$ are polynomials in
${\cp}_0 $. This kind of algebras can be mapped to a Heisenberg algebra
by introducing a function $F(C,{\cp}_0 )$ given by,
\be
F(C,{\cp}_0) =\fr{{\cp}_0 +\alpha}{C-g({\cp}_0 )}, \ee where
$\alpha$ is an arbitrary constant which has to be fixed by the
commutation relations of the polynomial algebra. The operator $C$
is the Casimir operator of the polynomial algebra given by(1.26).
By introducing an operator $\tilde{{\cp}}_+  $ , canonical
conjugate to ${\cp}_-$, given by
\be
\tilde{\cp}_+  = {\cp} _+  F(C.{\cp} _0 ). \lb{conj} \ee

The following commutation relation are obtained. \bea \lsb {\cp}_-
,\tilde{{\cp}}_+ \rsb &=& \lsb {\cp}_- , {\cp}_+ F(C,{\cp}_0 )\rsb
\nn \\
                           &=& {\cp}_- {\cp}_+ F(C,{\cp}_0 )- {\cp}_+ F(C,{\cp}_0) {\cp}_- \nn \\
                           &=& (C-g({\cp}_0)) F(C,{\cp}_0) - F(C,{\cp}_0 -1 )(C-g({\cp}_0 -1))\nn \\
                           &=& {\cp}_0 +\alpha - ({\cp}_0 -1+ \alpha )\nn \\
                           &=& 1 \nn \\
\lsb {\cp}_0 , {\cp}_- \rsb        &=& -{\cp}_- \nn \\
\lsb {\cp}_0 , \tilde{{\cp}}_+\rsb &=& +\tilde{{\cp}}_+ .
\eea
The sub-algebra satisfied by the operators ${\cp}_- $ and $\tilde{{\cp}}_+$ is a Heisenberg
algebra. The above mapping is realized on the representation space of the
polynomial algebra. Note that in the above formalism $\tilde{{\cp}}_+$ is not adjoint
to the ${\cp}_-$. A more symmetrical mapping to Heisenberg algebras given by,
\bea
\tilde{{\cp}}_- &=& [F(C,{\cp}_0)]^{1/2} {\cp}_- , \nn \\
\tilde{{\cp}}_+ &=& {\cp}_+[F(C,{\cp}_0)]^{1/2}
\eea
such that
\be
\tilde{{\cp}}_- = (\tilde{{\cp}}_+ )^{\dagger} . \ee Once we have
mapped the polynomial algebras  to a Heisenberg algebra , we can
find a realization of the number state in the representation space
of the polynomial algebra. We will now identify the number
operator as,
\be
\tilde{{\cp}}= \tilde{{\cp}}_+ {\cp}_-= {\cp}_+ F({\cp}_0 ,C)
{\cp}_- . \ee This will satisfy the Heisenberg algebra commutation
relation \bea \lsb \tilde{{\cp}} ,\tilde{{\cp}}_{\pm} \rsb &=& \pm
\tilde{{\cp}}_{\pm} \nn \\ \lsb \tl{{\cp}}_- , \tl{{\cp}}_+ \rsb
&=& -1 \eea Since,
 \bea \tl{{\cp}}_+ \tl{{\cp}}_- &=&  F({\cp}_0 -1,C)
{\cp}_+ {\cp}_- \nn \\
                      &=& \fr{ {\cp}_0 -1 +\alpha}{C-g({\cp}_0 -1)} (C-g({\cp}_0 -1)) \nn \\
                      &=& {\cp}_0 -1 +\alpha,
\eea thus ${\cp}_0 $ and $\tl{{\cp}}_0 $ differ by a constant
factor. Now the representation becomes \bea \tl{{\cp}}_0 \lmd k_i
,n\rra &=& n+k_i +\alpha -1 \lmd k_i ,n\rra\nn \\ \tl{{\cp}}_+
\lmd k_i ,n\rra &=& F({\cp}_0 ,C ){\cp}_+ \lmd k_i ,n\rra \nn\\
                    &=& \fr{n+k_i +\alpha }{[C-g(n+k_i )]^{1/2}}\lmd k_i ,n+1\rra \nn \\
\tl{{\cp}}_- \lmd k_i ,n\rra &=& [C-g(n+k_i -1)]^{1/2} \lmd k_i ,n\rra .
\eea
For an illustration, consider the quadratic algebra $Q^{-}(1,1)$ considered
in chapter 2, which
has an infinite dimensional representation. This can be mapped in to a Heisenberg
algebra by defining
\be
\tl{Q}_+ =Q_+ F(Q_0 ,C) , \,\, \,\tl{Q}_- =Q_- ,\,\,\,
\tl{N} = \tl{Q}_+ \tl{Q}_-
\ee
where
\be
F(Q_0 ,C) = \fr{Q_0 + \alpha }{C(l,k) -g(Q_0)}. \ee Now these
operators act on a Fock space  realization  given by $\lmd k_i
,n\rra$, \bea \tl{N} \lmd k,l,n\rra    &=& n+k+l +\alpha -1 \lmd
k,l,n\rra\nn \\ \tl{Q}_+ \lmd k,l ,n\rra &=& \fr{n+k +l+\alpha
}{[(n+1)(n+2k)(n+k-2l+n)]^{1/2}}\lmd k,l,n+1\rra \nn \\ \tl{Q}_-
\lmd k,l ,n\rra &=& [n(n+2k-1)(n+k-2l)]^{1/2} \lmd k,l,n-1\rra
\eea These polynomial algebras can also be mapped in to  $SU(1,1)$
or $SU(2)$ algebras. Consider a function $G(C,{\cp}_0 )$ and the
operator,
\be
\bar{{\cp}}_- = G(C,{\cp}_0 ) {\cp}_-
\lb{bar2}
\ee
such that
\bea
\lsb {\cp}_+ ,\bar{{\cp}}_- \rsb &=& \lambda 2{\cp}_0 \nn \\
\lsb {\cp}_0 ,\bar{{\cp}}_- \rsb &=& -2 \bar{{\cp}}_- .
\eea
Such a mapping is possible by choosing
\be
G(C,{\cp}_0 ) = \frac{{\cp}^{2}_0 -{\cp}_0  +
\epsilon}{C-g({\cp}_0 -1)}, \ee where $\epsilon$ is an arbitrary
constant which has to be fixed. For $\lambda =1$ and $-1$ one will
get $SU(2)$ and $SU(1,1)$ algebra respectively. This operator acts
on  the representation space of the quadratic algebra. In this
case , it is always possible to find a symmetric mapping by
choosing \bea \bar{{\cp}}_- &=& [G(C,{\cp}_0 )]^{1/2} {\cp}_- \nn
\\ \bar{{\cp}}_- &=& {\cp}_+ [G(C,{\cp}_0 )]^{1/2} \eea such that
$\bar{{\cp}}_- $ and $\bar{{\cp}}_- $ are adjoint to each other.

Thus, we have shown how two mutually commuting polynomial algebras
of order $m$ and $n$ can be combined to get two polynomial
algebras of order $m+n+1$.  The simplest example of this
construction is the Jordan-Schwinger construction in which $SU(2)$
and $SU(1,1)$, linear algebras corresponding to $m$ $=$ $1$, are
obtained from two commuting boson algebras which are algebras of
order $m$ $=$ $0$.  By combining a boson algebra with $SU(2)$ or
$SU(1,1)$ we get four classes of quadratic algebras.  Using a
boson algebra and these quadratic algebras one can generate
fourteen classes of cubic algebras.  Higher order algebras can be
generated in a similar way by combining lower order algebras.  It
should be noted that the above construction leads to polynomial
algebras in which the coefficients of the polynomial are central
elements which are defined in terms of the Casimir operators of
the original algebras with which one starts or a combination of
their generators ${\cp}_0^{(m)}$ and ${\cp}_0^{(n)}$.  Then, an
interesting open problem is the following: given a polynomial
algebra with numerical constant coefficients is it possible to
identify it with a particular type of polynomial algebra generated
in the above manner and corresponding to certain numerical values
of the central elements.  We hope that an answer to this question
would help understand the classification problem of polynomial
algebras.

We have also mapped the Polynomial algebras to lower order algebras by using the techniques learnt from the theory of deformed algebras.
This will aid us in constructing the coherent states presented in section 5.



\setcounter{equation}{0}
\chapter{Coherent States of the Polynomial Algebras}
\markboth{}{Chapter 5.   Coherent States of ....}
The study of coherent states has been one of the fastest developing areas
in physics during the last three decades. These states were first introduced
by Schroedinger to describe the non-spreading wave packets of the quantum harmonic oscillator\cite{sch}.
It was Sudarshan, Glauber and Klauder
who  resurrected these states in the context of  quantum optics \cite{sud,gb,kld}.
A good survey of the  important in this field  work can be found in \cite{kla}.
Since then many generalizations of the canonical (harmonic oscillator) coherent states have been
introduced on mathematical and physical grounds.
Of these, the notable generalizations have been the
group theoretic coherent states introduced by Perelomov \cite{p} and Gilmore \cite{glm},
 the generalized lowering operator states of Barut and Girardello\cite{Barut}, 
the minimum uncertainty states of Nieto and Simmons\cite{nieto} and the algebraic coherent states
introduced by Satyaprakash and Agarwal and Trifonov\cite{agr,trif}.

Coherent states associated with the symmetry group of a system with Hamiltonian H have the particularly useful property that the time evolution
of any coherent state remains a coherent state. This is the property of temporal stability.
Furthermore, the general coherent states associated with the dynamical symmetry group of
a quantum mechanical Hamiltonian  aid us in finding the eigenspectrum of the  physical problem.
Since it has been seen that
not only linear algebras but also the non linear algebras appear as the
dynamical algebra and the symmetry algebra of many quantum Hamiltonian
systems, it is of
physical and mathematical interest to construct the  various coherent states associated with non-linear algebra.
The aim of this chapter is to present these construction methods.

For completeness we begin this chapter with a brief review of the different types of coherent states associated with Lie algebras.

\section{A brief review of coherent states}
It is well known that electromagnetic field can be represented by harmonic
oscillator and the symmetry algebra of the harmonic oscillator is the
Heisenberg algebra. It was Glauber who showed that coherent states can be used to
describe the electromagnetic field.
Glauber's construction allows three types of coherent states, which are all equivalent for the
 harmonic oscillator.

Consider the Hesenberg algebra generated by $a,a^{\dag}, N=a^{\dag}a$ which obey the commutation relations
\be
\lsb a, a^{\dag} \rsb=1\;\;\;\; \lsb N,a \rsb=-a\;\;\; \lsb N, a^{\dag}\rsb=a^{\dag},
\ee
and the Hamiltonian$ H=\hbar \omega(N+\frac{1}{2}) \;\; H|n>=E_n|n>$,
then the coherent states of the Harmonic oscillator are defined by the following relations:
\begin{itemize}
\item{1}
\be
a \lmd \alpha \rra = \alpha \lmd \alpha \rra
\ee
In terms of the number states they are given by
\be
\lmd \alpha \rra = e^{\fr{- \mid \alpha \mid^2 }{2} } {\sum}^{\inf}_{n=0} \fr{\alpha^n}{(n!)^{1/2}} \lmd n \rra
\ee
These are called annihilation operator coherent states (AOCS).
\item{2} \\
These are obtained by applying the operator $D(\alpha) =e^{\alpha a - \alpha^* a^{\dagger}}$
on the ground state. \\
\be
\lmd \alpha \rra = e^{\alpha a - \alpha^* a^{\dagger}} \lmd 0 \rra
\ee

These are called displacement operator eigenstates (DOCS).\\
\item{3}\\
The states which minimize the uncertainty product
$\Delta x \Delta p $ .
For Harmonic Oscillator these are given by
\be
\psi_{cs} (x) = [2\pi(\Delta x)^2 ]^{-1/4} e^{-\fr{(x-<x>}{2(\Delta x)})^2 + \fr{i}{h} <p> x}
\ee
These are called minimum uncertainty coherent states (MUCS).\\
\end{itemize}
In the harmonic oscillator case all the three coherent states are equivalent,
 but in general they correspond to different states.
Coherent states have three important properties:\\
1. Resolution of identity:\\
   There exist a measure $d\mu(\alpha)$ for a coherent states such that
\be
\int d\mu(\alpha) | \alpha >< \alpha | = 1\,.
\ee
For the Harmonic oscillator,
 $d\mu(\alpha)= e^{-|\alpha|^2/2}d^2\alpha$, where $\alpha=\alpha_1+i\alpha_2$ and $d^2 \alpha=d\alpha_1d\alpha_2$.
This resolution of unity enables us to expand any arbitrary state $|\psi>$ in the Hilbert space in terms of coherent states.
\be
|\psi>=\int d\mu(\alpha)\psi(\alpha)|\alpha>
\ee
2. They are non-orthogonal normalized states which are overcomplete.
\be
<\alpha | \beta > \neq 0 \,\,\,\mbox{for} \,\,\,\alpha \neq \beta
\lb{cspr1}
\ee
and also
\be
<\alpha | \alpha > = 1 \, .
\lb{cspr2}
\ee
This property is called overcompleteness because the coherent state basis
contains more states than necessary to decompose any arbitrary vector.\\
3. Temporal Stability: \\
If $|\alpha>$ is a coherent states associated with hamiltonian $H$
then the time evolved state, $ e^{iHt}|\alpha>=|\alpha(t)>$, is also a
coherent state.\\
A generalization of the harmonic oscillator coherent states  to other types of potentials was
done by Nieto and Simmons, Jr \cite{nieto}.

Fock, Bargmann and Segal made an important contribution to the field by
using coherent states for representing state vectors in a Hilbert
space by entire analytic functions called the Fock Bargmann or Bragmann
Segal representations\cite{fock,bargman,segal}.
If $|\psi>$ is an arbitrary vector in the Hilbert space of the Harmonic
oscillator then we have seen that
It can be determine completely by the function $<\alpha|\psi>=e^{\frac{1}{2}|\alpha|^2}
\psi(\bar{\alpha})$
and $\psi(z)=\sum c_n u_n(z)$, where $u_n(z)=z^n/n!$, a series which
converges uniformly in the z plane showing that $\psi(z)$ is an entire analytic function in the
complex z plane
with an inner product
\be
|\psi\ |^2=\int e^{-|z|^2}|\psi(z)|^2d\mu(z)
\ee
Bargmann showed that this functional space is a Hilbert space and Fock showed that in this space
$a $ and $a^{\dag}$ have the differential realizations
$a=\frac{d}{dz}$ and $a^{\dag}=z$. The representation was studied by Bargmann for a finite number of creation and annihilation operaors and for an infinite number by Segal.
It is therefore known as the BFS (Bargmann Fock Segal) representation.
This proves very useful in generalizations of coherent states to arbitrary algebras.

The generalization of coherent states to the compact and non-compact Lie algebras $SU(2)$ and $SU(1,1)$ was done by Gilmore, Perelomov and Barut and Girardello.
For SU(1,1) Barut and Girardell constructed the lowering operator eigenstates which correspond to the SU(1,1) generalization of the AOCS of the Harmonic oscillator.
For an SU(1,1) algebra generated by the Cartan basis $K_+,K_-, K_0$ such that

\be
\lsb K_+, K_- \rsb=-2 K_0\;\;\;\; \lsb K_0,K_- \rsb=-K_-\;\;\; \lsb K_0, K_+\rsb=K_+
\ee
The B-G coherent states  $|\xi>$ are given by:
\be
K_- \lmd \xi \rra = \xi \lmd \xi \rra .
\lb{5-2-1}
\ee
In terms of the basis functions corresponding to the positive
discrete series representations of $SU(1,1)$ we get
\be
\lmd \xi \rra = N \sum_{n} \lsb \fr{\Gamma(n+2k )}{\Gamma (n+1)}
\rsb^{1/2} \xi^n \lmd k,n \rra
\ee
where$ N^2= _0F_2 (0,2k;|\xi|^2  ) $ is the normalization factor.

The states $|\xi>$ can be shown to be overcomplete and the resolution of unity is given by
\begin{equation}
\int d^{2}\xi \frac{2}{\pi} I_{k}(2|\xi|)K_{k}(2|\xi|)|\xi><\xi|=1,
\end{equation}
where $I_k$ and $K_k$ are the modified bessel function of first and third kind
respectively.

Analogous to the BFS representation for the Heisenberg group,
 there exist a realization for the Barut Girardello states  coherent states in the space of analytic functions.

For each $k$ In the Bargman space of the $SU(1,1) $, the linear operators acting on the Hilbert space of entire functions representing the generators
have a differential realization,
\bea
K_0 &=& z\ddz \nn \\
K_+ &=& z \nn \\
K_- &=& z\ddzt + 2k\ddz
\eea
The B-G coherent states in this realization are the solutions of the second order
differential equation,
\be
z \fr{d^2 \Psi}{dz^2} + 2k \fr{d\Psi}{dz} - \alpha \Psi = 0.
\ee
The solution is a Hypergeometric function given by,
\be
\Psi(z, \alpha ) = _0F_2 (0,2k;-\alpha z ) .
\ee
The lowering operator coherent states cannot be generalized to any
arbitrary Lie group but only those for which the Hilbert space on
which they are defined is infinite dimensional. Thus for SU(2),
which spin systems which have a finite dimensionality one has to
have an alternative definition. This was obtained for SU(2) by
generalizing the   displacement operator coherent
states(definition 2)  by Gilmore and other Lie groups by Perelomov
in the following way. Here we do not have to restrict the group to
a compact group but can encompass non-compact groups also.
 Let $G$
be a Lie Group and $g$ the corresponding Lie algebra and  let
$\lmd \psi_0 \rra$ in the  Hilbert space which carries a unitary
representation  $G$ . Let  $G_m$ be the maximal subgroup that
leaves the fixed vector invariant upto a phase factor. Since by
definition any element of g can be written as g=hm where
$h\in G_m$ and $m\in G/G_m$ and
$h|\psi_0>=e^{i\phi}|\psi_0>$, for Lie algebras which have a cartan
representation $m=e^{\sum_\alpha \gamma_{\alpha}E_\alpha -\gamma_{\alpha}^{*}E_{-\alpha}}$.
Then the vector,
\be
|\psi>_{G/G_m}=m |\psi_0 > \ee is defined the Lie algebra is
called the Perelomov or Generalized coherent state of G upto
normalization.

From the above definition  for the group $SU(1,1)$ these states
are given by
\be
e^{\alpha K_+ -\alpha^* K_-} \lmd k,0 \rra,
\ee
The coset space
in this case being given by SU(1,1)/U(1)( a two dimensional
hyperboloid).

For the $SU(2)$ group, the coset space in this case is $SU(2)/U(1)
$(a two dimensional sphere). The fixed state $|\psi_0>$ is usually
taken as the vector of lowest weight so that the coherent states
are given by
\be
e^{\beta J_+ - \beta^* J_- } \lmd j, -j |\rra \ee where $\alpha$
and $\beta$ are complex numbers and $\lmd k, 0 \rra $ and $\lmd j
-j \rra $ are the lowest weight basis states for the $SU(1,1)$ and
$SU(2)$ algebras. A more detailed study of these states can be
found in \cite{pe,g}. As a precursor to the next section (  at the
risk of being too detailed) we do mention here that an explicit
expression of the Generalized coherent states is obtained by using
the disentanglement formula based on the Baker Campbell Hausdorff
formalae of SU(1,1) and SU(2).
\section{Construction of coherent states for non-linear algebras}
Coherent states of Polynomial algebras can be constructed by  a
variety of methods. For the non-compact cases of the polynomial
algebras one can define Lowering operaor or Barut-Griradello
states in a straigthforward fashion by using the representation
theory that we have developed in the earlier chapters and we shall do
so in section 3. For constructing the generalized coherent states, however, we
are hampered by two facts peculiar to non-linear
algebras of the polynomial type. The first is that unlike the case
of the ordinary Lie group,
 an exponential mapping does not exist formally for the polynomial
algebra because they do not  admit a group structure because of
the non-linear terms coming in the algebra. For Lie
 algebras  explicit construction of the coherent states can be done because
 group multiplication is possible by using
BCH formulas. But in the case of polynomial algebras we do not
have  analogous BCH formulas. We circumvent this obstacle by using
a unified approach for the construction of the coherent states
(CS) of these algebras which is quite general and will greatly
facilitate the physical applications of these algebras to many
quantum mechanical problems. This method is a generalisation to
non-linear algebras of the procedure for constructing multiboson
states first presented by Shantha et.al\cite{shanta}. In this
method, for  ordinary Lie algebras, the construction of the CS for
 was shown to be a two step procedure. First, the canonical
conjugate of the lowering operator were found and then the BG CS
of these  algebras were obtained by the action of the exponential
of the respective conjugate operators on the vacuum
\cite{shanta,sc,pani}. Another CS, dual to the first one,which is
the generalized coherent state in the Perelomov sense, naturally
follows from the above construction.

More explicitly,
 let $F$ be an operator consisting of products of annihilation operators
$a$ and the number operator $N$.
Let $\lsb F,F^{\dag} \rsb=P(N)$, where P is a polynomial in N
(note here , this method has been used for q-deformed oscillators also , where P is not a polynomial function of N, we are concerned only with the case
when P is a polynomial).

 Let $\mid v_i > $ where $ i=0,1,2 ...)$
be the set of states annihilated by $F$. Let $G_{i}^{\dagger} $be
an operator satisfying the relation
\be
\lsb F, G^{\dagger}_i \rsb =1,
\lb{mcong}
\ee
in the space $S_i$ of states $=\{ (F^{\dagger} )^n \mid v_i > n=1,2... \}$.
The states
\be
\mid \alpha >_i = e^{\alpha G^{\dagger}_i } \mid v >_i \,\,\,\,i=1,2... ,
\ee
are eigen states of $F$ in the Barut-Girardello sense.
Thus for the construction of BG states the operator F is undeformed.
The corresponding
generalized (Perelomov) states are obtained by taking the dual of the above
construction in that $F^{\dag}$ is undeformed, but $F$is deformed. Here we define
$
G_i = (G^{\dagger}_i)^{\dagger} $ such that $\lsb G_i , F^{\dagger} \rsb =1.
$
Then the states defined by \be \mid \gamma > = e^{\gamma F^{\dagger }} \mid v>_i \,\, i=1,2, ...\ee are
eigen state of $G_i $ and upto a normalization equivalent to Perelomov states.

 We generalize the above
construction to non-linear polynomial algebras  which can be
considered as deformations of the SU(2) and SU(1,1) algebras and
exploit the fact that they have multiboson realizations by
mapping the polynomially deformed algebras to their undeformed
counterparts  as described in chapter 4. This mapping is utilized
to find the CS in the Perelomov sense \cite{p}.

Other coherent states in
the literature which are essentially special cases of this
construction are the `f-oscillator states' \cite{sudarshan} and
the non-linear states $
f(N_0)a|\lambda>=\lambda|\lambda>$,
 which have been shown to be useful for the trapped ion problem\cite{vogel}.
While these are non-linear {\it harmonic oscillator} coherent states, the CS that we construct
may be called {\it non linear} SU(1,1) (or SU(2)) coherent states.
The states  which we shall construct would give a multi-mode generalization of the type  $f(n_a,n_b)a^{n}b^{m}|\lambda>=
\lambda|\lambda>$, as {\it one} of the possible coherent states. Thus our construction encompasses existing non-linear states and allows  for the construction of new physical states.
One such state, for example, is the case n=1 and m=1, which is a two-mode realization of the non-linear coherent states.
 Our method is quite general and encompasses q-deformations 
of linear Lie algebras\cite{sun}.

Now we give an outline of our method of construction of the coherent states of polynomial algebras.
We have seen that the Polynomial algebras do not form a Lie algebra we
have seen that they satisfy all the properties, except the
closure property, of the Lie algebra. 
We concentrate on the three dimensional polynomial algebras with coset structure,
These resemble the Lie algebras $SU(2)$ and $SU(1,1)$ and we would like to exploit this resemblance.

 Consider a
3-dimensional polynomial algebra ${\cal P}^{(3)}$ generated by
${\cal P}_{0}^{(3)} ,{\cal P}_{+}^{(3)} $ and ${\cal P}_{-}^{(3)}$
satisfying,
\bea \lsb {\cp}^{(3)}_0 , {\cp}^{(3)}_{\pm} \rsb &=&
{\cp}^{(3)}_{\pm} \nn \\ \lsb {\cp}^{(3)}_+ , {\cp}^{(3)}_- \rsb
&=& f_{3} ({\cp}^{(3)}_{0} )
\lb{5-3-1}
\eea
Here,
$
 f_{3} ({\cp}^{(3)}_{0} )
=g_{3} ({\cp}^{(3)}_{0} )-
g_{3} ({\cp}^{(3)}_{0}-1 )$.

Then by generalizing the above construction of \cite{shanta}, we first construct the
Canonical conjugate $ \tilde{\cp} ^{(3)}_+$ of $ \cp ^{(3)}_-$ by

\be
\tilde{\cp} ^{(3)}_+ =\frac{ \tilde{\cp} ^{(3)}_0 +\delta}{C-g_{3} ({\cp}^{(3)}_{0} )}\, .
\ee
Depending on the order n of the polynomial algebra,  there will be n+1 degenerate states
annihilated by ${\cp}^{(3)}_-$. We denote these as $|j,0>_i$. For each, the value of $\delta=\delta_i$ is
appropriately chosen to project the ground state needed for the construction.
For each $\delta_i$
the B-G states in our construction are:
\be
\lmd \beta \rra  = e^{\beta \tilde{\cp} ^{(3)}_+} \lmd j_i ,0 \rra
\lb{5-2-12}
\ee
Now operating with ${\cp}^{(3)}_-$ on $\lmd \beta \rra $, one will get
\bea
{\cp}^{(3)}_- \lmd \beta \rra &=& {\cp}^{(3)}_- e^{\beta \tilde{\cp}^{(3)}_+} \lmd j_i ,0 \rra \nn \\
                    &=& \beta e^{\beta \tilde{\cp}^{(3)}_+} \lmd j_i ,0 \rra + e^{\beta \tilde{{cp}^{(3)}_+}}{\cp}^{(3)}_- \lmd j_i ,0 \rra
\lb{5-2-13}
\eea
since ${\cp}^{(3)}_-$ annihilate vacuum the second term vanishes . So one will get
\be
{\cp}^{(3)}_- =\lmd \beta \rra = \beta \lmd \beta \rra
\lb{5-2-14}
\ee
So the states defined by \ref{5-2-14} are  indeed B-G type coherent states.

 Now to construct the Perelomov type states which are dual to the above B-G states
 we observe that
for this algebra
we can take $U(1)$ as the stability group ($U(1)=e^{\gamma {\cp}^{(3)}_{0} }$).
Thus the operator
$e^{\alpha {\cp} ^{(3)}_+ -{\bar \alpha} {\cp}^{(3)}_{-} }$ is a
member of the coset space ${\cp}^{(3)} /U(1) $ , because of the
first two commutation relation of \ref{5-3-1}.
Therefore, a Perelomov
coherent state for a polynomial algebra is defined up to a
normalization factor as,
\be
|\alpha >= e^{\alpha {\cp} ^{(3)}_+ - {\bar \alpha } {\cp}_{-}^{(3)}} |k_i,0>,
\lb{5-3-2}
\ee
where $k_i$ represents all the labels of the polynomial algebra. The computation of such states are difficult because one cannot disentangle
the above displacement operator because of the non-linear terms coming in the commutation relation.
Instead of that we will make use of the mapping introduced in the chapter 4 to map these algebras to $SU(2)$
and $SU(1,1)$ algebras. For polynomial algebras having a finite dimensional representation
one can map these algebras to a $SU(2)$ algebra.  Here we will make use of
the mapping in which
${\cp}^{(3)}_{-} $  will change and ${\cp}^{(3)}_{+} $will remain the same.
This mapping is given explicitly by
\be
\bar{\cp} ^{(3)}_- = \cp ^{(3)}_-  G_1(C,{\cp}_0)\,\,\,\,,
\ee
Where $G_1$ is chosen such that
$
 \lsb {\cp}^{(3)}_+ , \bar{\cp}^{(3)}_- \rsb
= 2\lambda({\cp}^{(3)}_{0} ) $, where $\lambda=\pm1$ depending on whether we are mapping to $SU(2)$ or $SU(1,1)$.
The explicit form of this mapping is given in the last section of chapter 4.
Then we  define the Perelomov-type coherent states as
\be
|\alpha > = e^{\alpha \cp ^{(3)}_+ -\beta \bar{\cp}^{(3)}_- } |k_i,0>
\lb{5-3-3}
\ee
Note that the state $k_i,0> $ is annhilated by $\bar{{\cp}}^{(3)}_-$.
Since ${\cp}^{(3)}_+$ and $\bar{\cp}^{(3)}_-$ satisfy
 $SU(2)$ or $SU(1,1)$ algebra
we can use their   disentanglement formula
 to get up to a normalization factor, a Perelomov type coherent state:
\be
|\xi > = e^{\xi \cp ^{(3)}_+ } |k_i,0> \lb{5-3-4}
\ee where
$\xi$
is  a fuction of $\alpha $ and $\beta$ chosen appropriatelt for the physical application.
Since we know the action of
${\cp}^{(3)}_{+} $ on the state $|k_i,n>$ one can immediately
calculate the coherent states.

We now illustrate the above procedure by considering the quadratic algebra.
\be
[Q_0\,,\,Q_{\pm}]=\pm Q_{\pm}\,\,\,,\,\,\,[Q_+ \,,\,Q_-]=\pm 2bQ_0 + a Q_{0}^{2}+c\,\,\,\,.
\ee
In this case, $f_1(Q_0)=\pm 2bQ_0 + a Q_{0}^2+c=g_1(Q_0)-g_1(Q_0 -1)$.
with $ g_{1}(Q_{0})=\fr{a}{3} Q_{0}(Q_{0} +1)(Q_{0}+ \fr{1}{2}) + Q_{0}(c \pm b(Q_{0}+1))$.

In the non-compact case, i.e, for polynomial deformations of $SU(1,1)$,
the unitary irreducible representations (UIREP) are either bounded below or
above, we can construct the canonical conjugate $\tilde{Q_{+}}$ of $Q_-$
such that $[Q_-\,,\,\tilde{Q_+}]=1$. It is given
by $\tilde{Q_+}=Q_+F_{1}(Q,Q_0)$, with
\be
F_{1}(C,Q_0)=\frac{Q_0 + \delta}{C(Q_{0})- \fr{a}{3} Q_{0}(Q_{0} +1)
(Q_{0} + \fr{1}{2} ) - Q_{0}(c \pm b(Q_{0} + 1))} \,\,\,\,.
\ee
As can be seen easily, in the case of the finite dimensional UIREP,
 $\tilde{Q_+}$ is not well defined since $F_1(C,Q_0)$ diverges on the
highest state. The values of $\delta$ can be fixed
by demanding that the relation, $[Q_-\,,\,\tilde{Q_+}]=1$, holds in the
vacuum sector $\mid v_{i} >$,  where , $\mid v_{i} >$'s are annihilated by $Q_-$.
This gives $Q_-\tilde{Q_+}\mid v_{i}>=\mid v_{i}>$, which leads to
$(Q_0+\delta)\mid v_{i}>=\mid v_{i}>$. The value of the Casimir operator,
$C=Q_-Q_+ + g_1(Q_0)$, can then be calculated.
Hence, the unnormalized coherent state $\md \alpha >$, such that  $Q_-\md \alpha>=
 \alpha \md \alpha >$ is given by $e^{\alpha \tilde{Q_+}}\md v_{i}>$.
 We can define the canonical conjugate of $Q_{+}$ by $[\tilde{Q_{+}^{\dagger}}\,,\,Q_+]=1$.
The other coherent state
is $\md \gamma >=e^{\gamma Q_+}\mid \tilde{v}_i>$ , where
$\tilde{Q_{+}^{\dagger}}|\tilde{v}_i>=0$. This coherent state
is an eigen state of the operator $ \tilde {Q}^{\dagger}_{+}$. Depending
on whether the
UIREP is infinite or finite dimensional, this quadratic algebra can
also be mapped onto  the $SU(1,1)$ and $SU(2)$ algebras respectively. Leaving
aside the commutators not affected by this mapping, one gets,
\be
[Q_+\,,\,\bar{Q_-}] =- 2bQ_0\,\,\,\,,
\ee
where $\lambda=1$ corresponds to the $SU(1,1)$ and $\lambda=-1$ gives the $SU(2)$ algebra.
Explicitly,
\be
\bar{Q_-} = Q_- G_1(C,Q_0)\,\,\,\,,
\ee
and
\be
G_1(C,Q_0) = \frac{\lambda (Q_{0}^2 -  Q_0)+\epsilon}{C-g_1(Q_0 - 1)}\,\,\,\,,
\ee
$\epsilon$ being an arbitrary constant. One can immediately construct CS in
the Perelomov sense (see page 73-74 in ref\cite{p}) as
$|\xi>= U\mid v_{i}>$, where $U=e^{\eta Q_+ -\beta \bar{Q_-}}$,
with $\xi$ being a function of $\alpha$ and $\beta$..
For the compact case, the CS are analogous to the spin and atomic coherent
states\cite{atcs,spin}.\\

An explicit construction of the coherent states  for a special quadratic
algebra was presented in our paper \cite{vsk} will be given in the context
of its physical application in chapter6.\\

We now give an outline of the method of explicit construction of coherent states, which utilises the explicit matrix representations of the polynomial algebras given in the previous chapters.
We do this at the risk of being verbose as in some problems only the action of the generators on the basis functions is known and not the commutaion relations.
These explicit forms enable us to write down the various coherent states with great ease.
Consider the general algebra
$[N_0\,,N_{\pm}]=\pm N_{\pm}$ and $[N_+\,N_-]=g(N_0) -g(N_0 -1) $ \,.

The action on eigenstates of $N_0$ is given by
\be
N_0 \md j,m>=(j+m) \md j,m> \,,
\ee
\be
N_+ \md j,m>=\sq{C(j)-g(j+m)} \md j,m+1> \,,
\ee
\be
N_- \md j,m>=\sq{C(j)-g(j+m-1)} \md j,m-1>\,,
\ee
where $ C(j)=g(j-1) $.\\
 Now we construct the canonical conjugate operator
  $\tilde{N_+}=\frac{N_0+\delta}{C-g(N_0)}$ .
Depending on the order of the polynomial algebra n, there will be n+1 degenerate states
annihilated by $N_-$. We denote these as $|j,0>_i$. For each, the vaule of $\delta=\delta_i$ is appropriately chosen as shown earlier.

Explicitly the B-G  coherent state for a general polynomial algebra is thus given by: is given by
\bea
\md \beta >&=& A e^{\beta \tl{N_+}} \md j,0>_i \nonumber \\
&=&A\sm \beta ^n \fr{1}{\sq{(g(j-1)-g(j)) \ld (g(j-1)-g(j+n-1))}} \md j,n> \,.
\eea

A discussion of coherent states is incomplete without showing that these states do give a resolution
of the identity and that they are overcomplete. From the resolution of the identity we have:
\be
\int d\sigma(\beta^{*},\beta)\, |\beta\rangle\langle\beta| = {\bf 1}   \,.
\ee
Within the polar decomposition
ansatz
\be
d\sigma(\beta^{*},\beta) = \sigma(r) d\theta r dr
\ee
with $r=|\beta|$ and an yet unknown positive density $\sigma$  which provides the measure.
For the general case we have:
\be
2\pi\int_{0}^{\infty} dr\, \sigma(r) \, r^{2n+1} = (A)^{-2} (g(j-1)-g(j)).....(g(j-1)-g(j+n-1))
\ee
For the various cases the substitution of the explicit value of g(j) then reduces the expression on the R.H.S
to a rational function of  Gamma Functions and the measure $\sigma$ can be found by an inverse Mellin transform \cite{sol}.
For the general case the measure is a Meijer's G-function.
The fact that these states are overcomplete (i.e; $<\alpha|\beta>\ne 0$) can be shown for explicit examples.\\

For the corresponding Perelomov states
we deform $N_-$ keeping $N_+$ the same. For this we use the deformation to the SU(2) or SU(1,1) algebra to construct
  $\tilde{N_-}=N_-\frac{F(N_0,C)}{C-g(N_0-1)}$ .Then since $N_0,N_+,\tilde{N_-}$ satisfy the SU(2) or SU(1,1) algebra
  the state given by 
\be
\lmd \alpha \rra  = e^{\alpha N_+-\alpha^*\tilde{N_-}} \lmd j_i ,0 \rra=A e^{\gamma N_+} \lmd j_i,0 \rra
\ee
is the Perelomov state upto a normalization factor. Here we have used the disentangling formula for $N_0,N_+ and \tilde{N_-}$ which is well defined.

Explicitly this state is found by

\bea
\md \gamma >&=& A e^{\gamma N_+} \md j,0>_i \nonumber \\
&=&A\sm \gamma ^n\frac{\sq{(g(j-1)-g(j)) \ld (g(j-1)-g(j+n-1))}}{n!} \md j,n>
\eea

These states can be normalised in the ususl fashion we get the resolution of the identity by finding $\sigma(r)$ such that
For the general case we have:
\be
2\pi\int_{0}^{\infty} dr\, \sigma(r) \, r^{2n+1} = A^{-2}\frac{1}{ (g(j-1)-g(j)).....(g(j-1)-g(j+n-1))}
\ee
The inverse Mellin Transform will again be a Meijer's G function .

\section{Construction of B-G type coherent states using the analytic representations of polynomial algebras}
Although the method presented in section two is quite general and is a unified method, for the construction of BG states of general quadratic and cubic
algebras can be quite involved in actual practice.

For the quadratic algebras discussed in chapter2, we may use the differential realization of the generators
to construct the BG states in a straightforward fashion.
Note that B-G states  are only well defined for the infinite-dimensional non-compact cases of the polynomial algebras.
The algebra $Q^+ (1,1)$ (case 4) is one such case and serves nicely to illustrate the properties that the B-G states of quadratic algebras have
 The B-G states are
associated with the $(k,l)^{th}$ infinite dimensional unitary irreducible representation.
Now let $\lmd k,l,\alpha \rra $ be the eign state of the lowering
operator $Q_- $ so that
\be
Q_- \lmd k,l,\alpha \rra = \alpha \lmd k,l,\alpha \rra \,.
\lb{5-2-5}
\ee
Now extending the  state $\lmd k'l,\alpha \rra $ in terms of
the complete state of states $\lmd k,l,n \rra $ given in (\ref{q+11})
we will get
\be
\lmd k,l, \alpha \rra = \sum_n c_n (\alpha) \lmd k,l,n \rra
\lb{5-2-6}
\ee
where $c_n (\alpha ) = < k,l,n | \alpha > $\\
In this expansion the equation \ref{5-2-6} becomes
\be
\sum_n \lsb n(n+2k -1)(n+k-2l)\rsb^{1/2} c_n (\alpha ) \lmd k,l,n-1 \rra = \sum_n c_n (\alpha ) \lmd k,l,n \rra
\lb{5-2-7}
\ee
The $c_n (\alpha )$'s satisfies the recurrence relation
\be
\lsb (n+1)(n+2k)(n+k-2l +1)\rsb^{1/2} c_{n+1}(\alpha ) = \alpha c_n (\alpha )
\lb{5-2-8}
\ee
which can be solved for
\be
c_n (\alpha ) = \fr{\alpha ^n }{\lsb n!(n+2k-1 )! (n+k-2l)! \rsb^{1/2}} c_0 (\alpha ).
\lb{5-2-9}
\ee
we will take $c_0 (\alpha )=1 $ which will be compensated for choosing the correct
normalization factor for $\lmd k,l,\alpha \rra $. So the B-G states are given  algebraically,
\be
\lmd k,l, \alpha \rra = N \sum \fr{\alpha^n }{\lsb n!(n+2k-1)! (n+k-2l)! \rsb^{1/2}} \lmd k,l,n \rra
\lb{5-2-10}
\ee
where the normalization factor $N$ is given by
\be
N = \lsb \fr{(2k)! (k-2l +1)!}{0F_2 (-;2k,k-2l+1;\mid \alpha \mid ^2 )}\rsb^{1/2}
\lb{5-2-11}
\ee
In terms of the single variable realization the coherent state equation  
(\ref{5-2-5}) can be written as 
\bea 
  &   &  \lsb z^2 \ddzth + (3k-2l+2) z \ddzt 
         + \lrb 2k^2-4kl+2k \rrb \ddz \rsb \Psi_{k,l}(\al, z) \nn \\ 
  &   &  \qquad \qquad \qquad \qquad \qquad \qquad 
         = \al \Psi_{k,l}(\al, z)\,,
\eea
which is the differential equation for  
\be
\Psi_{k,l}(\al, z) = {}_0F_2(-;2k,k-2l+1;\al z)\,. 
\ee
The resolution of the identity is given by
\begin{equation}
\int d \sigma(\alpha,\alpha^{*};k,l) |\alpha;k,l \rangle\langle \alpha;k,l| = \hat{1} \,.
\lb{inte}
\end{equation}

With a polar decomposition ansatz $\alpha=re^{i\theta}$ we get
 \be
d\sigma(\alpha^{*},\alpha;k,l) =N_{l,k}(r^2)( \sigma(r^2)) d\theta r dr \,,
\ee
where
\be
N_{l,k}(r^2)= _0F_2(-;2k,k-2l+1;r^2)\,.
\ee
The integral (\ref{inte}) reduces to the following condition on $\sigma(r^2)$:
\begin{equation}
\frac{1}{2}\int_{0}^{\infty} d(r^2)\, \sigma(r^2) \, (r^2)^{(n+1)-1} = \frac{1}{2\pi}\Gamma(n+1)
\, \frac{\Gamma(2k+n)
\Gamma(k-2l+1+n)}
{\Gamma(2k)\Gamma(k-2l+1)}.
\end{equation}
 $\sigma(r^2)$ is found by an inverse Mellin transform to be:
\be
\sigma(r^2)=\fr{1}{ \pi \Gamma(2k)(\Gamma(k-2l+1 )}
G^{3\,0}_{0\,3} (r^2|^{-}_{0,k-2l+1,2k} ),
\ee
where $G$ is the Meijers G function \cite{grad}.

\section{Perelomov type coherent states for the polynomial algebras}
The method of construction of Perelomov type coherent states
for the general three dimensional polynomial algebra has been
given in section 5.2. In this section we will explicitly construct the Perelomov type coherent state
for illustration. 
For the non compact case the {\it Perelomov type} states are given by:
\be
| \beta > = e^{\beta {\bar {Q}}_+ } |k_1 ,k_2 ,l,0 >
\ee
After using the representation of the cubic algebra one will get the expression,
\be
|\beta >=N\sum_{n} (\beta)^n \sqrt{\frac{\Gamma(2k+n) \Gamma(k-2l+1+n)}{\Gamma(n+1)\Gamma(2k)\Gamma(k-2l+1)}}|l,k.n>,
\ee
where the normalization coefficient N is given by
\be
N=\lsb ^2F_0(2k,k-2l+1;(|\beta|^2)) \rsb^{-\frac{1}{2}}.
\ee
The resolution of the identity in this case reduces to finding $\sigma(r^2)$ such that,
\begin{equation}
\int_{0}^{\infty} d(r^2)\, \sigma(r^2) \, (r^2)^{(n+1)-1} = \frac{1}{\pi}\Gamma(n+1)
\, \frac{\Gamma(2k)\Gamma(k-2l+1)}
{\Gamma(2k+n)\Gamma(k-2l+1+n)}
\end{equation}
and the resultant $\sigma(r^2)$ is given by
\be
\sigma(r^2)=\fr{1}{ \pi} \Gamma(k-2l+1 )\Gamma(2k)
G^{1,0}_{2,1} (r^2|^{k-2l+1,2k-1}_{0} ),
\ee
where $G$ is the Meijers G function.

The corresponding states for the compact case are given by:
\be
|\alpha,k,l>=N\sum_{n=0}^{2l-k} (\alpha)^n \sqrt{\frac{(2l-k)! (2k-1+n)!}{n!(2l-k-n)!(2k-1)!}}|l,k.n>.
\ee
For the purposes of calculating the measure for the resolution of identity we define
$\gamma=\frac{1}{\alpha}$.
The coherent state $|\gamma,k,l>$ becomes
\be
|\gamma,k,l>=N \gamma^{k-2l}\sum_{n=0}^{2l-k} (\gamma)^n \sqrt{\frac{(2l-k)! (k+2l-1-n)!}{n!(2l-k-n)!(2k-1)!}}|l,k.n>,
\lb{csq-11}
\ee
with the normalization coefficient N  given by:
\be
N=\lsb \frac{\Gamma(k+2l)}{(|\gamma|^2)^{2l-k}}k(k-2l,1-2l-k;(|\gamma|^2))\rsb ^{-\frac{1}{2}},
\ee
where $k(a,b,x)$ is the Confluent Hypergeometric function $ (^1F_1)$.

The resolution of the identity
for the coherent states $|\gamma ,k,l\rangle$ is given by
\begin{equation}
\int d \mu(\gamma,\gamma^{*};k,l) |\gamma;k,l \rangle\langle \gamma;k,l| = \hat{1}_{l,k} , \label{id}
\end{equation}
where $\hat{1}_{l,k}$ is the projection operator on the subspace 
${\cal H}_{2l-k}$:
\begin{equation}
\hat{1}_{l,k} = \sum_{n=0}^{2l-k} |l,k,n \rangle \langle l,k,n| .
\end{equation}

Again defining $\gamma=re^{i\theta}$
we have,
\begin{equation}
\sum_{n=0}^{2l-k} \frac{ \Gamma(2l+k-n) }{n! (2l-k-n)! } \left[ 
\int_{0}^{\infty}( r^2)^{n} M(r^2;k,l) d (r^2) \right] 
|l,k,n \rangle \langle l,k,n |  
= \hat{1}_{l,k} , \nonumber
\end{equation}
where we have defined
\begin{equation}
M(r^2;k,l) \equiv \frac{\pi (2l-k)!}{\Gamma(2l+k)} \frac{ \sigma(r^2;k,l) }{
(\Phi(k-2l,1-2l-k;r^2))}
  .
\end{equation}
By using the integral \cite{grad}, 
\begin{equation}
\int_{0}^{\infty} r^{b-1} \Phi(a;c;-r) d r = \frac{ \Gamma(b)
\Gamma(c) \Gamma(a-b) }{ \Gamma(a) \Gamma(c-b) } ,
\end{equation}
we obtain 
\[
M(r^2;k,l) = \frac{ \Gamma(2l-k+2) }{ \Gamma(2l+k+1) }\, 
\Phi(2l-k+2;2l+k+1;-r^2) .
\]

This gives us the final expression for the integration measure: 
\begin{equation}
 d \mu(\gamma,\gamma^*;k,l) = \frac{1}{2\pi} \frac{(2l-k+1)}{(2l+k+1) } 
\Phi(k-2l;1-2l-k;r^2) 
\Phi(2l-k+2;k+2l+1;-r^2) d (r^2) d \theta .
\lb{mes}
\end{equation}

The resolution of the identity is important because it allows the use
of the coherent states as a basis in the state space.\\ 
\section{Coherent states of cubic algebras}
The cubic algebras that we have constructed can also be mapped be mapped to
a $SU(2)$ algebra so that one can construct a Pereleomove type coherent states
for them. Such a construction is also possible in all the 10 case that we
had come across in chapter 3. The coherent state is given by the action
of the operator$ e^{\alpha C_+ - \alpha ^* \bar{C}_- } $
on the ground state. Here $\bar C$ is the operator defined in (\ref{bar2}).
The resultant state is given by up to a normalization
\be
| \gamma > = e^{\gamma C_+ } | k_i ,0>
\ee
where $k_i$ for $i=1,2.3$ are the constants of the algebra and $\gamma=
\fr{\alpha }{| \alpha |} \tan{|\alpha |}$.
Since the method is same
and to avoid repetition the different cases are given in the table below.\\
\begin{table}
\centering
\caption{Coherent states and resolution of the identity for  Three Dimensional Cubic Algebras
$C_{\alpha}(a,b)$.}

\begin{tabular}{|l|l|l|} \hline
case& coherent state & $\sigma (r^2)$ \\
\hline
$C_+(11,2)$ &
$ N\sum_{n=0}^{j-2k-k_1} (\beta)^n $
&$\fr{\Gamma(2k+k_1-j)\Gamma (j+2k+k_1+1) \Gamma (2k)}{\pi}\times$\\
&$[\fr{1}{(n)_n} (-1)^n (2k+k_1 -j)_n $
& $G^{10}_{30} \lrb -r^2 |^{2k-k_1-j-1,j+2k+k_1,2k-1}_{1}\rrb $ \\
&$(j+2k+k_1 +1)_n (2k_1  )_n ]^{1/2}$
& \\
\hline
$C_- (11,2)$
&$ N\sum (\beta)^n$
&$\fr{\Gamma(-j)\Gamma (j+1) \Gamma (k_1-2k)\Gamma (1-2k-k_1)}{\pi}\times $ \\
&$ [ \fr{1}{(n)_n} (-1)^n (-j)_n (j+1)_n$
&$ G^{10}_{30} \lsb -r^2 |^{2k=k_1-j-1,j+2k+k_1,2k-1}_{1}\rsb $ \\
&$ (k_1-2k)_n  (1-2k-k_1 )_n ]^{1/2}$
& \\
& $  \lmd j,k_1,k,n \rra$
& \\
\hline
$C_- (q_- (1) ,h)$
&$ N\sum_{n=0}^{2k-k_1+l} (\beta)^n $
& $\fr{\Gamma(k_1-2k-l)\Gamma (2k_1)\Gamma (k_1-2l)}{\pi} \times $\\
&$ [ \fr{1}{(n)_n}(2k_1)_n (k_1-2l)_n $
&$G^{10}_{30} \lsb -r^2 |^{k_1-2k-l-1,2k_1-1,k_1-2l-1}_{1}\rsb $ \\
& $ (2k-k_1+l+1)_{-n}]^{1/2}$
& \\
& $ \lmd l,k_1,k,n \rra$
& \\
\hline
$C_+ (q_- (1) ,h)$
& $ N\sum_{n} (\beta)^n$
& $\fr{\Gamma(k_1-2k-l+1)\Gamma (2k_1)\Gamma (k_1-2l)}{\pi} \times $ \\
&$ [ (-1)^n \fr{1}{(n)_n}$
& $G^{10}_{30} \lsb -r^2 |^{k_1-2k-1,2k_1-1,k_1-2l-1}_{1}\rsb $ \\
& $  (k_1-2k-l+1)_n $ 
& \\
& $(2k_1)_n(k_1- 2l)_{n} ]^{1/2}$
& \\
&  $ \lmd l,k_1,k,n \rra$
& \\
\hline
$C_- (q_+ (1) , h) $
& $ N\sum_{n}(\beta)^n [ (-1)^n \fr{1}{(n)_n} $
& $\fr{\Gamma(k_1-2k-l)\Gamma (2k_1)\Gamma (k_1-2l)}{\pi} \times $ \\
& $    (k_1-2k-l)_n (2k_1)_n$
&$ G^{10}_{30} \lsb r^2 |^{k_1-2k-1-1,2k_1-1,k_1-2l-1}_{1}\rsb $ \\
& $(k_1-2l)_{n} ]^{1/2}$
& \\
& $\lmd l,k_1,k,n \rra$
& \\
\hline
$C_+(2,2)$
& $ \sum_n \beta^n $
&$\fr{1}{\pi} \fr{\Gamma (j_2 -j_1 -2k +1 )}{\Gamma (2j_1 +1 )\Gamma (j_2 +j_1 +2k +1 )} \times $\\
& $\fr{(j_2 -j_1 -2k +1+m)_n }{2j_1 +1)_{-n} (j_2 +j_1 +2k +1)_{-n} (1)_n }$
& $G^{10}_{03}\lrb -r^2 |^{1}_{j_2-j_1 -2k +1 ,-2j_1 ,-j_2 -j_1 -2k} \rrb$ \\
&$ |j_1 ,j_2 ,k,l > $
& \\
\hline
$C_- (11,11)$
& $ \sum_n \beta^n $
&$\fr{1}{\pi} \fr{\Gamma (2k_1)}{\Gamma (2k-k_1 -k_2 +1)\Gamma (2k+k_2 -k_1)} \times $ \\
& $\fr{(2k_1)_n}{(2k-k_1 -k_2 +1)_{-n} (2k+k_2-k_1 )_{-n} (1)_n }$
& $G^{10}_{03}\lrb -r^2 |^{1}_{2k_1 ,k_2 +k_1 -2k +1 ,1+k_1 -k_2 -2k } \rrb$ \\
&$ |k,k_1 ,k_2 ,n > $
& \\
\hline
$C_+ (11,11)$
& $ \sum_n \beta^n $
&$\fr{\Gamma (2k_1)\Gamma (k_1 -k_2 -2k+1)\Gamma (k_1 +k_2 -2k)}{\pi}\times $\\
& $\fr{(2k_1)_n (k_1 -k_2 -2k+1)_{n} (k_1 +k_2 -2k)_{n}}{ (1)_n }$
& $ G^{30}_{10}\lrb -r^2 |^{1}_{2k_1 ,k_1 -k_2 -2k +1 ,k_1 +k_2 -2k } \rrb$  \\
&$ |k,k_1 ,k_2 ,n > $
& \\
\hline
\end{tabular}
\end{table}


\setcounter{equation}{0}
\chapter{Applications of Polynomial Algebras to Physical Systems}
\markboth{}{Chapter 6.   Applications of....}
 In this chapter,  we give applications of the formalism that we have developed to many interesting physical systems.
 Not only do we present novel applications, but,  we also show how the work done in literature can be unified in our approach.

 In section 1, we shed new light on the
degeneracy structure of the quantum anharmonic oscillator that arises out
of the fact that the quadratic algebra is a symmetry algebra of
the system. We also give some applications to the quadratic oscillator system.

Since we have worked out the Boson realization of the algebras, a
natural application of non-linear algebra is to multiphoton
processes in quantum optics.
 In the section 2  we  give details of
some simple multiphoton processes where the formalism is useful.
We also compare our results with other works on multiphoton
processes \cite{brif,chumakov}.

 In the  section 3,  we give  an application of
cubic algebras  that exploits the fact that we get closed forms
for polynomial algebras only if  there are extra constants of
motion(constraints). This provides a natural framework to deal
with systems having accidental degeneracy and quantum versions of
superintegrable systems. We apply the formalism to two
representative systems of this type: the two dimensional singular
oscillator and the Calogero model. We also find the coherent
states of the Calogero model.

In section 4,  we generalize the work of Shifman using $SU(2)$
algebra to construct quasi-exactly solvable systems corresponding
to polynomial algebras.

\section{Some interesting properties of quadratic algebras}
In chapter 2 we have constructed four classes of
quadratic algebras and their unitary irreducible representations.
In the next subsection we apply some of the mathematical properties of the
quadratic algebra  to  establish a connection between the degeneracy
associated with the anisotropic quantum harmonic oscillator and the
theory of partitions.
\subsection{Quantum Anharmonic Oscillator}
Consider the Hamiltonian
\be
H = a_1^\da a_1 + a_2^\da a_2 + 2a_3^\da a_3 + 2 \lb{3daniso} \ee
which describes, in the units $\hbar = 1$ and $\omega = 1$, a
three dimensional anisotropic quantum harmonic oscillator with the
frequency in the third direction twice that in the perpendicular
plane.
We see that
 $Q^+(1,1)$ is the dynamical algebra of the system
.  From the representations , one can
easily arrive at the result that the spectrum of $H$ is the set of
all integers $\geq 2$.  Let us look at the invariance algebra of
$H$.
From the construction of $Q^-(1,1)$ we recognize that
\be
H = 4\cl + 1\,, \ee where $\cl$ is a central element of the
algebra generated by $\lrb Q_0, Q_\pm \rrb$ in eqn. [] or eqn[] of chapter .  Thus, $\lrb \ck , Q_0 , Q_\pm \rrb$ are the
integrals of motion for the system (\ref{3daniso}), or in other
words, $Q^-(1,1)$ is the invariance algebra of the system.  Since
$\cl$ has the spectrum
\be
\cl = l = n/4\,, \qquad n = 1,2,3,\ld\,, \ee it is clear that the
Hamiltonian (\ref{3daniso}) has the spectrum
\be
H = N+2\,, \qquad N = 0,1,2,\ld\,. \ee Each level can be labeled
by the eigenvalues of a complete set of commuting operators $\lrb
H-2 , \ck \rrb$.  It is interesting to compute the degeneracy of
the $N$-th level using the representation theory of the algebra
(\ref{q-11}).  For the $N$-th level the value of $\cl$ is $l =
(N+1)/4$. Calculating the corresponding values of $k$ for which
finite dimensional representations are possible we find that the
dimensions of the associated irreducible representations are
$(1,2,\ld\,,2m+1)$ if $N = 4m$ or $4m+1$\,, and $(1,2,\ld\,,2m+2)$
if $N = 4m+2$ or $4m+3$.  The degeneracy of the level is the sum
of the dimension of the $k = 1/2$ representation and twice the
dimensions of $k < 1/2$ representations.  One has to count the
dimensions of $k < 1/2$ representations twice in the sum since
there are two possible choices for the bases leading to the same
representation in these cases as already noted.  Now, the four
cases, $N = 4m, 4m+1, 4m+2$ and $ 4m+3$, are to be considered
separately.  The result is as follows\,: the degeneracies of the
levels, $N = 4m, 4m+1, 4m+2$ and $ 4m+3$, respectively, are
$(2m+1)^2$, $(2m+1)(2m+2)$, $4(m+1)^2$ and $2(m+1)(2m+3)$\,.  In
other words, the number of compositions of the integer $N$
(partitions with ordering taken into account) in the prescribed
pattern $n_1 + n_2 + 2n_3$, with the interchange of $n_1$ and
$n_2$ taken into account, is $(2m+1)^2$, $(2m+1)(2m+2)$,
$4(m+1)^2$ and $2(m+1)(2m+3)$, if $N = 4m, 4m+1, 4m+2$, and
$4m+3$, respectively.  It is to be noted that in this example the
sum of all the dimensions of the irreducible representations
associated with the given $l = (N+1)/4$ gives the number of
partitions of $N$ in the pattern $n_1 + n_2 + 2n_3$, disregarding
the interchange of $n_1$ and $n_2$.  This leads to the result that
the number of such partitions is $(m+1)(2m+1)$ for $N = 4m$ or
$4m+1$ and $(m+1)(2m+3)$ for $N = 4m+2$ or $4m+3$.  Thus, it is
interesting to observe this connection between a three dimensional
quadratic algebra and the theory of partitions.

It should be noted that if one can identify a given three
dimensional quadratic algebra as belonging to one of the four
classes we have considered then its representation theory can be
worked out immediately, at least partially.  For example, observe
that
\be
Q_0 = a^\da a\,, \qquad Q_+ = \fr{1}{\sqrt{3}}\lrb a^\da \rrb^3\,,
\qquad Q_- = \fr{1}{\sqrt{3}}\,a^3\,, \ee obey the algebra
\be
\lsb Q_0 , Q_\pm \rsb = \pm Q_\pm\,, \qquad \lsb Q_+ , Q_- \rsb =
-3Q_0^2 - 3Q_0 + 2\,. \ee This algebra is uniquely identified with
$Q^+(1,1)$ with $\cl$ $=$ $l = 1$ and $\ck$ $=$ $k(1-k) = -2$ (or
$k = 2$).  Correspondingly the algebra is seen to have the
infinite dimensional representation given by \bea Q_0 \lmd n \rra
& = & (n+1) \lmd n \rra\,, \nn \\ Q_+ \lmd n \rra & = &
(n+1)\sqrt{n+4} \lmd n+1 \rra\,, \quad Q_- \lmd n \rra =
n\sqrt{n+3} \lmd n-1 \rra\,, \nn
\\
  &   &  \qquad \qquad \qquad \qquad \qquad n = 0,1,2,\,\ld\,\,.
\eea The fact that the representations we have discussed are not
complete is clear from the following example. For a two
dimensional anisotropic quantum harmonic oscillator the
Hamiltonian is
\be
H = a_1^\da a_1 + 2a_2^\da a_2 + \fr{3}{2}\,, \ee in the units
$\hbar = 1$ and $\omega = 1$.  The invariance algebra of this
Hamiltonian is generated by
\be
Q_0 = \fr{1}{4}\lrb a_1^\da a_1 - 2a_2^\da a_2 + \half
\rrb\,,\quad Q_+ = \half\lrb a_1^\da \rrb^2 a_2\,, \quad Q_- =
\half a_1^2 a_2^\da\,, \ee and the algebra is given by \bea \lsb
Q_0 , Q_\pm \rsb & = & \pm Q_\pm\,, \nn \\ \lsb Q_+ , Q_- \rsb & =
& 3Q_0^2 + \half (H-3)Q_0
                           - \fr{1}{16} H(H+2) + \fr{3}{8}\,.
\lb{2daniso} \eea This algebra can be readily identified with
$Q^-(1,1)$ corresponding to $\cl$ $=$ $(H-1)/4$ and $\ck = k(1-k)
= 3/16$ or $k = 1/4$ or $3/4$.  It may be noted that the
corresponding representations cannot be presented in terms of the
three-boson Fock states since these, considered in Section 6,
correspond only to $k$ $=$ $1/2,1,3/2,\,\ld\,$.  To get the
representations of the algebra (\ref{2daniso}) one will have to
combine the representations of a boson algebra with the
representations of $su(1,1)$ for $k$ $=$ $1/4$ or $3/4$ (in terms
of single boson Fock states). A detailed discussion of the
algebraic approach to the two dimensional quantum system of an
anisotropic oscillator with an additional singular potential in
one direction is found in \cite{Le}.

\subsection{Quadratic Oscillator.}

An interesting possibility is suggested by the structure of the
algebra $Q^-(1,1)$.  Let us define
\be
N = Q_0\,, \quad A = \frac{1}{\sqrt{\cl(\cl+1)-\ck}}\,Q_-\,, \quad
A^\da = \frac{1}{\sqrt{\cl(\cl+1)-\ck}}\,Q_+\,. \ee Then the
algebra (\ref{q-11}) becomes \bea \lsb N , A \rsb & = & -A\,,
\quad \lsb N , A^\da \rsb = A^\da\,, \nn \\ \lsb A , A^\da \rsb &
= & 1 - \fr{2\cl-1}{\cl(\cl+1)-\ck}\,N
                            - \fr{3}{\cl(\cl+1)-\ck}\,N^2\,.
\lb{qo} \eea We may consider this as the defining algebra of a
quadratic oscillator, corresponding to a special case of the
general class of deformed oscillators (\cite{Ar}-\cite{D})\,:
\be
\lsb N , A \rsb = -A\,, \quad \lsb N , A^\da \rsb = A^\da\,, \quad
\lsb A , A^\da \rsb = F(N)\,. \ee The quadratic oscillator
(\ref{qo}) belongs to the class of generalized deformed
parafermions \cite{Q2}.  It should be interesting to study the
physics of assemblies of quadratic oscillators.  In fact, the
canonical fermion, with
\be
N = \lrb \ba{cc}
         0 & 0  \\
         0 & 1 \ea \rrb\,, \quad
f = \lrb \ba{cc}
         0 & 1 \\
         0 & 0 \ea \rrb\,, \quad
f^\da = \lrb \ba{cc}
          0 & 0  \\
          1 & 0  \ea \rrb\,,
\ee is a quadratic oscillator!  Observe that
\be
\lsb N , f \rsb = -f\,, \quad \lsb N , f^\da \rsb = f^\da\,, \quad
\lsb f , f^\da \rsb = 1 - \half N - \fr{3}{2} N^2\,.
\ee

\section{ Applications of polynomial algebras to multiphoton processes}
 Many quantum optical processes are described by multiphoton Hamiltonians of the form
 \begin{equation}
H=\frac{1}{m+n}(a_1^{\dagger}a_1+a_0^{\dagger}a_0) +
\kappa(a_0)^m(a_1^{\dagger})^n+c.c
\end{equation}
For the general class of these Hamiltonians the symmetry algebra
does not close on a finite dimensional Lie algebra hence ordinary
Lie algebras are too restrictive to find a complete solution for
the degenerate eigenvectors and the coherent states of the
process. Many authors {\cite {brif,chumakov,kara}, have used
 ordinary linear Lie
algebraic methods leading to approximate results for specific
cases. Infinite dimensional Lie algebraic techniques have also
been attempted and the physics has been extracted by a truncation
of these algebras, hence the results obtained have again been
approximate, with a number of assumptions \cite{brif}.

However, since we have found the Bosonic realization of Non-Linear
algebras we notice that if we define  $N_0,N_-,N_+$ in such a way
that
 \bea N_+&=&a_0^m(a_1^{\dagger})^n \nonumber\\
N_-&=&a_1^n(a_0^{\dagger})^m \nonumber\\
N_0&=&\frac{1}{m+n}(a_1^{\dagger}a_1-a_0^{\dagger}a_0) \eea then
n-dimensional polynomial algebras act as the symmetry algebra of
the Hamiltonian system provided
\be
H_0=\frac{1}{m+n}(a_1^{\dagger}a_1+a_0^{\dagger}a_0)
\ee
is a
constant of motion or invariant of the system.
In a large class of quantum optical systems , this does actually
occur as there exists a class of invariants of the system called
 Rowe's Invariants  which are combinations of the photon number
 operators multiphoton system.
The Hamiltonian $H_0$ can be
expressed as a linear combination of these invariants.
 In fact it is the existence of such an invariant for a two mode process
governed by the Hamiltonian
\begin{equation}
H=\frac{1}{2}(a^{\dagger}a+b^{\dagger}b) +
\kappa(ab^{\dagger})+\kappa^{*}(a^{\dagger}b)
\end{equation}
that allows the use of the SU(2) algebra as the symmetry algebra.
Here the invariant is $(a^{\dagger}a+b^{\dagger}b)$ which is the
polarization sphere of the two modes.

\subsection{Trilinear Bosonic Hamiltonian}
To illustrate how the Bosonic realizations of the quadratic
algebra developed by us is useful we consider first trilinear
Boson Systems governed by a tri-Boson Hamiltonian of the form
\be
H= \omega _{a} a^{\dag}a + \omega _{b} b^{\dag}b + \omega _{c}
c^{\dag}c + \kappa a b^{\dag}c^{\dag} + \kappa^{*} a^{\dag} bc .
\ee
 Raman and Brillouin scattering can be described by $H$, if $a$, $b$
and $c$ represent input, vibration and Stokes modes for a Stokes
process and anti-Stokes, input and vibration modes for an
anti-Stokes process($\bar h =1$units).
 This Hamiltonian also describes the parametric amplification
if $a$,$b$ and $c$  represent the pump, signal and idler modes.
 In a frequency conversion process $a$,$b$ and $c$ are
the idler, pump and signal modes .

For the Hamiltonian given in equation (4), let $a$ represent a pump system and $b$ and $c$ represent
the signal and idler variables. The interaction Hamiltonian between
the pump and signal-idler subsystem is given by
\be
H_{int} = \kappa a b^{\dag}c^{\dag} + \kappa^{*} a^{\dag} bc .
\ee
Energy conservation requires that
$\omega_{a} = \omega_{b} + \omega_{c}$.
If the signal and idler frequencies are equal then
$\omega_{b}=\fr{\omega _{a}}{2}$ and $\omega _{c} =\fr{\omega _{a}}{2}$.
Thus
\be
H_0 = \omega _a  (a^{\dagger } a +\frac{ b^{\dagger} b + c^{\dagger }c}{2})
\ee

The operators define the generators of the polynomial quadratic algebra
\be
Q_0
    =\fr{1}{2} (a^{\dag}a- K_{0})
\ee
\be
Q_{-}=\kappa a\, b^{\dag}c^{\dag}
     =\kappa a\, K_{+}
\ee
\be
Q_{+}=\kappa a^{\dag}bc
     =\kappa a^{\dag} K_{-}
\ee
where $K_{0}$, $ K_{-}$ and $K_{+}$ form  SU(1,1) generators.
The algebra closes only  if we define an additional conserved quantity  $\cl$ given by :
\be
L = \fr{a^{\dag}a + K_{0}}{2}=\frac{  (a^{\dagger } a +\frac{ b^{\dagger} b + c^{\dagger }{c}+1}{2})}{2}
\lb{inv}
\ee
Thus
\be
L=\frac{H_0}{2\omega_a}+\frac{1}{4}=\frac{\epsilon}{2}+\frac{1}{4}
\ee
where $\epsilon=\frac{H_0}{\omega_a}=2L-\frac{1}{2}$.

The Manley-Rowe invariants of the system are \bea M_{ab}&=&
a^{\dag}a + b^{\dag}b \\ \nn M_{ac}&=&a^{\dag}a +  c^{\dag}c \\
\nn
 M_{bc}&=& b^{\dag}b - c^{\dag}c
 \eea
Thus we see that as a result of the frequency relations generated
by energy conservation
\be
H_0=\omega_a\frac{M_{ab}+M_{ac}}{2}=\omega_a(a^{\dag}a +\frac{
b^{\dag}+ c^{\dag}c}{2}) \ee and is in fact an invariant of the
system . Hence the symmetry algebra of the system is a Quadratic
Algebra.
The algebra is given by:
\be
[Q_{+}, Q_{-}] = 3Q^{2}_{0} +(2 L -1) Q_{0}  + C_{bc}(K_{0}) -L(L
+ 1) \ee Where
$C_{bc}=\frac{1}{4}-\frac{(b^{\dag}b-c^{\dag}c)^2}{4}=\frac{1-Q^2}{4}$
is the Casimir operator for the idler-signal system , for which L
is a conserved quantity.

 In most physical cases, considered in literature,
$b^{\dag}b-c^{\dag}c=0$ , $C_{bc}=\frac{1}{4}.$ This is a special
case of the Quadratic algebra, $Q^- (1,1)$ for which $C_{bc} (K_0
)=k(1-k)=\frac{3}{4}$, thus the Bargmann index $k=\fr{1}{2}$. Now
it is straight forward to find the representation by substituting
$k=\fr{1}{2}$ in (\ref{q11-rep}). The basis states are given by $
\mid \fr{1}{2}, l,n > $, where $n=0,1,... 2l +\fr{1}{2} $. Here
$l$ is the eigen value of L. Since L and $\epsilon$ are related it
is instructive to label the representations as
$|\epsilon,\frac{1}{2},n>$. With these quantum numbers the
representations  are given by,
 \bea Q_0 \lmd
\frac{1}{2},\epsilon,n \rra &=&  ( n-\frac{\epsilon}{2}) \lmd \fr{1}{2} ,\epsilon,n \rra   \\
\nn Q_- \lmd \fr{1}{2} ,\epsilon,n \rra &=&
\sqrt{(n+1)^2 (\epsilon -n ) } \lmd
\fr{1}{2} ,\epsilon,n-1 \rra  \nn \\ Q_+ \lmd \fr{1}{2}
,\epsilon,n \rra &=& \sqrt{(n)^2 (\epsilon+1 -n ) } \lmd \fr{1}{2} ,\epsilon,n+1 \rra \eea
The Casimir operator for this representation is given by
\be
C= \lrb (\epsilon-1)\rrb ^{2} \lrb \epsilon -\frac{1}{2}\rrb \ee

They form a finite dimensional unitary irreducible representation of
dimension $\epsilon+1 $.

Using the analytic representation  of the generators given by (\ref{q11-an})
for $Q^{-}(1,1)$, the eigenstates of H can be found analytically as a solution of the differential equation
\be
(\kappa^{*}z \ddzt +(\kappa^{*}
-\kappa z^2 )\ddz + \epsilon( z +\omega)
)\psi(z)=E\psi(z)
\ee

The Perelomov type coherent states (in the sense of the ones
constructed in chapter 5) of this Boson Hamiltonian with
$k=\frac{1}{2}$ are given by the particularly simple form

\be
|\alpha,\epsilon>=N\sum_{n=0}^{\epsilon} (\alpha)^n
\sqrt{\frac{(\epsilon)! (n)!}{n!(\epsilon-n)!}}|\epsilon,n>. \ee

The Normalization coefficient $
N(|\alpha|^2,\epsilon)=(\frac{1}{|\alpha|^2\epsilon!})^{\frac{1}{2}}
e^{\frac{-1}{|\alpha|^2}}
(\Gamma(\epsilon,\frac{1}{|\alpha|^2}))^{-\frac{1}{2}}$

The resolution of the identity is given by:
\begin{equation}
\int d \sigma(\alpha,\alpha^{*};\epsilon) |\alpha;\epsilon
\rangle\langle \alpha;\epsilon| = \hat{1} .
\end{equation}

With a polar decomposition ansatz $\alpha=re^{i\theta}$ we get,
\begin{equation}
\frac{1}{2}\int_{0}^{\infty} d(r^2)\, \sigma(r^2) \;
(r^2)^{(n+1)-1} = \frac{1}{2\pi} \frac{ \Gamma(\epsilon-(n+1)} {
\Gamma(\epsilon+1)}.
\end{equation}
 $\sigma(r^2)=\frac{(r^2-1)^{n-1}}{\Gamma(n)(r^2)^{n+\epsilon}}$ is found by an inverse Mellin transform
 by the method discussed in chapter 5.

These are useful in studying the time evolution of the states of
the system.

\subsection{The Dicke model and quadratic algebras}
 The Dicke model in quantum optics
 describes the interaction of the
radiation field with a collection of identical two-level atoms
located within a distance much smaller than the wavelength of the
radiation.  In the particular case when the atoms interact
resonantly with a single mode coherent cavity field, the
(Tavis-Cummings) Hamiltonian, under the electric dipole and
rotating wave approximations, is given by (in the units $\hbar =
1$)
\be
H = \omega\lrb J_0+a^\da a \rrb + gJ_+a + g^*J_-a^\da\,.
\lb{tch}
\ee
where $\omega$ is the frequency of the field mode (and the atomic transition),
$J_0+a^\da a$ is the excitation number operator, and $g$ is the coupling
constant.  The annihilation and creation operators $a$ and $a^\da$ correspond
to the single mode radiation field.  The operators
\be
J_0 = \sum_j = \sigma^j_0\,, \qquad J_\pm = \sum_j =
\sigma^j_\pm\,, \ee with $\sigma^j_{0,\pm}$ as mutually commuting
triplets of the Pauli matrices, obey the $su(2)$ algebra and
define the collective atomic operators.  From purely physical
arguments it is possible to construct the matrix representation of
the Hamiltonian (\ref{tch}) and find its spectrum exactly or
approximately (using numerical methods) depending on whether the
number of excited atoms is small or large \cite{TC}.  However,
from an algebraic point of view, it seems that the spectrum
generating, or dynamical, algebra of the Hamiltonian (\ref{tch})
has not yet been identified precisely.  We now recognize the
dynamical algebra of the Hamiltonian (\ref{tch}) as $Q^-(2)$. The
generators are : \bea Q_0 &=&\frac{ a^{\dagger }a-J_0}{2} \nn
\\ Q_+ &=& a^{\dagger} J_+  \nn \\ Q_- &=& a J_- \nn \\ \cl &=&
\frac{a^{\dagger }a + J_0 }{2}\lb{6-2-3} \eea where,
\be
J_{0,\pm } = \sum_j S^{(j)}_{3,\pm }, \ee The quadratic algebra
satisfied by the above generators is \bea \lsb Q_0 , Q_{\pm } \rsb
&=& \pm Q_{\pm } \nn \\ \lsb Q_+ , Q_- \rsb &=& 3Q^{2}_{0}+(1-2\cl
)Q_0 + \cl (\cl -1 ) -j(j+1) \eea The total Hamiltonian in terms
these generators is given by
\be
H= 2\omega \cl + \kappa(Q_+ + Q_-) \ee. Thus the  quadratic
algebra given by $Q^-(2)$ in chapter 2  (\ref{6-2-3}) is the
dynamical algebra of the Dicke model .
\bea Q_0 \lmd j,l,n \rra &=& n \lmd j,l,n \rra  \nn \\ Q_+ \lmd
j,l,n \rra &=& \lsb (n+1)(j-l+n )(j+l+1-n )\rsb^{1/2} \lmd j,l,n1
\rra \nn \\ Q_- \lmd j,l,n \rra  &=&  \lsb (n+1)(j-l+1+n )(j+l-n
)\rsb^{1/2} \lmd j,l,n-1 \rra  \nn \\ \cl  \lmd j,l,n \rra &=& l
\lmd j,l,n \rra \eea. $l$ is the eigenvalue of the excitation
number operator $\cl=a^{\dag}a+J_0$. This is a $l+j+1$ dimensional
representation. Hence if $s$ is the eigenvalue of $a^{\dag}a$ we
are restricted to the (s+1) dimensional subspace spanned by the
basis vectors .

 The two cases
of representations (
\ref{q+2})-
\ref{q2-rep1})
we have found are
exactly the ones identified in the literature from a physical
point of view.  Further, the excitation number operator
(corresponding to our $\cl$) and $J^2$ are known integrals of
motion.  However, the fact that the third generator to be
augmented to $J_+a$ and $J_-a^\da$ to construct a closed quadratic
algebra is $J_0-a^\da a$ (corresponding to our $Q_0$) has not been
recognized so far . We hope that the precise identification of the
dynamical algebra of the Dicke model would help its further
understanding. In this regard, our earlier proposal of a general
method for constructing the Barut-Girardello-type and
Perelomov-type coherent states of any three dimensional polynomial
algebra \cite{Su} should be useful.

 The time evolution of the state is given by
\be
\lmd \psi (t)= \rra e^{-iHt}   \lmd \psi _0 \rra  \nn \\ \ee where
$\lmd \psi _0 \rra $ is the eigen state of the initial Hamiltonian
$H_0 $. i.e.
\be
H_0 \lmd \psi _0 \rra = E_0 \lmd \psi _0 \rra \ee . We have \bea
\lmd \psi (t) \rra &=& e^{-it(H_0 +H_I )} \lmd \psi _0 \rra \nn \\
&=& e^{(-itE_0 )} e^{(-i\kappa t) Q_+ -(-it)^* Q_- } \lmd \psi _0
\rra \eea. If $\lmd \psi_0 \rra $  is the lowest weight state of
the dynamical algebra then up to a phase factor one can write as
\bea \lmd \psi (t) \rra  &=& \lmd \beta \rra      \nn \\
                    &=& U(\beta ) \lmd 0 \rra   \nn \\
                    &=& e^{\beta Q_+ -\beta ^* Q_- } \lmd 0 \rra
\eea where $\beta =-i\kappa t $ . Which upto a normalization
factor can be seen to be the Perelomov type state given in chapter
5.

\section{Superintegrable Systems}

 Superintegrable systems in N dimensions have more than N
independent classical constants of motion, while maximally
superintegrable systems in N dimensions have 2N$-$1 independent
classical constants of motion, N of which are in involution.
Mathematically,  for a super integrable system,  not only are
there N independent integrals of motion $\{X_i,i=1..n\}$ in
involution $\{X_i,X_j\}=0$ , but, there exist an additional m
integrals of motion $\{Y_j,j=1..m\}$ which have vanishing Poisson
brackets with H but not with each other. If $m=n-1$ the system is
superintegrable. Another interesting fact about superintegrable
systems is that they are separable two or more co-ordinate
systems. The concepts of complete integrability and
superintegrability have their analogue in quantum mechanics. In
quantum mechanics , the Poisson brackets are replaced by
commutators and a superintegrable quantum mechanical system is
described by m+n quantum observables.
 These
additional constants of motion introduce additional degeneracies
in their quantum mechanical versions, which are known as
accidental degeneracies. Often , for such systems the spectra are
not linear , but quadratic or have higher orders and thus the
Casimir operator is taken as a Hamiltonian and the advantages of
using Lie algebras as a dynamical symmetries are lost. To use the
dynamical symmetry properties effectively, it is necessary to
extend the Lie algebra to polynomial algebras in which the
Hamiltonian is one of the diagonal generators. In this way, one
can relate the degree of degeneracy to the dimensions of the
representation and find out how the additional symmetry generators
transform one degenerate eigenstate to another. For this it is
useful also that additional symmetry generators close on a finite
algebra, which in most cases is a quadratic algebra.

In the construction of cubic algebras from products of SU(1,1) and
SU(2) generators, we find that,  the algebra closes on a cubic
algebra only when we impose an extra constraint (corresponding to
an additional invariant) of the system. Thus,  the cubic algebra
dynamical symmetry is present only for systems which do have
additional accidental degeneracy and thus this is related to the
classical concept of superintegrability. As an illustrative
example of the application of these algebras we consider the two
dimensional singular oscillator which has 4 quadratic constants of
motion. We show that a cubic algebra is the symmetry algebra of
this system and that we can construct the additional invariants
from a combination of generators of the cubic algebra which among
themselves satisfy a quadratic Hahn algebra. We also construct the
spectrum generating algebra of this system.

We consider the two dimensional singular oscillator with the Hamiltonian $ H$ is given by
\begin{equation}
H = -1/2 (\delta_x ^2 + \delta^2_y ) + \frac{w^2}{2} (x^2 + y^2 ) + \frac{\mu _1}{x^2} + \frac{\mu _2}{y^2}.
\end{equation}
For simplicity we take the case $m=\omega=h=1$

This constitutes one of the four quantum (and also classical) Hamiltonians characterized by 2 integrals of motion quadratic in momenta other then H.
This is thus a maximally super integrable system.
We show that its symmetry algebra is a cubic algebra and that we can construct the invariants from a combination of the generators of the cubic algebra, hence we know how
the degenerate eigenstates transform under the action of the symmetry operators.

Using the creation and annihilation operators  corresponding to the two individual oscillators we have:
\begin {eqnarray}
H & = & a_1 ^{\dagger} a_1 + 1/2 + \frac{2\mu _1 }{a_1 + a_1 ^{\dagger})^2 }
+ a_2 ^{\dagger} a_2 + 1/2 + \frac{2\mu _2 }{a_2 + a_2 ^{\dagger})^2 } \\ \nn
  & = & H_1 + H_2
\end{eqnarray}
One can associate a cubic algebra by
defining
\begin{eqnarray}
L_+ & = & 1/2 ( (a^{\dagger}_1)^2 - \frac{2 \mu _1 }{(a_1 + a^{\dagger}_1)^2} \\ \nn
L_- & = & 1/2 ((a_1)^2 -\frac{2 \mu _1 }{(a_1 + a^{\dagger}_1)^2} \\ \nn
L_0 & = & H_1 /2 \\ \nn
M_+ & = & 1/2 ( (a^{\dagger}_2)^2 - \frac{2 \mu _2 }{(a_2 + a^{\dagger}_2)^2} \\ \nn
M_- & = & 1/2 ((a_2)^2 -\frac{2 \mu _2 }{(a_2 + a^{\dagger}_2)^2} \\ \nn
M_0 & = & H_2 /2 \\ \nn.
\end{eqnarray}
L and M satisfy the $ SU(1,1 )$ algebra
\begin{equation}
[L_{\pm} , L_0 ]=\pm L_{\pm} , [L_+ , L_- ] = -2 L_0
\end{equation}
Now consider the operators
\begin{eqnarray}
C_+ & = & L_+ M_- \\ \nn
C_- & = & L_- M_+ \\ \nn
C_0 & = & 1/2(L_0 - M_0 ) \\ \nn
K=  & = & 1/2 (L_0 + M_0 ) \\ \nn
\end{eqnarray}
They close to give the cubic algebra
\begin{equation}
[C_+ ,C_- ] = -4C_0^3 + (4K^2 -3/2 +  \mu_1 + \mu_2 )C_0 +\\
(\mu_2 -\mu -1 )K
\end{equation}
Where the Bargmann index $k_1$ and $k_2$ are given by
\begin{equation}
k_1 =\frac{1\pm \sqrt{1/4 + \mu_1 }}{2} ,k_2 = \frac{1\pm \sqrt{1/4 + \mu_ 2 }}{2}
\end{equation}
Note that $k =H/4 $, the total Hamiltonian.

One can generate a quadratic algebra from the above cubic algebra . The
generators of the quadratic algebra is given by
\begin{eqnarray}
Q_1 & = & \frac{C_++C_-}{2} +  C_0^2  \\ \nn
Q_2 & = & C_0 \\ \nn
Q_3 & = & \frac{C_--C_+}{2}
\end{eqnarray}
\begin{eqnarray}
[Q_1 , Q_2]  & = & Q_3 \\ \nn
[Q_2 , Q_3 ] & = & -Q_1 + Q_2^2 \\ \nn
[Q_3 ,Q_1]   & = &  [Q_2 , Q_1]_+ +(3/4 -(\mu_1 + \mu_2 )/2 -2K^2 )Q_2 + (\mu _2 -\mu_1)/2 K
\end{eqnarray}

The operators $Q_1,Q_2 and Q_3$ are the additional invariants of
the system beside the Hamiltonian. This is called the Hahn algebra
( or often the quadratic Askey -Wilson algebra)

To see that these indeed correspond to the additional invariants of the system
we write $Q_1,Q_2,Q_3$ in co-ordinate space representations.
\begin{eqnarray}
Q_2 & = & H_1-H_2= \\ \nonumber
& =& -1/2 (\delta_x ^2)  + \frac{w^2}{2} (x^2)  + \frac{\mu _1}{x^2}
+1/2 (\delta_y ^2)  - \frac{w^2}{2} (y^2)  - \frac{\mu _2}{y^2}
\end{eqnarray}
Which is the quantum symmetric version of the traditional classical constant
\be
A=p_x^2+\omega^2 x^2+\frac{\mu_1}{x^2}
\ee
Since K is a commuting generator we can subtract from above multiples of K to get
the conventional conserved quantities corresponding to the two dimensional singular oscillator,
and in fact $A=Q_2-K$

\bea
Q_1 &=&
\left(x p_y - y
p_x\right)^2+r^2\left(\frac{\mu_1}{x^2}+\frac{\mu_2}{y^2}\right)
+K^2
\eea
Which again is the quantum symmetric version of the classical constant of motion
\be
B=\left(x p_y - y
p_x\right)^2+r^2\left(\frac{\mu_1}{x^2}+\frac{\mu_2}{y^2}\right)
\ee
For the singular oscillator these $Q's $ satisfy the quadratic algebra.

Since for most quantum systems the structure of the symmetry
algebra is fairly easy to identify, it becomes a simple matter to
construct the invariants of systems with a polynomial algebra
symmetry. The problem on the classical level is much more
difficult and many papers have been devoted to the search for
these invariants  \cite{evans}.

This quadratic Hahn algebra structure has the interesting property
that the overlap functions between the eigenstates of $Q_1$ and
$Q_2$ can be expressed in terms of the Hahn polynomials \cite{zed}
. Since $Q_2$  is diagonal in the polar co-ordinate system and
$Q_1$ is diagonal in the elliptical co-ordinate system we obtain
an overlap between the wave functions in the polar and elliptical
co-ordinate system in terms of the Hahn polynomial.

\section{ Dynamical symmetry and coherent states of  the two body Calogero model}
 The Calogero model represents the interaction of N particles in
a line by a Harmonic and inverse square potential. It has become a
paradigm for integrable models in Physics and In the last decade a
large volume of literature has been devoted to it. For references
see \cite{gurrappa}. The dynamical algebra of this model as an
$S_n$ extended Heisenberg Algebra has been studied. A cubic
polynomial symmetry as an invariance algebra of the Two body
Calogero sutherland model has already been established
\cite{vinet}, here the generators of the cubic algebra commute
with the Hamiltonian. We show another cubic polynomial symmetry of
the two body Calogero model in which the Hamiltonian is one of the
generators of the algebra and thus this is the dynamical symmetry
of the two- body Calogero model and the generators are ladder
operators that take one energy level to another.
 We study the
Applications of Polynomial algebra to this model.

The Hamiltonian for the two body Calogero Sutherland model
is given by
\be
H_c = \fr{1}{2} \sum_{i=1}^{2} (\fr{\partial^2}{\partial x^2 }
+ \omega ^2 x^{2}_i +\sum_{j<i} \fr{\lambda (\lambda -1)}{(x_i -x_j )^2 }
\ee

 This Hamiltonian has been shown to be factorized by
defining the creation and annihilation operators, \bea
a^{\dagger}_i &=& \fr{1}{\sqrt{2}} (-D_i + \omega x_i )  \nn
\\ a_i &=& \fr{1}{\sqrt{2}} (D_i + \omega x_i ) \eea
Where $ D_i$ is the Dunkel derivative
\be
D_i = \fr{\partial}{\partial x } + \lambda \sum_{i\neq j}\fr{1}
{(x_i -x_j )} (1-\sigma_{ij}) \ee and $\sigma_{ij}$,is the
exchange operator which satisfies the following relations
\be
\sigma_{ij} x_j =x_i \,\,\,\sigma_{ij} , \,\,\,\,
\sigma_{ij}=\sigma_{ji} ,\,\,\,\, (\sigma_{ij})^2 =1 , \ee and
generates the symmetry group $S_N$. In terms of the $a_i $ and
$a^{\dagger}_{i}$
\be
H= \fr{1}{2} \sum_{i=1}^2 [a_{i}^{\dagger} , a_i]_+ \ee In the
centre of mass coordinates defined by \bea y_i    &=& M_{ij} x_j
,\\ M_{ij} &=& \fr{1}{\sqrt{2}} \left[ \begin{array}{cc} -1 & 1 \\
 1 & 1
\end{array}
\right] \eea the corresponding creation and annihilation operators
are given by
\be
\tilde{A}_i = M_{ij} a_j , \,\, \tilde{A}^{\dagger}_i = M_{ij}
a^{\dagger}_j \ee they satisfy the algebra

 \bea \lsb A_1 ,
A^{\dagger}_1 \rsb &=& 1+2\lambda \sigma \;\;\; \lsb A_2 ,
A^{\dagger}_2 \rsb = 1 \\ \nn  \lsb A_1 , A^{\dagger}_2 \rsb &=&
0 \;\;\; \lsb A_2 , A^{\dagger}_1 \rsb = 0 \\ \nn \lsb A_1 , A_2 \rsb
&=& 0 \lsb A^{\dagger}_1 , A^{\dagger}_2 \rsb =0 \eea

But one can get two Bosonic algebra by redefining
\bea
A_1 &=& \fr{\tilde{A}_1 }{\sqrt{1+ 2\lambda \sigma }} \nn \\
A^{\dagger}_1 &=& \fr{\tilde{A}^{\dagger}_1 }{\sqrt{1+ 2\lambda \sigma }} \nn \\
A_2 &=& \tilde{A}_2 , \,\, A^{\dagger}_2 =\tilde{ A}^{\dagger}
\eea
The Hamiltonian in terms of the $A_i$s are given by
\be
H= \fr{1}{2} \lrb [ A_1 , A^{\dagger}_1 ]_+ (1+2\lambda \sigma ) +
[ A_2 , A^{\dagger}_2 ]_+ \rrb \ee Now consider the operators \bea
C_0 &=&\fr{1}{2} (A^{\dagger}_1 A_1 -A^{\dagger}_2 A_2 ) \nn \\
C_+ &=& \fr{1}{2} (A_{1}^{\dagger} A_2 )^2 \nn \\ C_- &=&
\fr{1}{2} (A_{2}^{\dagger} A_1 )^2 \nn \\ J   &=&  \fr{1}{2}
(A^{\dagger}_1 A_1 + A^{\dagger}_2 A_2 ) \eea The Hamiltonian in
terms of the operator can written as
\be
H= (2\lambda \sigma ) C_0 + (1+\lambda \sigma )(2J +1) \ee The
operator $C_+ ,C_- ,C_0 $and $J$ satisfying the algebra \bea \lsb
C_0 , C_{\pm} \rsb &=& C_{\pm} \nn \\ \lsb C_+ , C_- \rsb &=& -2
C^{3}_0 + (2C(J)-1 )C_0 \eea.

This is significantly different from the algebra generated in reference
\cite{vinet}, as the Hamiltonian is one of the generators. Here,
C(J)=J(J+1) is the Casimir of the underlying $SU(2)$ algebra. An
interesting feature of the above algebra is that it has the same
structure as the cubic Higgs algebra, which was one of the first
polynomial algebras to be studied. Thus there is a similarity in
the symmetry structure of the Calogero system and the Coulomb
system on curved space.

  Note that the above
algebra has a one mode $SU(2)$ realization, given by \bea C_0 &=&
J_0 \nn
\\ C_+ &=& \fr{1}{2} J^2_+ \nn \\ C_- &=& \fr{1}{2} J^2_- \eea
In this case the unitary irreducible space will be the $SU(2)$
even states, $|j,2m
>$.  Explicitly the
representation is given by \bea C_+ |j,m>&=& [
(j-m)(j+m+1)(j-1-m)(j+2+m) ]^{1/2} |j,m+2> \nn \\ C_- |j,m> &=&
[(j+m)(j-m+1)(j-1+m)(j+2-m)]^{1/2}|j,m-2> \nn \\ C_0 |j,m> &=&
m|j,m > \eea.
 Since  we can map the above algebra to a $SU(2) $
algebra  \cite{gurrappa} , using the algorithm given in the
previous chapter,  the coherent states in this representation are
given by \bea | \beta > &=& Ne^{\beta C_+ } |j, -j> \nn \\
         &=& N\sum_n \fr{(16 \beta )^n}{n!} \lsb \fr{(\fr{j}{2} +1)_n (\fr{j+1}{2} )_n }{(j+1)_{-n} (j+\fr{1}{2} )_{-n} } \rsb^{1/2} |j, -j+2m>.
\eea. As in the previous cases one can find by inverse Mellin
transform the resolution of the identity in polar coordinates
where the  measure $\sigma (r^2 ) $ is given by
\be
\fr{1}{\pi}\sigma (r^2 ) = \Gamma (-j-1  ) \Gamma (-j-\fr{1}{2}
),\Gamma (\fr{j}{2}) \Gamma (\fr{j-1}{2}) G^{0,2}_{4,0}\lsb -(4
r)^2 |-j-1 , -j-\fr{1}{2} ,\fr{j}{2} , \fr{j-1 }{2} \rsb \ee

Like the two dimensional singular oscillator , one can also find a
quadratic algebra structure in the Calogero model by defining the
operators
\begin{eqnarray}
Q_1 & = & \frac{C_++C_-}{2} +  gC_0^2  \\ \nn Q_2 & = & C_0 \\ \nn
Q_3 & = & \frac{C_--C_+}{2}
\end{eqnarray}
where $g=(\frac{2C(J)-1}{4})^{\frac{1}{2}}$. These satisfy the
quadratic algebra
\begin{eqnarray}
[Q_1 , Q_2]  & = & Q_3 \\ \nn [Q_2 , Q_3 ] & = & -Q_1 + g Q_2^2 \\
\nn [Q_3 ,Q_1]   & = & g [Q_2 , Q_1]_+ + Q_2.
\end{eqnarray}
This again is a special case of the quadratic Askey Wilson algebra
or the Hahn algebra.

\section{Algebraic coherent states of polynomial algebras and related
quasi-exactly solvable models}

Some time ago, it was shown by Shifman \cite{shifman}, Turbiner
\cite{Turbiner} and others that if a Hamiltonian $H_{"G"}$ can be
expressed as the polynomial  combination of the generators  of a
Lie algebra ,  whose representations are known, then by using
\ref{hg} and \ref{gp} one can find the potential, which are
solvable. If the Lie algebra has  only finite dimensional
representations, in particular $SU(2)$, then the corresponding
potentials not completely solvable but partially solvable. They
are known as quasi exactly solvable systems in literature. The
details of the method can be find in \cite{shifman}.
 We have seen in chapter 2 and chapter 3
that  for the polynomial algebra, the generators themselves
products of Lie algebra generators and possess finite and infinite
dimensional representations. So one can use LINEAR COMBINATIONS of
 polynomial algebra generators to construct  quasi exactly solvable
systems and polynomial combinations of polynomial algebra
generators to construct new  quasi exactly solvable systems going
beyond second order systems .

The construction of QES systems in the context of Polynomial
algebras, reduces the arbitrariness involved in the above
construction because the known QES systems appear as
 algebraic coherent states. Newer exactly QES systems can also be
 constructed in this approach.
 Thus it is useful to show the method of construction of algebraic
 coherent states of polynomial algebras. Many examples can be
 worked out with all the 4 cases of quadratic algebras we have
 presented and the 10 cases of cubic algebra.
 Listing of all possible QES systems we can construct would entail
 another thesis, hence we demonstrate the method with an example.

 First we briefly describe the method of Shifman.
 Given a Hamiltonian $H_G$ which is a polynomial combination of
 Lie algebra generators ,using the differential representation of
 the generators  (upto second order derivatives )it can be brought to
 the form
 \be
H_{"G"}= -\fr{1}{2} \fr{d^2}{dx^2 } + A(x) \fr{d}{dx } + \Delta V,
\lb{hg} \ee where
\be
\Delta V = V(x) + \fr{1}{2} \fr{dA}{dx} -\fr{1}{2} A^2 \lb{gp} \ee

$H_{"G"} $ can be written as \bea H_{"G"} &=& -\fr{1}{2}
\lrb\fr{d}{dx} -A(x) \rrb ^2 + V(x) \nn \\ \eea.
 Then  an imaginary
phase transformation,
\be
\Psi = \tilde{\Psi} e^{-a(x)}, \ee  with $ a(x) =\int A(x) dx$
will bring  the Hamiltonian to the form, \be H = -\fr{1}{2}
\fr{d^2}{dx^2} + V(x). \ee.

Now we go on to how the algebraic coherent states of cubic
algebras can be used to construct QES models.
 The algebraic
coherent state was first defined for the $SU(1,1)$ algebra by
Trifonov \cite{trif}. They are defined as the eigen state of the
linear operator $u K_- + v K_+ + w K_0 $, where $K_0,K_+$ and $K_-
$ are the generators of the $SU(1,1)$ algebra. The constants $u,v$
and $w$ are complex in general. For $SU(1,1)$, the algebraic
coherent states(ACS) are given by
\be
(uK_- + vK_+ +wK_3)|\psi\ra = \beta |\beta , u,v,w \ra, \ee By
using the differential representation of SU(1,1) (BG
representation) the eigenvalue equation becomes a second order
linear differential equation for analytical functions $\phi(z)$ of
growth $(1,1)$, where $\phi_z(z;u,v,w)=\la k;z^*|z;u,v,w;k\ra$ can
be found exactly and turns out to be the confluent hypergeometric
function \cite{trif}.

The case u=0,v=0 gives the Barut Girardello states and we can
recover the Perelomov CS as a subset of $|\beta,u,v,w=0;k\ra$: if
we put \be w = 0, \quad {\rm and}\quad   \beta=-k\sqrt{-uv} \ee in
$|\beta,u,v,w;k\ra$ then we get the CS $|\tau;k\ra$,
$\tau=\sqrt{-v/u}$. At $w=0$ the conditions are reduced to
$|v/u|<1$ so that the whole family of Perelomov CS is recovered by
the ACS $|\beta,u,v,w=0;k\ra\equiv |\beta,u,v;k\ra$.\\

As our demonstrative case,we will extend this construction to the
three dimensional polynomial algebras with cubic algebras as a
special case. Consider the cubic algebra of case $C_+(11)$ of
chapter 3. For the cubic algebra the algebraic coherent states are
given by

\be
(uC_- + vC_+ +wC_0)|\psi\ra = \beta |\psi\ra.
\ee

By using the differential representation for the case $C_+(11,11)$
cubic algebra, the above equation becomes, \bea
u\mu(z\frac{d^2}{dz^2} +2k_1\frac{d}{dz})\psi(z,\beta)+v\mu
z(z\frac{d}{dz} -2k-k_2)(z\frac{d}{dz}-2k+k_2+1)\psi(z,\beta) \nn
\\ +w(z\frac{d}{dz}+k-k_1)\psi(z,\beta)  =\beta\psi(z,\beta) \eea

This equation is of the form
\be
H\psi(z,E)=-\frac{1}{2}P_2(z)\frac{d^2}{dz^2} \psi(z,E) +P_1(z)
\frac{d}{dz}\psi(z,\beta)+P_0(z)\psi(z,E)=E\psi(z,E),\label{h} \ee
Where $P_2(z)=-2(uz +vz^3)$, $P_1(z)=2vz^2(2k+k_2-1) +2uk_1+wz)$,
$P_0(z)=-vz((2k+k_2)(2k-2k_2-1)+w(k-k_1)$

 This can be reduced to a
Schr$\ddot{o}$dinger type form by a method
  of gauge transformation developed by Shifman in the
context of quasi-exactly solvable system\cite{shifman}. Returning
to the Hamiltonian given in equation (\ref{h}), we find that for
general u,v and w solution is difficult (it can be done in terms
of elliptic functions) to solve, but, for special cases, we can
explicitly solve the equations in a simple fashion to get exactly
solvable quantum mechanical systems for which the potential takes
on a simple forms, these are the cases when either u=0,v=0 or w=0.
Furthermore, the  algebraic coherent states can be used to get the
exact eigenfunctions for these systems. For $v=0$ we have
\be
(uz)\frac{d^2}{dz^2} \psi(z,\beta) +( +2uk_1+wz)\frac{d}{dz}\psi(z,\beta)+(+w(k_1-k))\psi(z,\beta)=\beta\psi(z,\beta)
\ee
whose solution is
\be
\psi(\frac{1}{z},\beta)= N e^{\frac{c}{z}}M(a(\beta),b,\frac{c_1\beta}{z})
\ee

\bea
a = (k_1-k)+\frac{\beta}{w},\,\,\quad b=2k_1,\nonumber\\
c=-\frac{w}{2u},\quad c_1=\frac{w}{u}.
\eea

With $E=\beta$ this is a solution to the equation
$H\psi(E,z)=E\psi(E,z)$
with
\be
H=-\frac{1}{2}P_2(z)\frac{d^2}{dz^2}  +P_1(z)\frac{d}{dz}+P_0(z)
\ee Where $P_2(z)=-2(uz)$, $P_1(z)= +2uk_1+wz)$,
$P_0(z)=+w(-k+k_1)$.\\ The above operator can be transformed into
a $H_{"G"}$ type operator given by,
\be
H_{"G"} = -\fr{1}{2} \fr{d^2}{dx^2 } + (\fr{wx}{2} + (\fr{1}{2} -
2k_1 )\fr{1}{x} ) \fr{d}{dx} + (k_1 -k ) w \ee

 The gauge potential can be identified as
\be
A(x) =  \fr{wx}{2} + (\fr{1}{2} - 2k_1 )\fr{1}{x} \ee The
imaginary phase factor is given by

\bea a(x) &=& \int dx A(x) \nn
\\
     &=& (\fr{1}{2} -2k_1 ) lnx +   \fr{wx^2}{2}
\eea

The solution of the eigen value equation
\be
H\psi(x,E)= E\psi(x,E) \ee
 can be constructed from the algebraic coherent state and the
 "gauge transformation"to be
\be
\psi(\frac{1}{z},\beta)= N
x^{(2k_1-\frac{1}{2})}M(k_1-k+\frac{E}{w},2k_1,\frac{-wx^2}{2})
\ee The potential for which the algebraic coherent state is eigen
state is given by
\be
V(x) = wk + \fr{w^2 x^2}{8} + \fr{4k^{2}_1 -\fr{1}{4}}{2x^2 }
\ee
which is a singular oscillator.

This  potential belongs to the categories of potentials
constructed by Shifman  by using combinations of $SU(2)$
generators. What we have shown is that it can be constructed in a
natural fashion by using the algebraic coherent states of the
cubic algebra. If we apply the same procedure using the algebraic
coherent states of the quadratic algebra then one can find QES
potentials which  are not in the list given in Shifmans list  as
the quadratic algebra generators are not products of SU(2) or
SU(1,1) generators but that of a Heisenberg algebra and SU(2) or
SU(1,1) algebra.

 A plethora of potentials can be obtained from the
algebraic coherent states of  all the cubic  and quadratic
algebras presented in chapter 2 and 3 algebra. A catalogue of
these would be illuminating and is the subject of our future
work.\\

To conclude this , we have presented a number of physical systems
in which the  explicit matrix ,differential representations and
coherent states of quadratic and cubic algebras find a useful
application. We have not exhausted all the possibilities of
applications due to volume of their large number, but, we have
demonstrated on a few representative systems , how the rich and
varied structure of these polynomial algebras can be put to
fruitful use. In the concluding chapter we will give future
directions inspired by these applications.


\setcounter{equation}{0}
\chapter{CONCLUSION}
\markboth{}{Chapter 7.   Conclusion}
We conclude this thesis by briefly summarizing the work done
and suggest further avenues of investigation opened up by our preliminary study.

We have seen that polynomial algebras emerge as the dynamical
symmetry , invariance and spectrum generating algebras of many
interesting physical systems . Some of these have been discussed
in the last chapter. Among these,  a special position is occupied
by three dimensional polynomial algebras with a coset structure.
Thus, a systematic study and the proper classification of these
algebras and their irreducible representations was warranted. Such
a comprehensive study was carried out in detail in this thesis.

The three dimensional polynomial algebras  are classified as
separate entities  according to their order as the quadratic
algebra, cubic algebra, quartic algebra etc. The investigation of
the relation of the quadratic and cubic algebras to ordinary three
dimensional Lie algebras revealed the fact that they  can be
constructed from the generators of the Lie algebras subject to
additional constraints required to close the algebra. This
construction is not unique but gives rise to  many different
classes of quadratic and cubic algebras.
 The
unitary irreducible representations (UIRs) of the quadratic and
cubic algebras were found by  taking a product state of the
representations of the corresponding Lie algebras and  putting
 constraints on it  coming from the construction of the polynomial algebra.
Such a construction leads to  both finite and infinite dimensional
representations. Different classes of the cubic algebra can also
be constructed using products of a Heisenberg algebra and a
quadratic algebra. Differential realizations in the space of
analytic functions follow immediately from the construction.

The similarities between our construction of quadratic algebras
and the well known Jordan Schwinger realization of the $SU(2)$ and
$SU(1,1)$ algebras throws some light in the direction of
generalization of our construction to higher order algebras. Such
a generalization allowed us to generate a chain of polynomial
algebras starting from a Lie algebra.

The immediate application of polynomial algebras was to construct
the  various coherent states associated with these algebras. Since
it has been observed that the polynomial algebras appear as the
dynamical symmetry algebra or the spectrum generating algebra  of
many interesting physical systems,  the corresponding coherent
states provide tools for studying the dynamical evolution of these
systems. For the cases that admit infinite dimensional UIRs,  the
Barut-Girardello (Lowering operator) type coherent states are
easily constructed by using the analytic representations of the
algebra. However, for finite dimensional representations we are
hampered by the fact that BCH (Baker- Campbell Hausdorff) type
formulas for the polynomial algebras are highly difficult to
calculate . We are able to overcome this limitation by using a
unified approach for the construction of various types of coherent
states by constructing Lie algebra generators from the polynomial
algebra in such a way that one can exploit BCH formulae available
for Lie algebras. This can be done both for the finite and
infinite dimensional cases and many applications have been given.

It has been seen that the physical systems like the two body
Calogero Sutherland model,  the Dicke model of N two level atom
interacting with single mode radiation field and  the anisotropic
singular oscillator  contain a polynomial algebraic structure
which can be mapped  to the quadratic and cubic algebras that we
have studied . The coherent states of such system can easily be
found. These algebras  also shed new light on the degeneracy
structure of superintegrable systems  and allow for a natural
construction of additional invariants in these systems.

As with all research investigations, more questions than answers
are generated at the end of the study. The end itself is the
beginning of new directions. Some generalizations immediately come
to mind. The first is to go beyond three dimensional polynomial
algebras and study higher dimensional polynomial algebras. One
that would find physical application is that arising from the
polynomial extension of SU(3). This would be an eight dimensional
polynomial algebra which would contain the three dimensional
algebra as its subalgebra. Some steps in this direction have been
taken by De Boers et. al. \cite{tjin} , but further investigations
of the representations and physical applications will be
illuminating. Since the representations of the SU(3) algebra have
been extensively used in particle and nuclear physics, many body
problems in these areas would require the study of such algebras
.

Another interesting application of these algebras is the
construction of gauge theories based on these algebras. A simple
gauge theoretic investigation based on a finite $W_3$ algebra (a
quadratic algebra) has been done by Schoutens et. al.
\cite{shou1, shoutens}. The preliminary study of these new gauge theories
has been done only at the classical level. Additional gauge
couplings will be introduced because of the non-linearity inherent
in these algebras and this might lead to interesting one loop and
quantum effects. It will also be of interest to study the gauge
theories based on a cubic algebra, in particular integrable
theories like the Chern Simons theories and self dual Yang Mills
theories, since the cubic algebra structure is intimately related
to the quantum versions of classically superintegrable theories.
For detailed studies, one would need the finite dimensional
irreducible representations of polynomial algebras given in this
thesis.

An area in which these algebras have found an application which we
have not presented in this thesis is   supersymmetric quantum
mechanics. Isospectral potentials of the Harmonic oscillator
obtained by the factorization method have ladder operators that
obey a quadratic algebra \cite {F,  Ro, sukh}. By the same
argument,  potentials isopectral to a system with SU(2) or SU(1,1)
symmetry would have ladder operators that obey a higher
dimensional polynomial algebra. The investigations of these
families of potentials would require the use of the UIR's of the
cubic algebra worked out in this thesis. Furthermore, in the
supersymmetric generalization of polynomial algebras (non-linear
supersymmetric polynomial algebra) , the spectrum need not contain
a boson for every fermion, giving rise to new speculations beyond
the standard model.

The list of future problems is by no means exhaustive.  What we
have learnt in the process of this investigation is that we need
no longer be bound to the linear concept of symmetry. Since
non-linearity is inherent in most physical systems, one has to
expand the tools by which we exploit the symmetry of a system to
encompass non-linear symmetries. Although many complicated
non-linear structures have now been studied, the polynomial
algebra is one of the simplest non-linear extensions of the Lie
algebra. Yet, albeit its simplicity, it still yields a plethora of
new structures and simplifies the study of a large class of
physical phenomena. We hope that our study will prove useful in
providing the tools for more extensive studies.


\end{document}